\documentclass{aa}  
\usepackage{graphicx}
\usepackage{txfonts}
\usepackage[colorlinks,citecolor=blue,urlcolor=blue,filecolor=blue,linkcolor=blue]{hyperref}
\usepackage{amsmath}
\usepackage{amssymb}
\usepackage{upgreek}
\usepackage{natbib}
\usepackage{xcolor}
\usepackage[export]{adjustbox}
\usepackage{booktabs}
\usepackage{multirow}
\usepackage{subcaption}

\makeatletter
\renewcommand\paragraph{%
  \@startsection{paragraph}{4}{\z@}{1.5dd \@plus1\p@ \@minus0.5\p@}%
  {-\fontdimen2\font \@plus-\fontdimen3\font \@minus-\fontdimen4\font}%
  {\normalsize\sffamily}}
\renewcommand\section{%
  \@startsection{section}{1}{\z@}{-10dd \@plus-3\p@ \@minus-3\p@}{4dd}%
  {\large\sffamily\bfseries\boldmath}}
\renewcommand\subsection{%
  \@startsection{subsection}{2}{\z@}{-9dd \@plus-3\p@ \@minus-3\p@}{3dd}%
  {\normalsize\sffamily\itshape}}
\renewcommand\subsubsection{%
  \@startsection{subsubsection}{3}{\z@}{-8dd \@plus-2\p@ \@minus-2\p@}{3dd}%
  {\normalsize\sffamily}}
\makeatother

\begin{document} 
    \title{LOFAR follow-up of sources in the First LHAASO Catalogue}
   
    \author{M.\ Arias\inst{1,2}
        \and
        T.\ Shimwell\inst{2}
        \and
        M.J. Hardcastle\inst{3}
        \and
        R.\ Timmerman\inst{4,5}
    }

    \institute{
        Instituto de Astrof\'isica de Andaluc\'ia (IAA-CSIC), Glorieta de la Astronom\'ia s/n, 18008, Granada, Spain\\
        \email{marias@iaa.csic.es}  
        \and
        ASTRON Netherlands Institute for Radio Astronomy, Oude Hoogeveensedijk 4, 7991\,PD Dwingeloo, The Netherlands
        \and
        Centre for Astrophysics Research, Department of Physics, Astronomy and Mathematics, University of Hertfordshire, College Lane, Hatfield, AL10 9AB, UK
        \and
        Centre for Extragalactic Astronomy, Department of Physics, Durham University, Durham, DH1 3LE, UK
        \and
        Institute for Computational Cosmology, Department of Physics, Durham University, South Road, Durham, DH1 3LE, UK
    }

    \date{Received \today; Accepted }

    \abstract{The First LHAASO Catalogue lists 90 sources of very-high-energy gamma-ray emission, many without an established counterpart at other frequencies, reopening the question of which Galactic accelerators produce the highest-energy cosmic rays. We search for low-frequency radio counterparts to the 55 1LHAASO sources that fall within the Galactic fields of the LOFAR Two-metre Sky Survey Data Release Three (LoTSS-DR3), using 144~MHz continuum maps at $6''$ and $20''$ resolution, complemented with
archival data from other radio surveys where available. We present an overview of the radio emission towards all 55 sources, and identify the structures potentially associated with the gamma-ray emission. We report 24 new supernova remnant (SNR) candidates, eight flat-spectrum radio shells with faint or no infrared counterpart, and six sources with no clear Galactic radio emission present in the LOFAR maps. A bootstrap analysis shows that 23 of the 50 sources in the Galactic plane footprint (46\%, for $|b| \leq 5$) overlap a catalogued SNR, whereas random placement yields only $10.0\pm2.8$ matches even if the underlying Galactic SNR population is ten times larger than currently catalogued; in the outer Galaxy, where chance alignment is rarest, 4 of 12 sources coincide with an SNR against a null expectation of $\lesssim2$. We estimate that at least a quarter, and up to half, of the 1LHAASO Galactic-plane sources genuinely reside in SNR environments, consistent with a population dominated by pulsar wind nebulae, SNR--PWN composites, and cosmic-ray-illuminated clouds. Individual results include a new SNR candidate inside the TeV shell HESS~J1912+101, a radio candidate for the SNR hypothesised to explain the GeV emission of 1LHAASO~J1945+2424, and the first proposed counterparts of 1LHAASO~JJ0056+6343u and 1LHAASO~J2200+5643u.}

    \keywords{ISM: supernova remnants -- Gamma rays: ISM -- Radio continuum: ISM -- Surveys -- Pulsars: general -- Acceleration of particles}

    \maketitle
    
\section{Introduction}

Ultra-high-energy (UHE) gamma-rays, those with energies exceeding 100~TeV, have become observable thanks to large particle air-shower facilities such as the Tibet Air Shower Array \citep{amenomri08}, the High Altitude Water Cherenkov observatory \cite[HAWC,][]{abeysekara17}, and the Large High Altitude Air Shower Observatory \cite[LHAASO;][]{aharonian21}. Whereas a few years ago this region of the electromagnetic spectrum was essentially unexplored, detections in this band have now become routine.

The First LHAASO Catalog of Gamma-ray Sources \cite[][hereafter C24]{cao24} lists 90 sources that are detected above 0.1~TeV (the very high energy, VHE, domain), 43 of which display emission extending into the UHE domain. The LHAASO field of view covers the Milky Way between approximately $l=10^\circ-220^\circ$ in Galactic longitude; finding so many of these sources in the relatively sparse northern Milky Way proved to be a surprise.
The results first presented in \cite{cao21} triggered a renewed debate over the picture of cosmic-ray (CR) production in the Galaxy, which has traditionally considered 
supernova remnants (SNRs) as the primary PeVatrons, or engines capable of accelerating CRs up to a few PeV \citep{ginzburg64,hillas05}.
Yet, despite the wealth of MeV–TeV gamma-ray detections associated with SNRs \cite[e.g.,][all for SNR~W44]{abdo10,giuliani11,uchiyama12,ackermann13,peron20,abe25}, several other classes of astrophysical systems have now emerged as competitive candidates for Galactic PeVatrons.
Pulsar wind nebulae \cite[PWNe;][]{abeysekara20,cao21}, extended TeV halos around pulsars \citep{linden17,lopez-coto22}, large structures such as superbubbles \citep{abeysekara21}, microquasars \citep{lhaaso25}, and even millisecond pulsars \citep{cao25} could all meaningfully contribute to the Galactic CR population.
A comprehensive, multiwavelength characterisation of LHAASO-identified sources has become essential for clarifying the origin of their gamma-ray emission and the nature of the underlying particle accelerators.

The LOFAR Two-metre Sky Survey\footnote{\url{http://lofar-surveys.org/}} \cite[LoTSS,][]{shimwell17,shimwell19,shimwell22,shimwell26} has observed and imaged 88\% of the northern sky, including a substantial fraction of the Milky Way between $l=30^\circ-210^\circ$. 
This makes LoTSS an ideal tool with which to look for counterpart radio emission to the LHAASO-identified gamma-ray sources in C24, from here on referred to as the 1LHAASO sources, as many of the proposed counterparts produce synchrotron emission, which dominates in the LOFAR bands. In fact, the existing LoTSS Galactic fields cover 55 of the 90 1LHAASO sources (and 22 out of the 43 UHE sources). Of the 35 sources not covered, 14 lie at $l<30^\circ$, at declinations below the LoTSS limit of $0^\circ$; the remaining 21 are at accessible declinations but fall in regions not yet imaged, and can be observed once LOFAR 2.0 comes online. In particular, the 13 that lie in the Galactic plane will be covered by the upcoming LOFAR 2.0 Galactic Plane Survey (L2GPS), which will contiguously cover $l=30^\circ-210^\circ$.

This paper aims to search for low-frequency radio counterpart candidates to the 55 sources in the 1LHAASO catalogue that are covered in LoTSS in order to help identify their physical counterparts. In Sect. \ref{sec:data} we present the LoTSS Galactic fields, the existing observations of the Milky Way that have been made public via the LoTSS survey. In Sect. \ref{sec:results} we search the LoTSS data for coincident radio emission for all 55 1LHAASO sources that are within the footprint of the survey. In Sect. \ref{sec:discussion}, we discuss the results, which include the identification of 24 new SNR candidates within or near the 1LHAASO source regions, and a number of extended sources with a seemingly flat radio spectrum that lack infrared counterparts. We also quantify the statistical significance of the coincidence between the 1LHAASO sources and the Galactic SNR population via a bootstrap analysis, finding that the SNR alignment cannot be explained by chance under any realistic assumption about the incompleteness of the SNR census. In Sect. \ref{sec:conclusions} we summarise our findings.

\section{Data}
\label{sec:data}

Here we present the LoTSS Galactic fields, as well as other public ancillary data products we make use of in this work.

\subsection{The LoTSS DR-3 Galactic fields}

This work is heavily anchored on the LoTSS dataset. LoTSS is a survey of the Northern sky carried out with the LOFAR High Band Antennas (HBA) at $120-168$~MHz. By the end of LOFAR operations in summer 2024 (on account of the upgrade of the array's electronics that will result in LOFAR 2.0), LoTSS had surveyed 88\% of the Northern sky. The full LoTSS coverage data has been made available, in the form of images with $6''$ and $20''$ resolution among other data products, as Data Release Three \cite[DR-3][]{shimwell26}.

Out of the $\sim$$12$\% of the Northern sky that has not been observed as part of LoTSS, the majority of the missing regions are at Galactic latitudes $|b|<20^\circ$. However, a substantial fraction of the Milky Way has been observed as part of LoTSS DR-3, including two long contiguous segments between Galactic longitudes $l=30^\circ-77^\circ$ and $l =123^\circ-180^\circ$, as well as smaller ones around $l=97^\circ-106^\circ$, and $l=190^\circ-209^\circ$. We refer to these regions, presented in this paper, as the LoTSS Galactic Fields.

The images in LoTSS-DR3 have a $uv$-minimum of 100~m; this corresponds to a largest angular scale of $\sim$$1^\circ$ at LOFAR HBA frequencies. This means that for many of the sources considered here we cannot report a value of the flux density. While in the future we plan to re-image some of the most interesting sources that come out of this exploration with dedicated imaging strategies aimed at recovering the diffuse emission with as much fidelity as possible, for the purpose of this paper we will limit ourselves to reporting the presence of sources that could potentially be associated with the 1LHAASO emission.

\begin{figure*}[!tp]
    \centering
    \includegraphics[width=\textwidth,height=0.155\textheight,keepaspectratio]{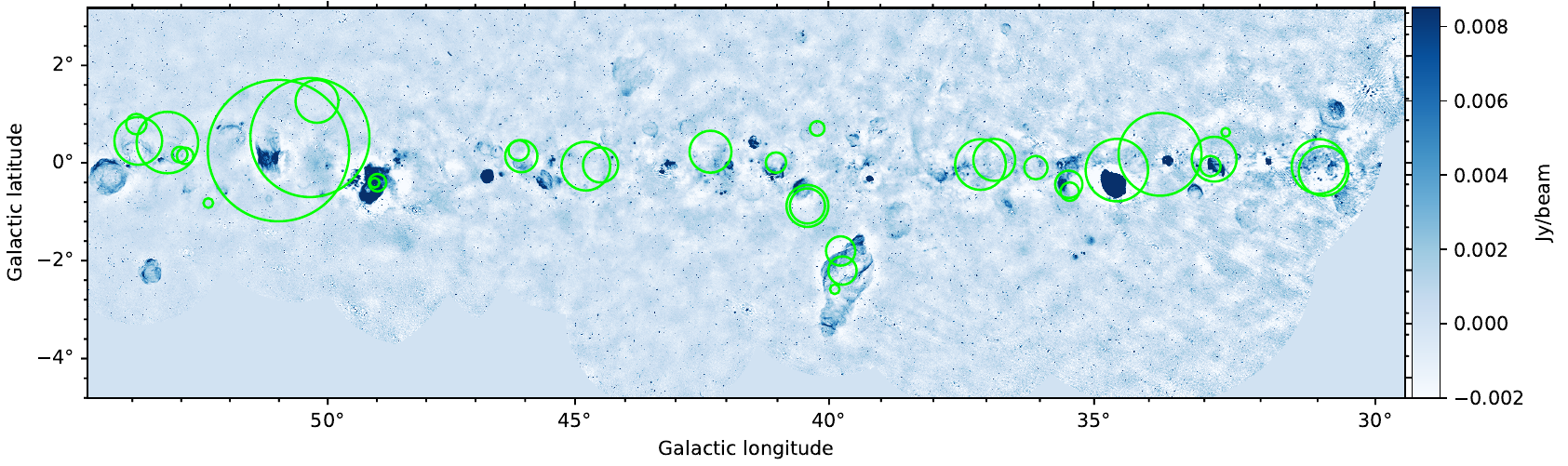} \\
\includegraphics[width=\textwidth,height=0.155\textheight,keepaspectratio]{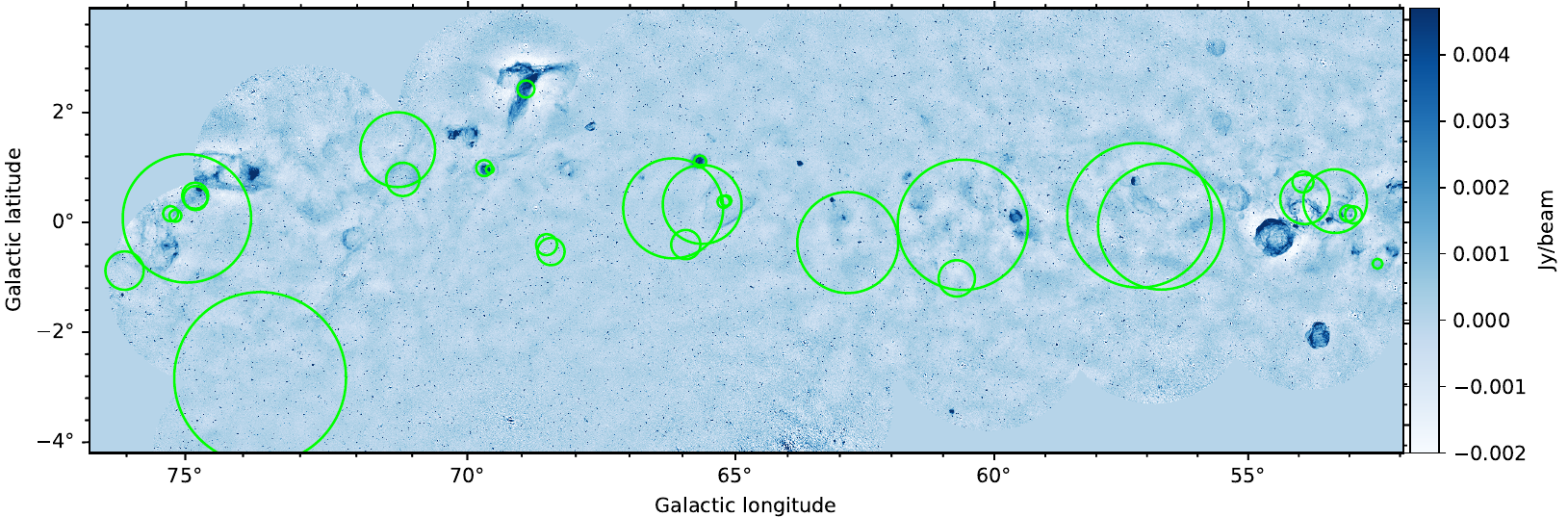} \\
\includegraphics[width=\textwidth,height=0.155\textheight,keepaspectratio]{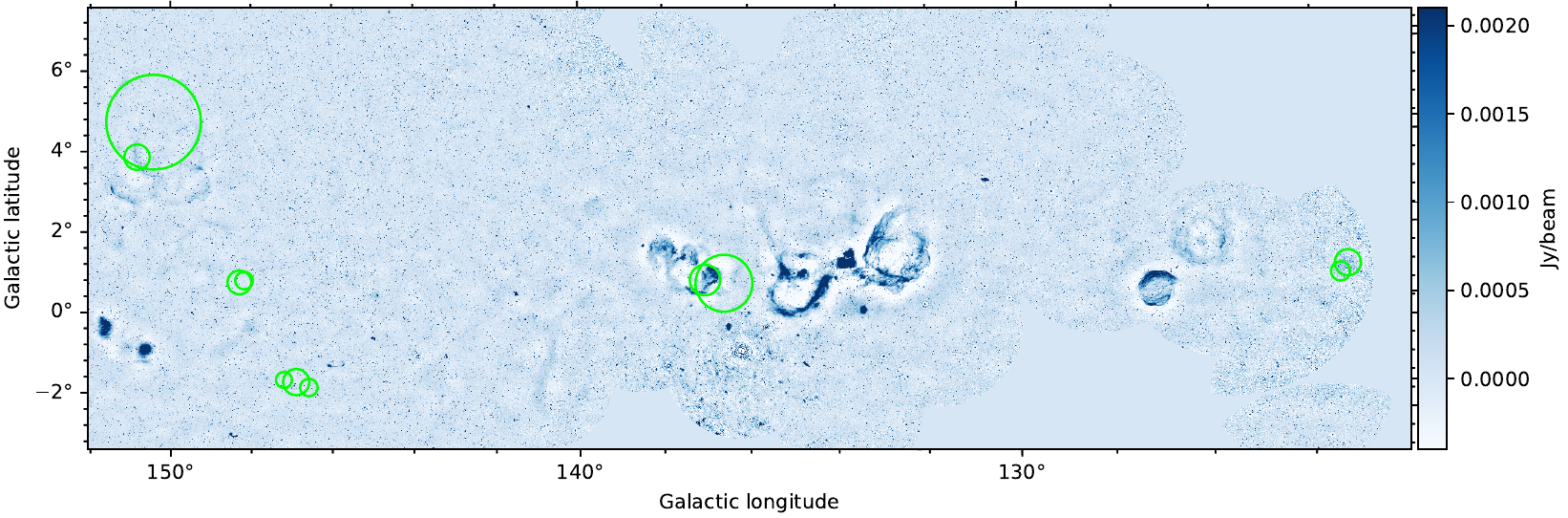} \\
\includegraphics[width=\textwidth,height=0.155\textheight,keepaspectratio]{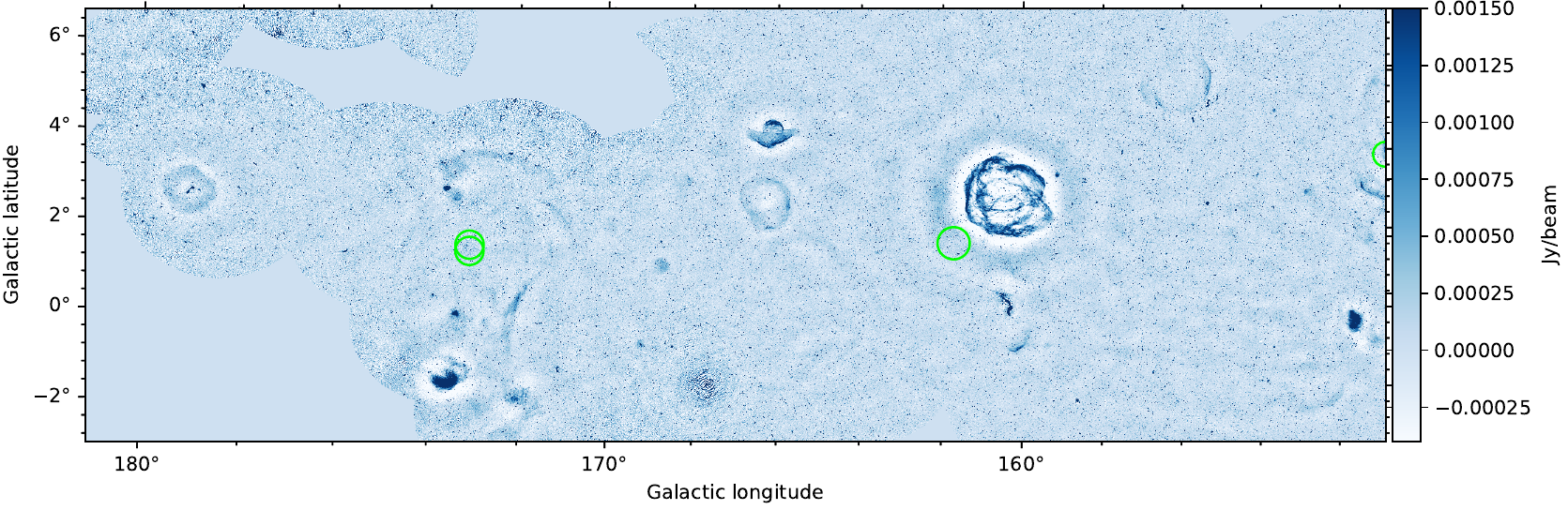} \\
\includegraphics[width=\textwidth,height=0.155\textheight,keepaspectratio]{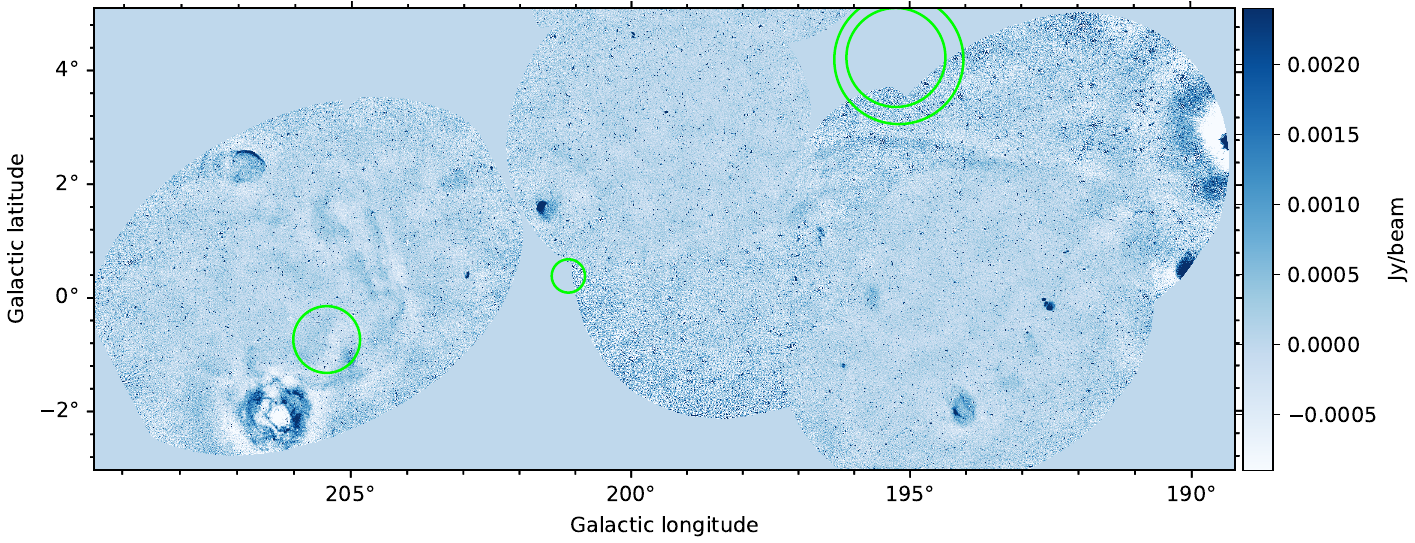}
    \caption{Galactic Fields in the LOFAR Two-metre Sky Survey (LoTSS), displayed to $20''$ resolution (although the region has been imaged to $6''$ resolution). The top two panels show the region of contiguous coverage in the first Galactic quadrant, the third and fourth panels the contiguous coverage in the second Galactic quadrant, and the bottom panel the coverage of the third Galactic quadrant. The green circles are the 1LHAASO sources catalogued in C24, the radius is $r_{39}$, the 39\% (or 1$\sigma$) containment radius of the two-dimensional Gaussian model (Table 2 of C24). Sources detected by both LHAASO detectors contribute one circle per component, of generally different radius and centre. }
    \label{fig:GF_cont}
    \label{fig:GF_thirdQ}
\end{figure*}

\subsubsection{The first Galactic quadrant}

The region of contiguous Galactic coverage in the first quadrant, between $l=30^\circ-77^\circ$, is shown in Figure \ref{fig:GF_cont}, top two panels. The coverage in Galactic latitude is uneven: for this band, the narrowest coverage occurs at $l\approx46^\circ$, with $b$ only reaching values between $-2.5^\circ<b<4.5^\circ$, and the broadest is at $l\approx69^\circ$, with $-7.5^\circ<b<4.0^\circ$. This difference is due to the observational strategy of LoTSS, which did not prioritise an even coverage in Galactic latitude.

The most complex regions of emission, corresponding to the lowest Galactic longitudes surveyed ($l\gtrsim30^{\mathrm{o}}$), are also close to the LoTSS declination limit of zero degrees. This means that the most challenging emission to image (at low Galactic longitudes, where there is abundant, bright emission, with complex structure on many scales that is often not encompassed in the sky models available for calibration) is also subject to low declination effects, such as an elongated synthesised beam, a lower instantaneous sensitivity due to station projection, an uncertain primary beam response, and a less contiguous $uv$-coverage due to the impossibility of conducting a full 8 hour synthesis \cite[see e.g.][]{hale19}. This implies that the regions around $l\gtrsim30^{\mathrm{o}},\ b \approx 0$ have some of the highest rms noise in the survey, with $500-800$~$\mu$Jy~bm$^{-1}$, unlike the $\sim$$80$~$\mu$Jy~bm$^{-1}$ of typical extragalactic fields \citep{shimwell22}.

The region surveyed here has been covered by multiple Galactic plane surveys at a variety of frequencies. The radio surveys whose footprint overlaps with the LoTSS Galactic fields are summarised in Table \ref{tab:surveys}. The Very Large Array Galactic Plane Survey \cite[VGPS,][]{stil06} observed the Galactic plane between $18^\circ<l<66^\circ$ at 1.4~GHz. More recently, this region has been observed by The H I, OH, Recombination line survey \cite[THOR,][]{beuther16} and the SARAO MeerKAT Galactic Plane Survey \cite[SMGPS,][]{goedhart24}; combining the low-frequency observations carried out by the LoTSS with the breadth of available centimeter-range data will allow for detailed spectral index studies (provided reimaging is done with consistent imaging parameters for the different surveys) at high resolution --in particular, the MeerKAT's resolution in the SMGPS, $8''$, is close to that provided by the LoTSS data products at $6''$.

At the longitudes that our work covers, in the first quadrant, our line of sight intersects with the Sagittarius and Perseus Arms of the Galaxy \citep{hou14}. Even at the lowest Galactic longitude that we observe, at $l=30^\circ$, the emission that we see in the LoTSS Galactic Fields shows less complexity of structure than seen in other surveys in and around the Galactic centre \cite[e.g.,][]{goedhart24}. \cite{stil06} noted that the rapidly increasing complexity of structure towards lower Galactic latitudes reflects the increased star formation rate towards the inner Galaxy. In our continuum mosaics, the complexity of emission and the density of extended sources generally diminishes as we move from the inner Galaxy to the Galactic anti-centre, with some of the Galactic fields looking very similar to the LoTSS extragalactic fields. Given the 100~m $uv$-cutoff applied when reducing the LoTSS data and the high resolution of the LoTSS images, the diffuse, large scale extended emission from the Galactic thin disk \citep{beuermann85} is resolved out, and hence not captured in our images.

\renewcommand{\arraystretch}{0.95}
\begin{table*}[t]
\centering
\caption{Radio Galactic plane surveys whose footprint overlaps the LoTSS Galactic fields.}
\label{tab:surveys}
\begin{tabular}{@{}ccccc@{}}
\toprule
Survey & Overlapping coverage & Frequency & Resolution & Reference \\
\hline
CGPS & \begin{tabular}[c]{@{}c@{}}all $52<l<192$\\ $-6.5<b<+8.5$\end{tabular} & 408~MHz & $3'$ & \cite{tung17} \\
\hline
CGPS & \begin{tabular}[c]{@{}c@{}}all $52<l<192$\\ $-3.5<b<+5.5$\end{tabular} & 1.4~GHz & $1'$ & \cite{taylor03} \\
\hline
VGPS & \begin{tabular}[c]{@{}c@{}}$30^\circ$–$67^\circ$\\ $b$ varies\end{tabular} & 1.4~GHz & $1'$ & \cite{stil06} \\
\hline
MAGPIS\tablefootmark{a} & \begin{tabular}[c]{@{}c@{}}$l=30^\circ$–$48^\circ$\\ $|b|<0.8$\end{tabular} & 1.4~GHz & $5''$ & \cite{helfand06} \\
\hline
THOR & \begin{tabular}[c]{@{}c@{}}$l=30^\circ$–$67^\circ$\\ $|b|<1$\end{tabular} & $1$–$2$ GHz & $20''$ & \cite{beuther16} \\
\hline
SMGPS & \begin{tabular}[c]{@{}c@{}}$l=30^\circ$–$61^\circ$\\ $|b|<1.5$\end{tabular} & 1.3~GHz & $8''$ & \cite{goedhart24} \\
\hline
GLOSTAR\tablefootmark{b} (pilot) & \begin{tabular}[c]{@{}c@{}}$30^\circ$–$36^\circ$\\ $|b|<1$\end{tabular} & $4$–$8$ GHz & $1.5''$ & \cite{brunthaler21} \\
\hline
GLOSTAR (planned) & \begin{tabular}[c]{@{}c@{}}$30^\circ$–$60^\circ$\\ $|b|<1$\end{tabular} & $4$–$8$ GHz & $1.5''$ & \\
\hline
Sino-German $\lambda$6~cm survey & all $l$, $|b|<1$ & 5~GHz & $9.5'$ & \cite{sun07} \\
\hline
CORNISH\tablefootmark{c} & \begin{tabular}[c]{@{}c@{}}$l=30^\circ$–$65^\circ$\\ $|b|<1$\end{tabular} & 5~GHz & $1.5''$ & \cite{hoare12} \\
\bottomrule
\end{tabular}
\tablefoot{
\tablefoottext{a}{Multi-Array Galactic Plane Imaging Survey.}
\tablefoottext{b}{Global View of Star Formation in the Milky Way.}
\tablefoottext{c}{Co-Ordinated Radio `N' Infrared Survey for High-mass star formation.}
}
\end{table*}

\subsubsection{The second Galactic quadrant}

The second region of continuous Galactic coverage, between $l=123^\circ-180^\circ$, is shown in Fig. \ref{fig:GF_cont}, second and third panels. There is another smaller region observed in the second Galactic quadrant at $l=97^\circ-106^\circ$ (not shown here). For the second Galactic quadrant the coverage in $b$ of the LoTSS Galactic fields is patchier than in the first quadrant, with, for instance, $l=125^\circ$ having only the thickness of a single LoTSS pointing. 

The area towards the Galactic anti-centre has not been observed as extensively in the radio as the first Galactic quadrant, but there is one important legacy resource that covers it: the Canadian Galactic Plane Survey \cite[CGPS,][]{taylor03}, which includes rare survey observations below 1~GHz with an angular resolution of a few arcminutes (see Table \ref{tab:surveys} for details).

In the second Galactic quadrant we can see features associated with the continuation of the Perseus Arm, and with the Outer Arm \citep{hou14}. The fields are relatively quiet, with some large structures of emission present, such as the W3/W4/HB3 complex at $l=144^\circ,\ b=+1^\circ$, or the large SNR HB9 at $l=160^\circ,\ b=2.5^\circ$. The rms noise of most fields is consistently $<100$~$\mu$Jy~bm$^{-1}$.

\subsubsection{The third Galactic quadrant}

The LoTSS only observed a small region in the third Galactic quadrant, at $l=190^\circ-209^\circ$. The coverage here is patchy, discontinuous, and consists only of a handful of pointings, displayed in Figure \ref{fig:GF_thirdQ}, bottom panel. This area is again close to the LoTSS declination limit of zero degrees, and the noise is in the $200-700~\mu$Jy~bm$^{-1}$ range.

\subsection{Other archival data}

\cite{goedhart24} conducted a Galactic Plane Survey at 1.3~GHz (SMGPS) with the MeerKAT telescope. This survey overlaps with the LHAASO field of view for $l<61^\circ$, $|b|<1^\circ$. For the fields in the LHAASO-LoTSS sample that have MeerKAT coverage, we also make use of the MeerKAT data to investigate the presence of sources, and make $144~\mathrm{MHz}-1.3~\mathrm{GHz}$ spectral index maps\footnote{We use the convention $S_\nu \propto \nu^{\alpha}$.}. The data have been convolved to a common beam, but not $uv$-matched before making the spectral index maps. \cite{goedhart24} caution that only angular scales up to $\sim$$10'$ are well-recovered in the SMGPS; this means that the spectral index maps can only be taken as indication of the thermal or non-thermal nature of the emission. Although spectral index maps that are not $uv$-matched are not ideal, we find that they are still useful for the purpose of this work. In general, since the data at 1.3~GHz miss more extended emission than those at 144~MHz, the spectral index map will show steeper (more negative) values than the true spectral index value of the emission.

We further checked other existing radio and infrared surveys and catalogues including: CGPS, THOR, The Milky Way Imaging Scroll Painting Survey \cite[MWISP,][]{yang25}, the WISE Catalog of Galactic H II Regions \citep{anderson14}, the catalogue of extended sources in the SMGPS \citep{bordiu25}, and the Australia Telescope National Facility Pulsar Catalog \cite[ATNF][]{manchester05, manchester16}. We also often refer to SNR candidates presented by \cite{anderson25} (from here on A25).

\section{Results}
\label{sec:results}

We now discuss source-by-source the LOFAR or LOFAR-MeerKAT emission potentially associated with the 1LHAASO sources. We follow the order of Table 2 of C24 of ascending right ascension. LHAASO catalogues sources according to the instrument with which they are detected: the Water Cherenkov Detector Array (WCDA, sensitive to energies between $0.10-30~\mathrm{TeV}$), and the Kilometer Square Array (KM2A, which operates  
in the $30~\mathrm{TeV}-30$~PeV band). Each of these components is labelled separately in our images\footnote{With respect to the 1~LHAASO notation, the presence of a \lq u\rq\ in the name of a source indicates that it shows emission at UHE. In the figures we have kept an asterisk in the source names to note that the association between the detections in both energy bands (the WCDA and the KM2A components) is ambiguous as per C24.}. All the 1LHAASO source regions overlaid in this work show the 39\% (or 1$\sigma$) containment radius of the 2D-Gaussian model ($r_{39}$ in table 2 of C24).

\subsection{1LHAASO sources in the second Galactic quadrant (I)}

The LOFAR maps for the following sources are shown in Fig. \ref{fig:secondQ_1}.

\begin{figure*}[t]
\centering
\includegraphics[width=0.46\textwidth]{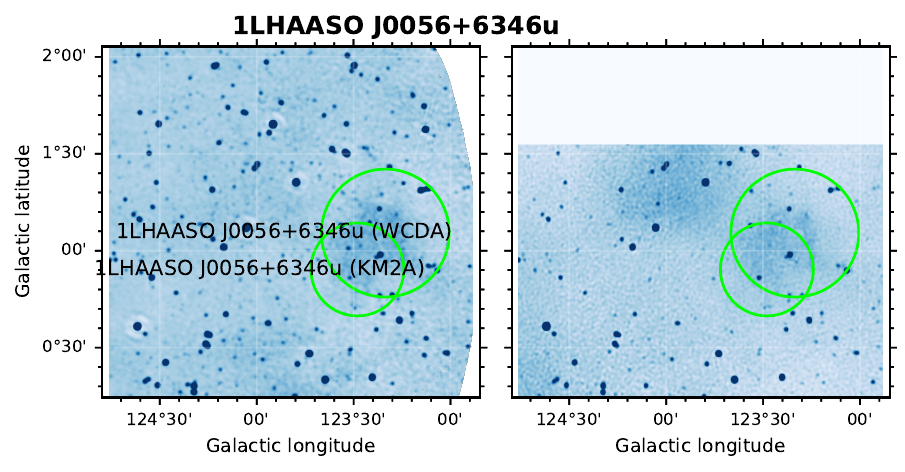}\hfill
\includegraphics[width=0.26\textwidth]{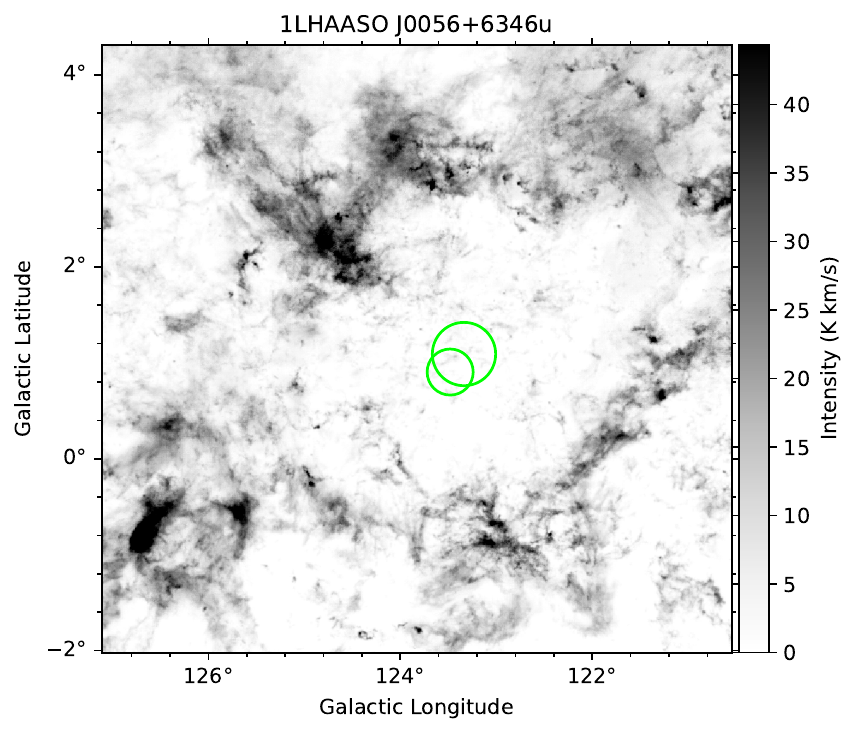}\hfill
\includegraphics[width=0.26\textwidth]{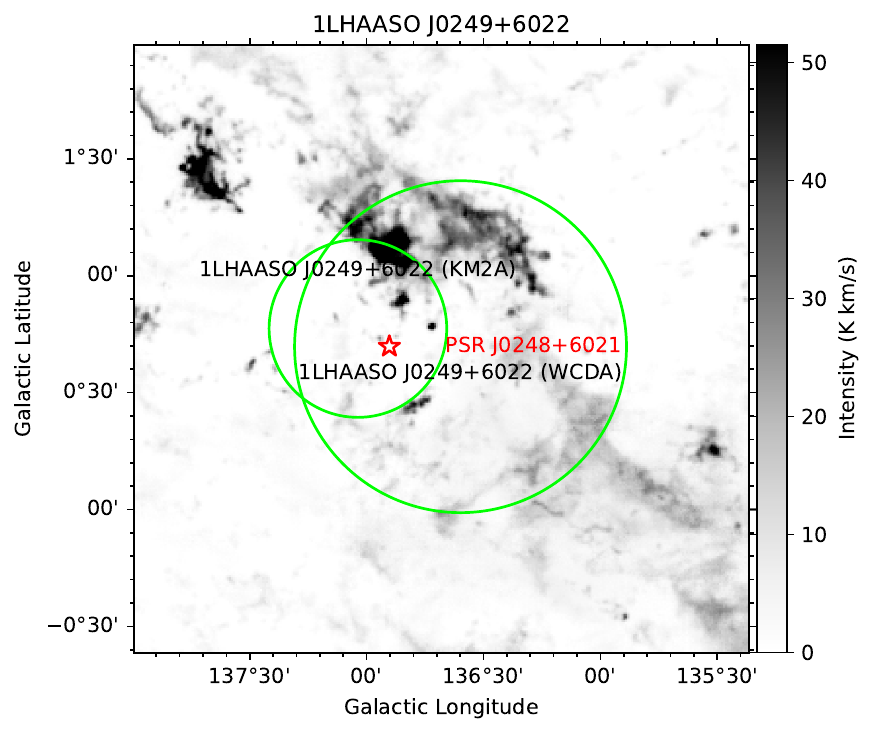}\\[3pt]
\includegraphics[width=0.27\textwidth]{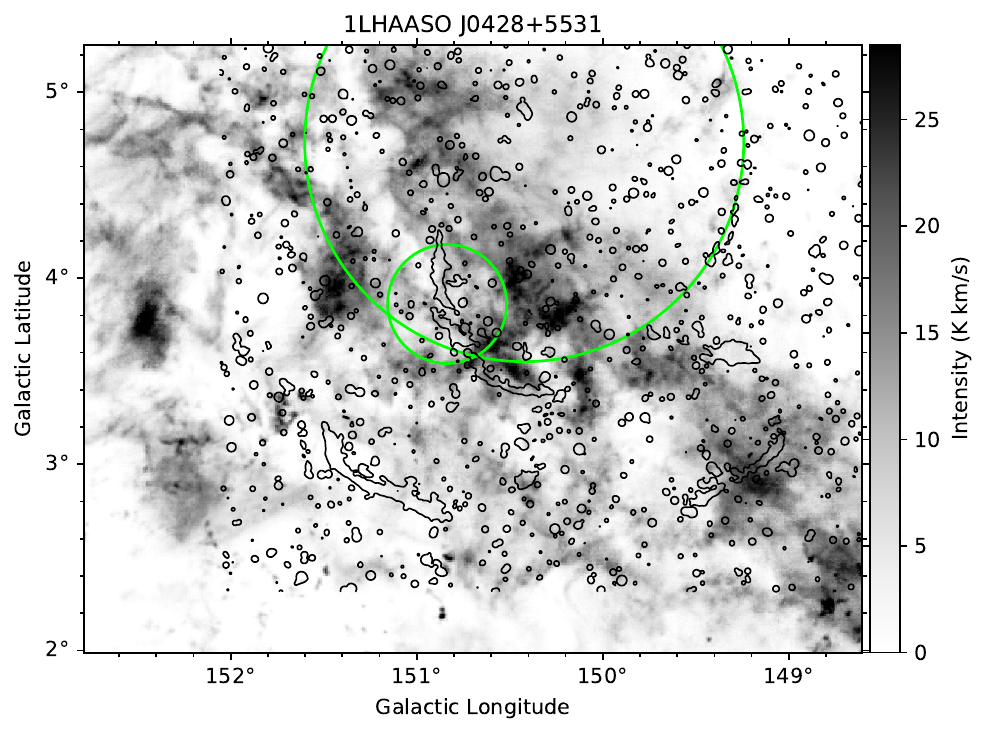}\hfill
\includegraphics[width=0.24\textwidth]{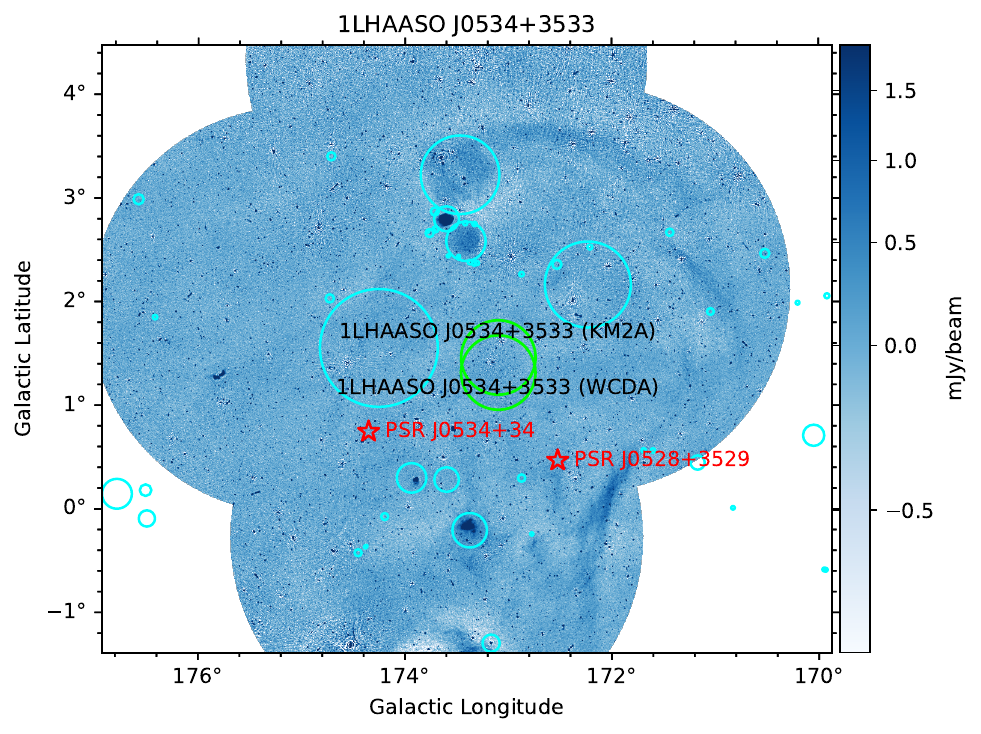}\hfill
\includegraphics[width=0.24\textwidth]{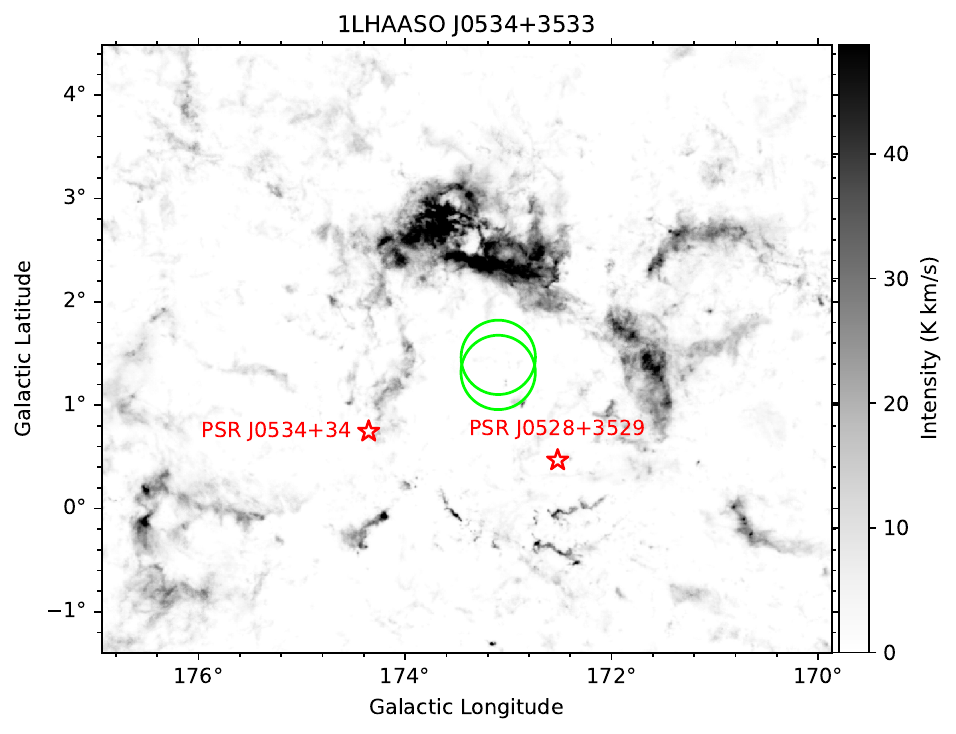}\hfill
\includegraphics[width=0.23\textwidth]{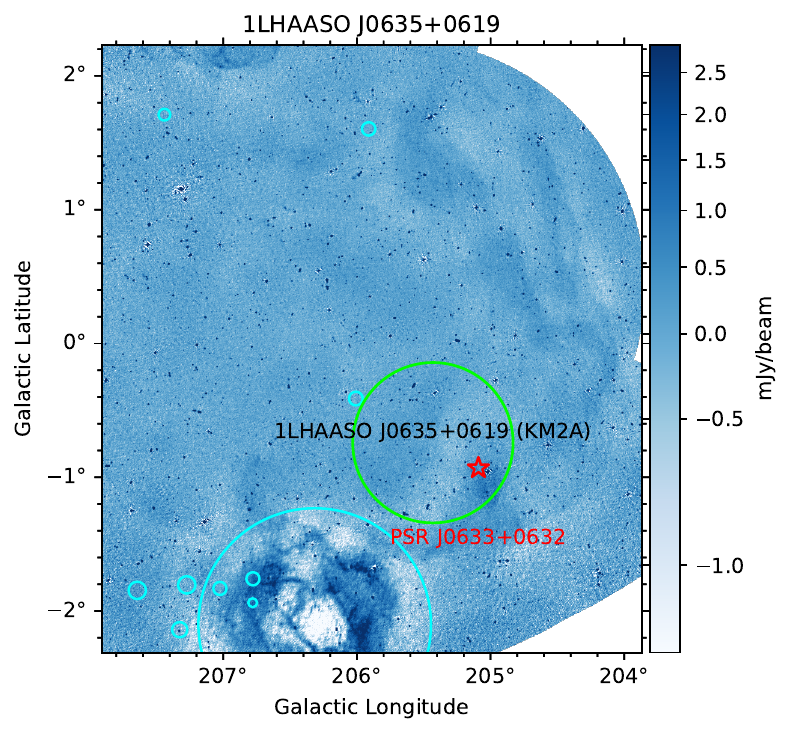}
\caption{Top row (left to right). (i) Region of 1LHAASO~J0056+6346u as seen at 144~MHz with LOFAR, (ii) at 1.4~GHz in the CGPS, (iii) and in $^{12}$CO. (iv) 1LHAASO~J0249+6022 in $^{12}$CO. Bottom row. (i) Region towards 1LHAASO~J0428+5531 as seen in $^{12}$CO. The contours correspond to 0.4~mJy~bm$^{-1}$ in the LOFAR map (see Fig. \ref{fig:secondQ_1}). (ii) Region towards 1LHAASO~J0534+3533 in LOFAR and (iii) $^{12}$CO. (iv) Region of 1LHAASO~J0635+0619 as seen at 144~MHz with LOFAR. The large radio shell is SNR G205.5+0.5. All the molecular maps (shown in greyscale) are taken from the MWISP survey, and correspond to the $^{12}$CO $J=1-0$ transition in the integrated velocity range of $-94$ to 24~km~s$^{-1}$. In all cases the overlaid cyan circles are H\,\textsc{ii} regions from \cite{anderson14}. Red stars mark pulsars discussed in the text \citep[from ATNF,][]{manchester05}. The radius of the 1LHAASO regions is the 39\% containment radius of the two-dimensional Gaussian model in C24.}
\label{fig:all_regions_outer}
\end{figure*}

\vspace{0.1cm}
\noindent
\paragraph{1LHAASO~J0056+6346u}
this is a UHE source seen in the two LHAASO detectors. The LOFAR maps show faint, diffuse ($\sim$$30'$) emission coincident with both components of this source. The CGPS images also show some faint extended radio emission at the same location (see Fig. \ref{fig:all_regions_outer}, top row, first two panels).
There is no infrared emission at this location in any of the WISE bands, and no catalogued H\,\textsc{ii} region \citep{anderson14}. Furthermore, the LHAASO emission is located at the centre of a $\sim$$3^\circ$ molecular cavity as seen in $^{12}$CO --see Fig. \ref{fig:all_regions_outer}, top row, third panel; this structure was noted by \cite{chen23}. The morphology of the extended radio continuum emission is consistent with an aged SNR or PWN. We therefore propose G123.4+1.0 as a new SNR/PWN candidate.

\vspace{0.1cm}
\noindent
\paragraph{1LHAASO~J0206+4302u, 1LHAASO~J0212+4254u, and 1LHAASO~J0216+4237u}
these three high Galactic latitude ($b\sim-17^\circ$) sources are discussed in C24 as possibly linked; J0206+4302u, J0212+4254u are described as \lq dark\rq\ (no known counterpart), and J0216+4237u is possibly associated with millisecond pulsar PSR J0218+4232. Although the pulsar emits brightly in the radio continuum ($560\pm56$~mJy at 144~MHz) we see no hints of a nebula, and none of the 1LHAASO sources shows extended emission in the LOFAR maps (Fig. \ref{fig:secondQ_1}, bottom right panel).

\vspace{0.1cm}
\noindent
\paragraph{1LHAASO~J0249+6022}
the two LHAASO components are coincident with abundant thermal radio emission, and overlap (to the $r_{39}$ containment radius) with a total of 12 H\,\textsc{ii} regions from \cite{anderson14}.
Comparing the LOFAR and CGPS maps yields no new synchrotron components.  C24 discuss the young, energetic pulsar PSR~J0248+6021, noting that an astrophysical system associated with it, such as a composite SNR, could be the origin of the gamma-ray emission. PSR~J0248+6021 is bright at LOFAR frequencies ($136 \pm 14$~mJy at 144~MHz) but is compact in the $6''$ map. We see no hints of a possible PWN, nor of a SNR shell. There is bright, plentiful molecular material tracing the northern outline of the WCDA component (Fig. \ref{fig:all_regions_outer}, top row, fourth panel; the brightest molecular material was studied thoroughly by \cite{issac24}). We note that this source is close ($\sim$$45'$ from the centre of the WCDA component) to gamma-ray binary LS~I~$+61^\circ$303, which has, since the publishing of C24, been detected in UHE emission \citep{cao25b}.

\vspace{0.1cm}
\noindent
\paragraph{1LHAASO~J0339+5307 and 1LHAASO~J0343+5254u}
\cite{dikerby25} discovered XMMU~J034124.2+525720, a diffuse X-ray source spatially coincident with the WCDA component of J0343+5254u. In the LOFAR image there is some clear, structured radio emission coincident with XMMU~J034124.2+525720. In \cite{edler26} we argue against the PWN interpretation, and propose that the X-ray and the radio emission are a line-of-sight merging galaxy cluster unrelated to the gamma-ray emission. There is no further extended radio emission that we can link to the 1LHAASO sources, including in the area of the Fermi Large Area Telescope point source, 4FGL~J0340.4+5302 \citep{desarkar24}, nor those of the molecular clouds identified in \cite{tsuji25}.

\vspace{0.1cm}
\noindent
\paragraph{1LHAASO~J0359+5406}
this extended source is detected by both the WCDA and the KM2A. C24 discuss the gamma-ray emission as a possible PWN or TeV halo from one of two catalogued pulsars that are spatially coincident with it.
PSR~B0355+54 is a bright radio point source for which we measure $66.8\pm6.7$~mJy at 144~MHz and see no hints of extended emission in either the $6''$ or the $20''$ map. We reimaged the source using international baselines to exploit LOFAR's full $0.3''$ resolution capabilities \citep{morabito22}, and it remained compact at this scale, with no evidence for a radio PWN.
PSR~J0359+5414 appears as a fainter point source,
$0.58\pm0.11$~mJy at 144~MHz. It has a $\sim$$30''$ X-ray nebula \citep{zyuzin18} which is not visible in the LOFAR $6''$ map (to an rms sensitivity of $114~\mu$Jy~bm$^{-1}$). The MWISP map shows sparse, faint, small clumps of emission coincident with the 1LHAASO sources, and no clear shell structure surrounding them.

\vspace{0.1cm}
\noindent
\paragraph{1LHAASO~J0428+5531}
C24 indicate that the association between both components of this source is uncertain, but that they overlap with the large complex including SNR candidate G150.3+04.5 and SNR G150.8+03.8 \citep{gao14,devin20}. The LOFAR observations of this region reveal a source that is larger, and more intricate than seen in previous observations (see Fig. \ref{fig:secondQ_1}). It is difficult to determine the extent of individual objects within this large structure, but the KM2A component overlaps a (candidate) SNR shell, whereas the WCDA appears to be in a blowout region. The appearance of a more cohesive structure in the LOFAR map than seen in the CGPS image (Fig. \ref{fig:ancillary_new}) suggests a non-thermal origin (as LOFAR renders visible faint structures of negative spectral index). There is abundant molecular material as seen in MWISP, including what appears to be a $\sim$$1^\circ$ circular shell surrounding the KM2A component, with the centres of both structures having a $10'$ offset (see Fig. \ref{fig:all_regions_outer}, bottom row, first panel).

\vspace{0.1cm}
\noindent
\paragraph{1LHAASO~J0500+4454}
The centre of this source is located $40''$ from the edge of the bright and extended SNR HB9. We see no diffuse radio emission at the 1LHAASO location. The MWISP data (Fig. \ref{fig:mwisp_all}) shows a bright molecular cloud adjacent to this source in the direction of higher Galactic latitude, and several small, faint, clumps of emission overlapping its footprint, although \cite{alford26} disfavour a SNR/MC origin of the emission and instead propose that this LHAASO source could be associated with magnetar SGR 0501+4516.

\vspace{0.1cm}
\noindent
\paragraph{1LHAASO~J0534+3533}
the two LHAASO components are inside the large ($>4^\circ$) SNR candidate G172.5+1.5 that is coincident with an  H\,\textsc{ii} complex \cite[see Fig. \ref{fig:all_regions_outer}, bottom row, second panel]{kang12}. C24 suggest that if the 1LHAASO emission is indeed related to the SNR candidate then it must be due to \lq the central pulsar wind nebula;\rq\ however, querying the ATNF pulsar catalogue gives two pulsars within the shell of the SNR candidate (PSR J0528+3529 and PSR J0534+34), and we see no evidence for a radio PWN in either. The MWISP map shows a molecular ring surrounding the 1LHAASO source (Fig. \ref{fig:all_regions_outer}, bottom row, third panel). 

\subsection{1LHAASO sources in the third Galactic quadrant}

Maps for the following sources are shown in Fig. \ref{fig:secondQ_1}. 

\vspace{0.1cm}
\noindent
\paragraph{1LHAASO~J0635+0619}
this source is also within a very extended SNR ($\sim$$3^\circ$), in this case G205.5+0.5, the Monoceros Loop (see Fig. \ref{fig:all_regions_outer}, bottom row, fourth panel). C24 do not mention SNR G205.5+0.5 as possibly associated with 1LHAASO~J0635+0619, but they do list PSR~J0633+0632 \cite[which][proposed was born in the Rosette Nebula]{danielenko15} as a potential origin of the gamma-ray emission. We see no emission, neither point source nor extended, at the location of the pulsar to an rms of $230~\mu$Jy~bm$^{-1}$ (unsurprisingly, since these are \textit{Fermi} blind search pulsars).

\vspace{0.1cm}
\noindent
\paragraph{1LHAASO~J0703+1405}
this source, seen by both the WCDA and the KM2A detectors, is not explicitly discussed in C24. In the LOFAR data there is no extended emission coincident with the LHAASO regions, and no visible large-scale shell structure surrounding it.

\subsection{Extragalactic 1LHAASO sources}

\vspace{0.1cm}
\noindent
\paragraph{1LHAASO~J1653+3943}
C24 establish an association between this source and nearby active galactic nucleus (AGN) Mrk 501. In the LOFAR map the AGN is seen as a bright point source, and there is no additional extended radio emission. 

\subsection{1LHAASO sources in the first Galactic quadrant}

Many of the sources with $l<61^\circ$ are also visible in the SMGPS. We show LOFAR, MeerKAT, and spectral index maps for these sources in Figs. \ref{fig:Q1_1}, \ref{fig:Q1_2}, and \ref{fig:Q1_3}.

\vspace{0.1cm}
\noindent
\paragraph{1LHAASO~J1848$-$0153u}
this source, best known as HESS~J1848$-$018, is associated with the W43 system \citep{lhaasocol25}, one of the most active star forming regions in the Galaxy \citep{motte03}. The source is at the edge of the LOFAR low-declination coverage, and the LOFAR map is of poor quality, but the MeerKAT map shows plenty of thermal sources, with 136 catalogued H\,\textsc{ii} regions within the footprint of the LHAASO components.

\vspace{0.1cm}
\noindent
\paragraph{1LHAASO~J1848-0001u}
C24 discuss PSR~J1849-0001 \cite[which has a PWN visible in the X-rays;][]{kim24} as potentially associated with this source. We do not see the pulsar as a point source in either the LOFAR (with an rms noise of 1.3~mJy~bm$^{-1}$) nor the MeerKAT (rms $51~\mu$Jy~bm$^{-1}$) image \cite[unsurprisingly, given that this is an X-ray discovered, radio-quiet pulsar,][]{gotthelf10}, although the pulsar is coincident with some larger-scale extended emission that could obscure a faint radio PWN. The 1LHAASO source is surrounded by an arc of synchrotron emission, especially clear in the north-western direction which we propose is a new SNR candidate, G32.64+0.55. This source is not flagged by A25, but the LOFAR detection, the synchrotron spectrum, and a lack of corresponding emission in any of the WISE bands all point to a possible SNR nature.

\vspace{0.1cm}
\noindent
\paragraph{1LHAASO~J1850-0004u}
both components of this source (which C24 classify as dubiously merged) have a clear overlap with the bright SNR~G32.8$-$0.1 \cite[also known as Kes~78, and known to be interacting with molecular clouds;][]{kes68,miceli17}, and the WCDA component further overlaps with SNR G32.4+0.1, SNR candidate G032.458$-$0.112 (identified in A25, although the LOFAR data points against a SNR classification), and luminous blue variable (LBV) candidate G032.914+00.208 \citep{gvaramadze10}. A total of 49 H\,\textsc{ii} regions are within the LHAASO components footprint for this source.

\vspace{0.1cm}
\noindent
\paragraph{1LHAASO~J1852+0050u}
C24 indicate that the two components might be dubiously merged; indeed in the radio map each is overlapping a different SNR. The WCDA component coincides with SNR G34.7$-$0.4 \cite[W44, a SNR known to be interacting with molecular clouds;][]{seta98}, which contains the PWN of PSR~B1853+01 \citep{frail96}, clearly visible in the MeerKAT map, and SNR candidates G034.524$-$0.761 and G034.619+0.240 (A25). The KM2A component overlaps with SNRs G33.6+0.1 \cite[Kes~79, also interacting with molecular material;][]{zhou16}, G33.2$-$0.6 \citep{reich82}, and SNR candidate G033.847+00.062 (A25). Furthermore, there are 148 catalogued H\,\textsc{ii} regions within the $r_{39}$ of both components. C24 suggest a TeV halo around PSR~J1853+0056 as the origin of the LHAASO emission; the MeerKAT map shows extended emission surrounding this pulsar, but this is likely due to the line-of-sight H\,\textsc{ii} region G034.026$-$00.058. 

\vspace{0.1cm}
\noindent
\paragraph{1LHAASO~J1857+0203u}
\cite{cao25c} propose SNR~G35.6$-$0.4 (seen most clearly in the spectral index map in blue, surrounded by the contour line) or H\,\textsc{ii} region G35.6$-$0.5 (seen in red atop the SNR) as possible PeVatrons responsible for the gamma-ray emission (purportedly emitted as the accelerated particles interact with molecular clouds present in the region). We note the additional presence of a partial shell and a complete shell of non-thermal emission within the footprint of the KM2A component, which we propose are new SNR candidates G35.38$-$0.27 and G35.42$-$0.08 (although the latter is noted as an unconfirmed H\,\textsc{ii} region in the WISE catalogue). 

\vspace{0.1cm}
\noindent
\paragraph{1LHAASO~J1857+0245}
C24 mention PSR~J1856+0245 as the potential origin of the gamma-ray emission. The pulsar is visible as a point source in the MeerKAT image, but not in the LOFAR one. The LOFAR map does show extended emission at the location of the pulsar in what could be its PWN; this emission is not present in the MeerKAT map, although there is a negative bowl at this location.
We note a large, very faint shell surrounding 1LHAASO~J1857+0245, which we list as new SNR candidate G36.02$-0.03$, although it is also consistent with being the superbubble described in \cite{petriella21}.

\vspace{0.1cm}
\noindent
\paragraph{1LHAASO~J1858+0330}
C24 list this source as one of eight with only GeV counterparts. The radio maps show a region rich in thermal emission, with 53 catalogued H\,\textsc{ii} regions within the $r_{39}$ of both components. There appears to be a partial shell on the lower Galactic latitude edge of the WCDA component, seen most clearly at Galactic coordinates $l,\ b =37.09^\circ,\ -0.27^\circ$, which is more prominent in the LOFAR map and we propose is a new SNR candidate (G37.05$-$0.49, for the coordinates of the circle encompassing the partial shell).  Moreover, there is an extended region of thermal emission, G36.67$-$0.14, showing substructure in the MeerKAT map, that is not catalogued as an H\,\textsc{ii} region and that shows no IR emission, which we report as one of several flat spectrum shells showing faint or no IR emission presented in this work (see Table \ref{tab:irdark_thermal}). 

\vspace{0.1cm}
\noindent
\paragraph{1LHAASO~J1902+0648}
another source for which C24 only identify GeV counterparts, J1902+0648 is spatially coincident with H\,\textsc{ii} region G040.154+00.648 and overlaps SNR candidate G040.449+00.540 (A25). The LOFAR map shows further synchrotron emission at the location of the 1LHAASO source, which we propose is a new SNR candidate G40.20+0.77.

\vspace{0.1cm}
\noindent
\paragraph{1LHAASO~J1906+0712}
the LHAASO emission is in a complex region of radio emission with 5 H\,\textsc{ii} regions, and adjacent to the very bright SNR G41.1$-$0.3 \cite[3C397, another SNR with a well-established association with its neighbouring molecular environment;][]{jiang10}. There is also a MeerKAT-identified SNR candidate tangent to the LHAASO emission (G040.866+00.155; A25), and we propose an additional LOFAR-identified one, G41.02+0.06, that surrounds and contains the LHAASO emission.

\vspace{0.1cm}
\noindent
\paragraph{1LHAASO~J1907+0826}
this source overlaps with a complex region of thermal and non-thermal radio emission. Not all the thermal sources are catalogued H\,\textsc{ii} regions \citep{anderson14}; we note two more flat spectrum shells showing no IR emission (Table \ref{tab:irdark_thermal}). SNRs G42.0$-$0.1 and G42.8+0.6 \citep{alves12, furst87} are adjacent to the LHAASO emission, and we propose new SNR candidate G42.09$-$0.21.

\vspace{0.1cm}
\noindent
\paragraph{1LHAASO~J1908+0615u}
this source is MGRO J1908+06, one of the most studied PeVatron candidates in the sky. It overlaps SNR G40.5$-0.5$ \citep{downes80}, and two H\,\textsc{ii} regions. C24 identify PSR J1907+0602 as potentially related to the gamma-ray emission; the pulsar does not appear as a point source in either of the radio images \cite[not unexpectedly,][]{abdo10} and we see no evidence of a PWN. This region was thoroughly studied in radio continuum and line emission by \cite{duvidovich20,crestan21}.

\vspace{0.1cm}
\noindent
\paragraph{1LHAASO~J1910+0516, and 1LHAASO~J1913+0501}
these sources are related to the X-ray binary SS433 \citep{lhaaso25}, which has been previously observed with LOFAR \citep{broderick18}. There is no SMGPS data at these Galactic latitudes, the LOFAR map is shown in Fig. \ref{fig:Q1_5}. The gamma-ray emission sits atop SNR G39.7$-2.0$ (also known as W50). 

\vspace{0.1cm}
\noindent
\paragraph{1LHAASO~J1912+1014u}
this source is HESS~J1912+101, a TeV shell with no known counterpart. It is
surrounded by many catalogued H\,\textsc{ii} regions and SNR candidate G044.106+00.595 (A25), clearly visible in the LOFAR map. The LHAASO emission contains what we propose is a new SNR candidate G45.26+0.17. C24 suggest a possible association with PSR~J1913+1011; we see it as a point source in the LOFAR and MeerKAT maps ($7.56\pm0.7~$mJy at 144~MHz; $888\pm111~\mu$Jy at 1.3~GHz), but see no sign of a radio PWN \cite[in conflict with the 6~GHz finding in][]{duvidovich23}. There is a 10$'$, edgeless synchrotron source at $l,\ b=45.35^\circ,\ -0.38^\circ$, in close proximity to PSR~J1915+1045.

\vspace{0.1cm}
\noindent
\paragraph{1LHAASO~J1914+1150u}
this source partially overlaps the MeerKAT-identified SNR candidate G046.304+00.352, and is in close proximity to SNR candidates G46.18$-$0.02 \citep{anderson17}, G46.6+0.2 \citep{tsalapatas24}, and G046.420$-$00.169 (A25). PSR~J1915+1150, noted by C24 as potentially associated, is not visible as either a point or extended source by either MeerKAT nor LOFAR. 

\vspace{0.1cm}
\noindent
\paragraph{1LHAASO~J1919+1556}
this source, only listed as having a KM2A component in C24, shows no clear features in either of the radio maps. It is entirely contained within the KM2A component of 1LHAASO~J1924+1609.

\begin{figure*}[!tp]
\centering
\includegraphics[width=\textwidth]{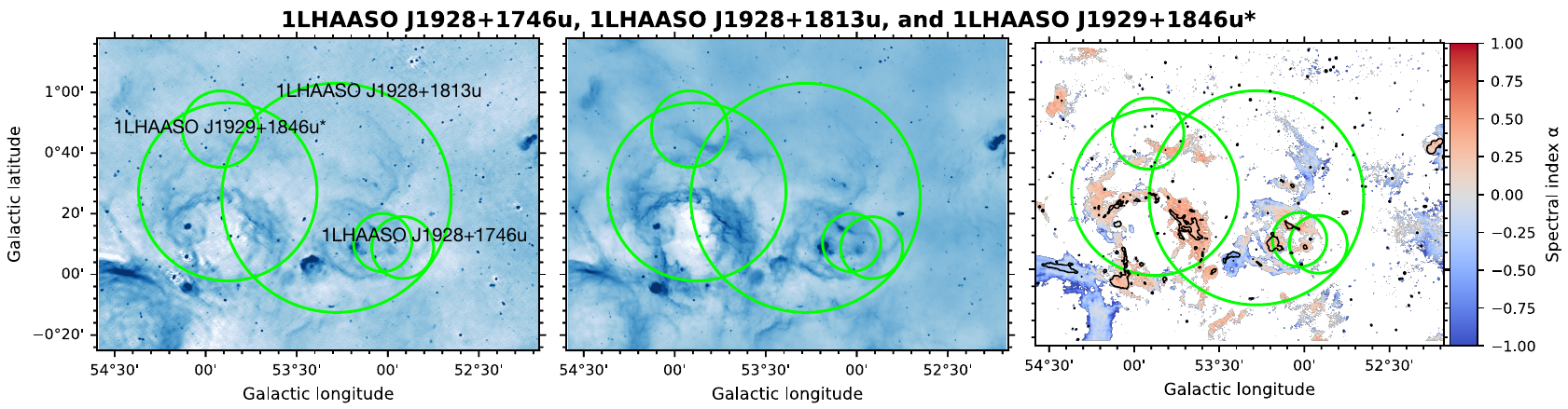}\\[6pt]
\includegraphics[width=0.40\textwidth]{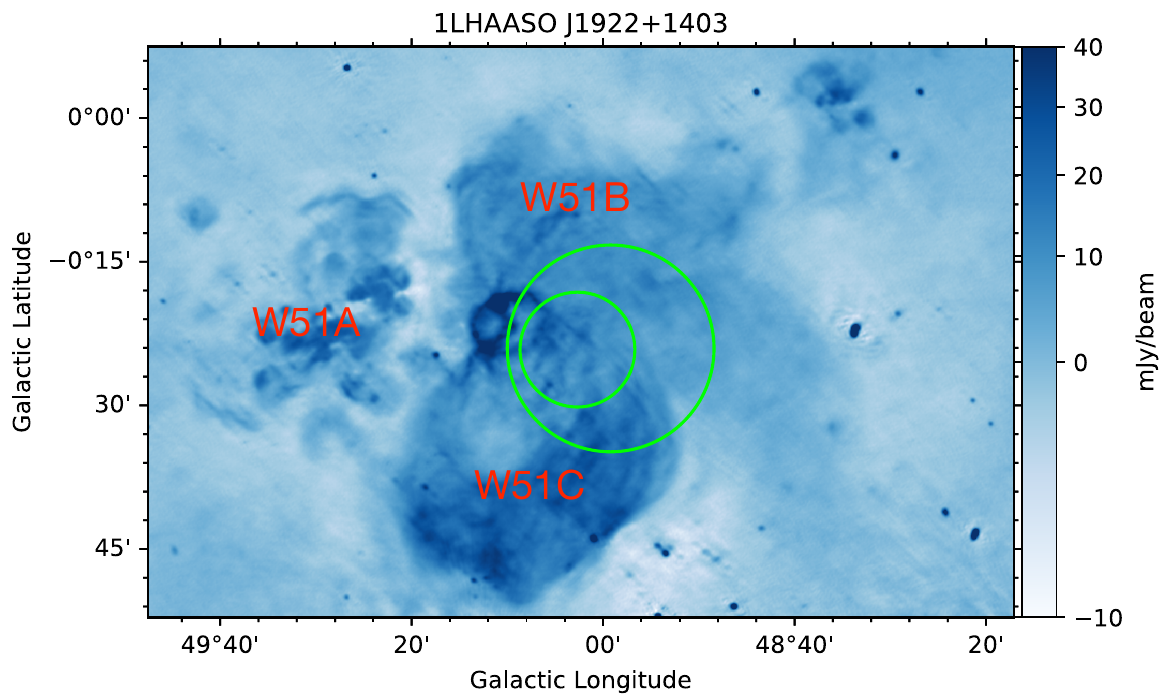}\hfill
\includegraphics[width=0.29\textwidth]{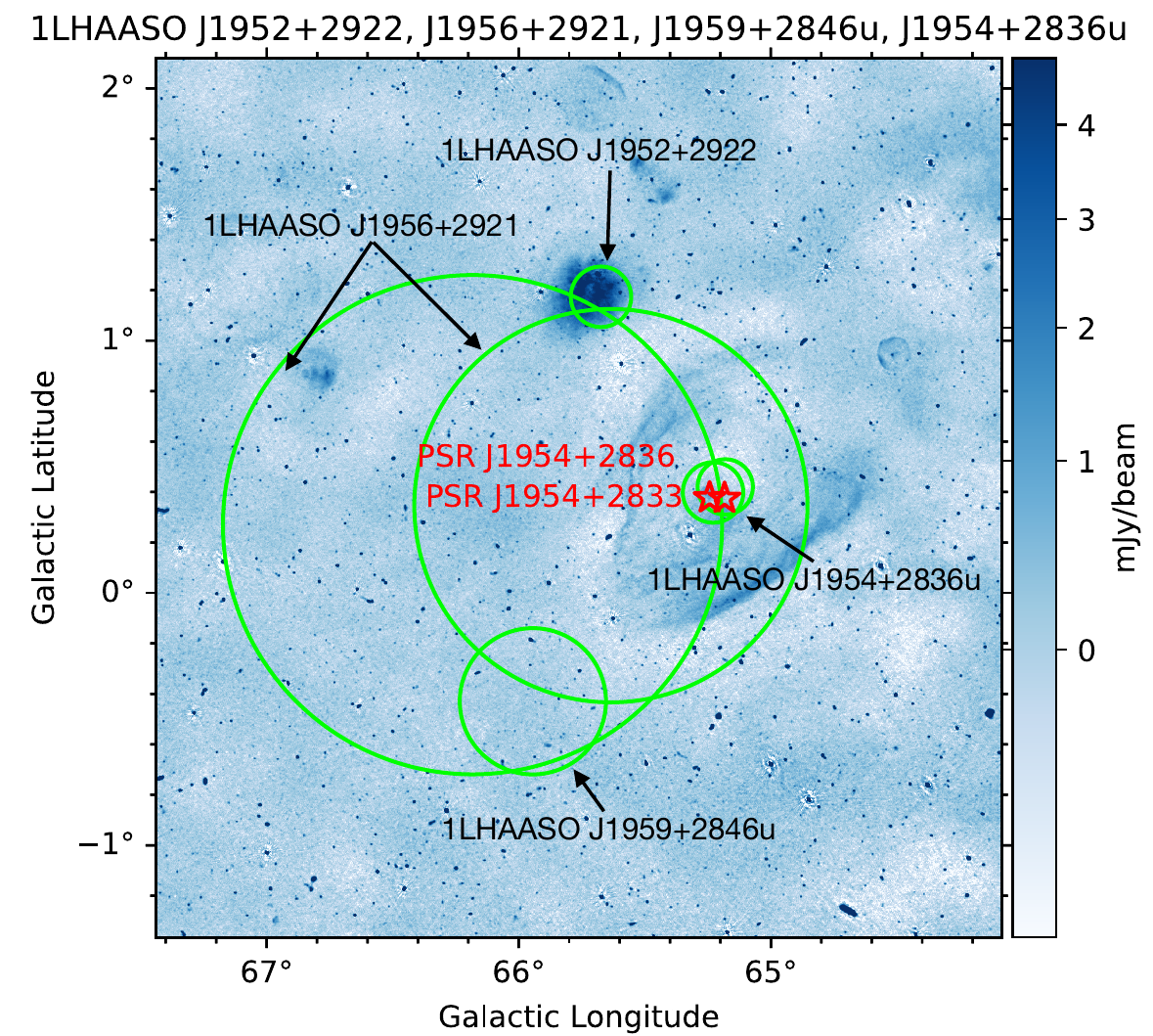}\hfill
\includegraphics[width=0.30\textwidth]{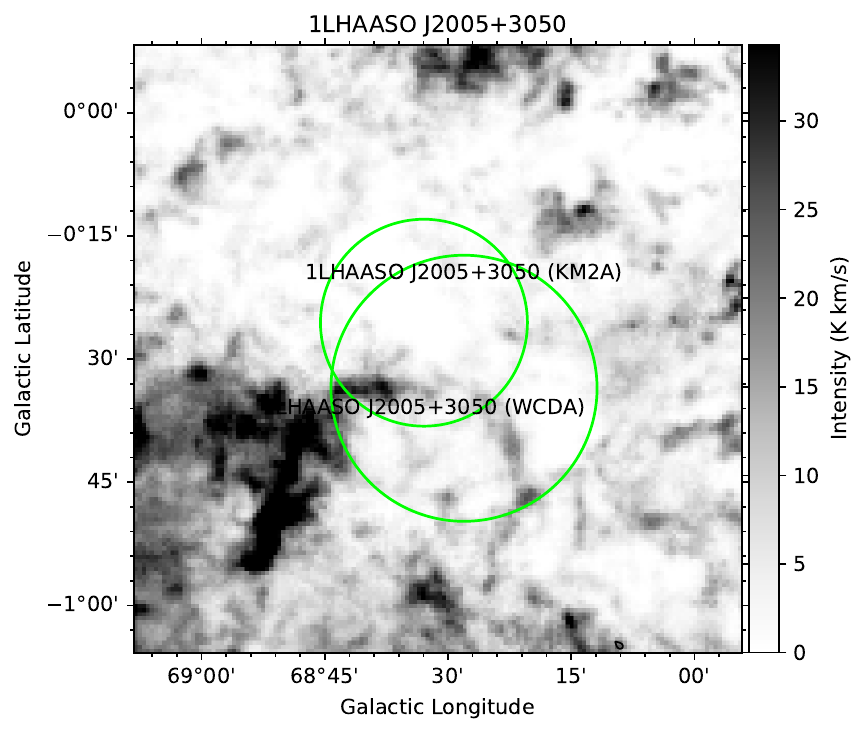}\\[3pt]
\includegraphics[width=0.32\textwidth]{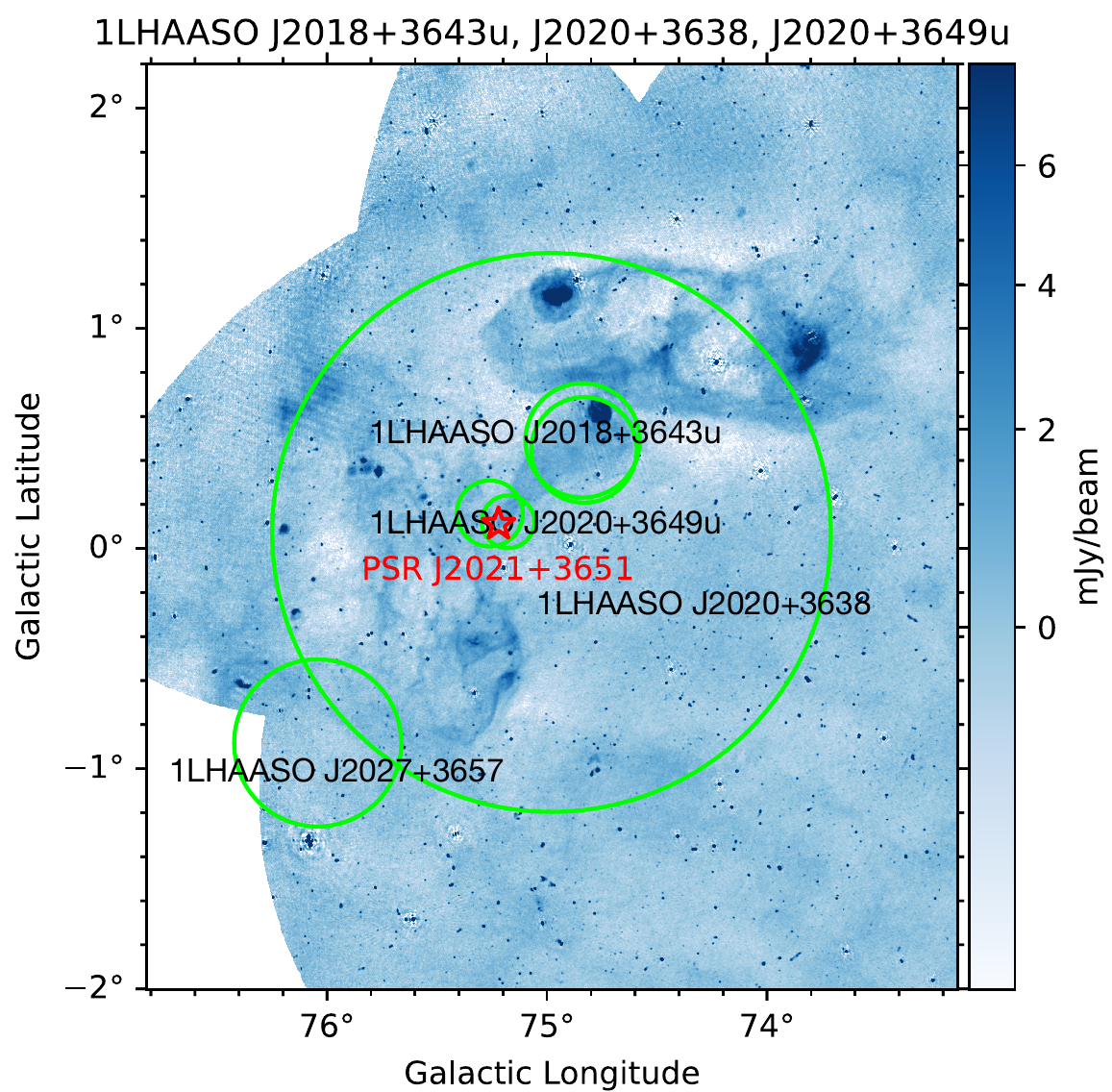}\hfill
\includegraphics[width=0.34\textwidth]{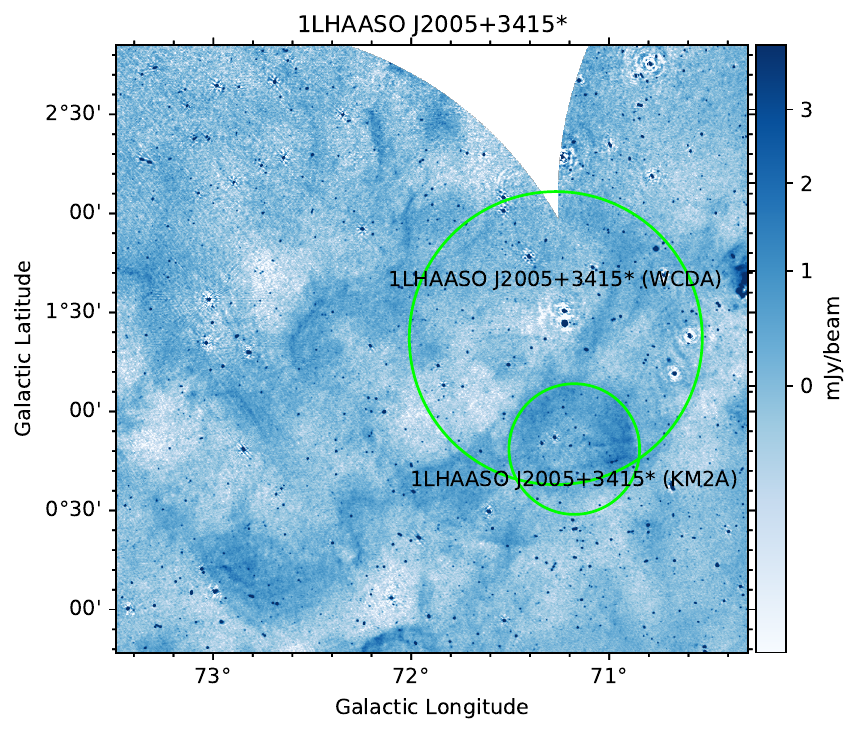}\hfill
\includegraphics[width=0.34\textwidth]{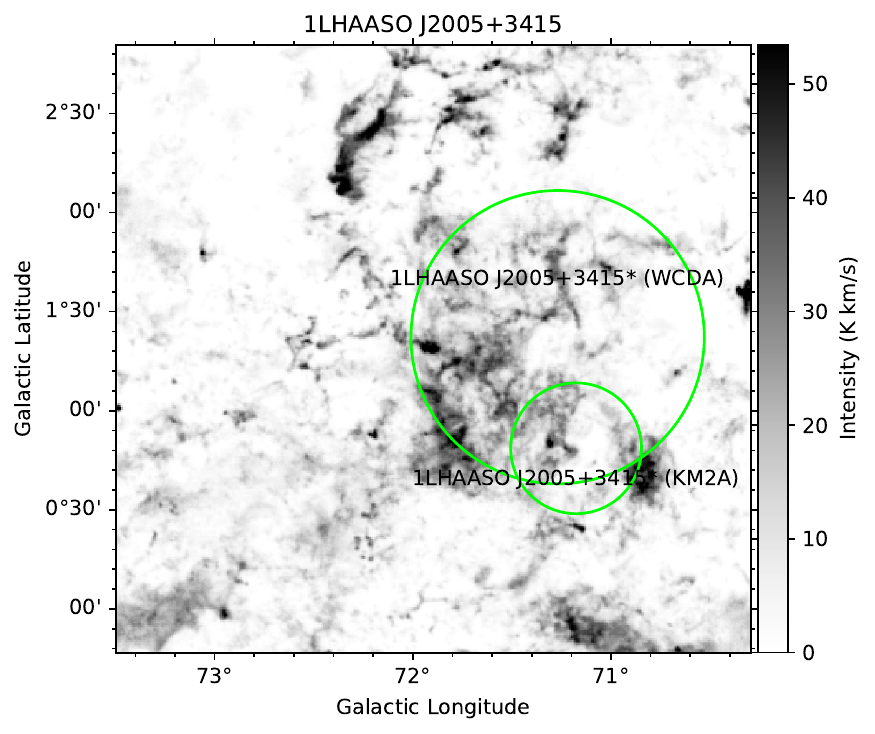}\\[3pt]
\includegraphics[width=0.9\textwidth]{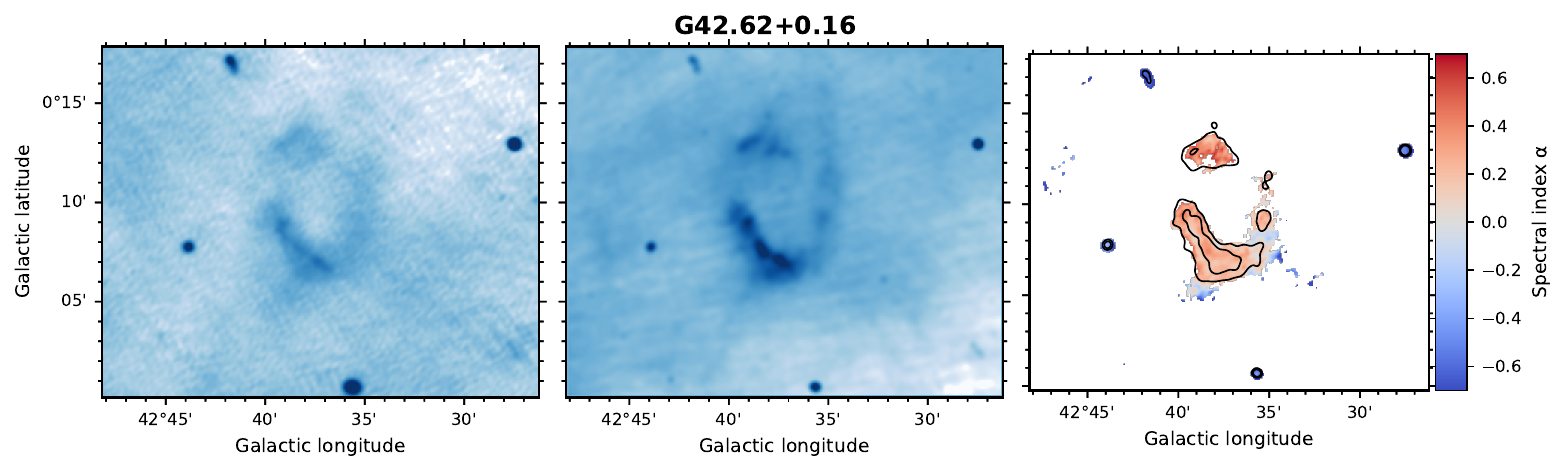}
\caption{Top: LOFAR $20''$ (144~MHz), MeerKAT (1.3~GHz), and spectral index maps of the region towards three 1LHAASO sources, 1LHAASO~J1928+1746u, 1LHAASO~J1928+1813u, and 1LHAASO~J1929+1846u. Note that these sources are within a complex region of emission, mostly of a thermal nature. Second row: (i) LOFAR map of the vicinity of 1LHAASO~J1922+1403: the main components of the W51 complex are labelled in red. (ii) LOFAR map of the region towards 1LHAASO~J1952+2922, 1LHAASO~J1956+2921, 1LHAASO~J1954+2836u, and 1LHAASO~J1959+2846u. (iii) MWISP integrated velocity ($-94$ to +24~km~s$^{-1}$) map of the region towards 1LHAASO~J2005+3050. Third row: (i) LOFAR map of the region towards 1LHAASO~J2018+3643u, 1LHAASO~J2020+3649u, and 1LHAASO~J2020+3638. (ii)~LOFAR and (iii) MWISP $^{12}$CO maps of the region towards 1LHAASO~J2005+3415. Bottom row: example of a flat spectrum shell with faint or no IR emission, G42.62+0.16. Red stars mark pulsars discussed in the text \citep[positions from ATNF,][]{manchester05}. The radius of the 1LHAASO regions is the 39\% containment radius of the two-dimensional Gaussian model in C24.}
\label{fig:1928_etal}
\end{figure*}

\vspace{0.1cm}
\noindent
\paragraph{1LHAASO~J1922+1403}
this source has received substantial attention due to its link with the W51 complex, a well-known gamma-ray emitter \citep{lhaaso24}. The W51 region consists of (see Fig. \ref{fig:1928_etal}, second row, first panel): W51A, a massive star-forming region, W51B, a complex of H\,\textsc{ii} regions, and W51C, a SNR interacting with molecular clouds \citep{brogan13}. The LHAASO emission coincides with the region of SNR-MC interaction, and contains ten H\,\textsc{ii} regions catalogued in \cite{anderson14}.

\vspace{0.1cm}
\noindent
\paragraph{1LHAASO~J1924+1609}
C24 list this source as only having known GeV counterparts. The source is positionally coincident with at least four SNR candidates: G050.489$-$00.410 (A25), G51.21+0.11 \citep{anderson17,driessen18}, G51.04+0.07 \citep{supan18}, and G051.061+0.563 \citep{dokara21}, as well as 125 catalogued H\,\textsc{ii} regions. We additionally propose a new SNR candidate, G51.21$-$0.18, also within the footprint of the LHAASO emission. \cite{supan18} identify molecular material that could presumably be interacting with the SNR candidates in this line-of-sight. 

\vspace{0.1cm}
\noindent
\paragraph{1LHAASO~J1928+1746u}
both components of this 1LHAASO source are contained within 1LHAASO~J1928+1813u (KM2A, see Fig. \ref{fig:1928_etal}). C24 discuss an association with PSR~J1928+1746,
which is visible as an unresolved source in the MeerKAT map ($57\pm25~\mu$Jy), but not with LOFAR.
The gamma-ray emission also overlaps a cluster of 13 H\,\textsc{ii} regions. There is a shell-like source that is clearly visible at $l=53.167^\circ,\ b = 0.189^\circ$, with a radius of $13.5'$, in both the LOFAR and MeerKAT images, which is not catalogued as an H\,\textsc{ii} region or SNR candidate. Its spectral index map reveals predominantly thermal emission, with some evidence of steepening at the rim, suggestive of non-thermal emission at the shell boundary, consistent with a wind-blown bubble around a massive star or stellar cluster. A SIMBAD cone search within 30$'$ reveals several small WISE infrared bubbles and two stellar clusters, but no catalogued WR star or OB association at this position. 

\vspace{0.1cm}
\noindent
\paragraph{1LHAASO~J1928+1813u}
this source, listed as having only a KM2A component, overlaps with both components of 1LHAASO~J1928+1746u and 1LHAASO~J1929+1846u (Fig. \ref{fig:1928_etal}), plus 24 catalogued H\,\textsc{ii} regions, and SNR G53.41+0.03 \citep{driessen18}. We further identify a partial synchrotron shell as new SNR candidate G53.23+0.60.

\vspace{0.1cm}
\noindent
\paragraph{1LHAASO~J1929+1846u}
This source encompasses the composite SNR G54.1+0.3, which contains a PWN powered by PSR J1930+1852. A link between the SNR and the UHE emission was proposed by \cite{xia23,shi25}. 
The source further covers H\,\textsc{ii} region G053.935+00.228, and two partial flat spectrum shells showing faint or no IR emission \cite[listed as unclassified in][]{bordiu25}. 

\vspace{0.1cm}
\noindent
\paragraph{1LHAASO~J1931+1653}
this source, with only the KM2A component, is located atop some faint extended emission that is visible in both maps, seemingly of a synchrotron nature. Surrounding the source to the direction of lower Galactic longitude and latitude, there is a larger faint, edgeless shell, this time thermal and showing no IR emission ($l,\ b=52.37,\ -0.93,\ \theta\sim15'$). 

\vspace{0.1cm}
\noindent
\paragraph{1LHAASO~J1937+2128}
both large components of this source sit atop a large number of radio shells: 47 H\,\textsc{ii} regions; SNR~G57.2+0.8 \citep{sieber94} which hosts magnetar  SGR~1935+2154 and is interacting with molecular material \citep{zhou20}; six SNR candidates from A25; and four new SNR candidates we identify in this work: G55.74+0.22, G56.71+0.35, G57.11+0.38, G56.49$-$0.78, all of which except G56.71+0.35 show a morphology with flatter emission in the centre and steeper synchrotron emission towards the rim, also compatible with a wind bubble scenario. 

\vspace{0.1cm}
\noindent
\paragraph{1LHAASO~J1945+2424}
this source overlaps with 47 H\,\textsc{ii} regions, SNR G59.5+0.1 \citep{taylor92}, and what we propose is new SNR candidate G60.2+0.67. This SNR candidate could be the one hypothesised in \cite{araya24} to explain an extended GeV counterpart to the LHAASO source discovered in \textit{Fermi} data. 

\vspace{0.2cm}
\noindent
The following regions are outside of the range of the SMGPS. We show maps for these sources in Fig. \ref{fig:Q1_4}.

\vspace{0.1cm}
\noindent
\paragraph{1LHAASO~J1951+2608}
this source overlaps with 16 known H\,\textsc{ii} regions. There is diffuse emission surrounding H\,\textsc{ii} region G062.921+00.079 with no clear IR emission; it is also visible at 1.4~GHz in the THOR map, but the low signal-to-noise precludes us from distinguishing a thermal or non-thermal origin. It is further coincident with some CO emission visible in MWISP (Fig. \ref{fig:mwisp_all}). 

\vspace{0.1cm}
\noindent
\paragraph{1LHAASO~J1952+2922}
this source has excellent positional coincidence with 
PWN G65.7+1.2 \cite[also known as DA~495]{kothes06}, which \cite{coerver19} discuss in relation to the gamma-ray emission. 
Its environment is
relatively sparse in CO (Fig. \ref{fig:mwisp_all}), consistent with a proposed leptonic origin of the emission \citep{lhaaso26}.

\vspace{0.1cm}
\noindent
\paragraph{1LHAASO~J1954+2836u}
both components of this source are contained within the larger 1LHAASO~J1956+2921 (see Fig. \ref{fig:1928_etal}), and within SNR G65.1+0.6 \citep{landecker90}. Pulsars PSR~J1954+2836 and PSR~J1954+2833 are within the 1LHAASO source; the former is located within some faint extended emission that is also visible at 1.4~GHz in the CGPS (Fig. \ref{fig:ancillary_new}, second panel). 

\vspace{0.1cm}
\noindent
\paragraph{1LHAASO~J1954+3253}
this source is coincident with the famous SNR CTB 80, which hosts PSR~B1951+32 and its PWN \citep{castelletti03}.

\vspace{0.1cm}
\noindent
\paragraph{1LHAASO~J1956+2921}
this source overlaps with three other 1LHAASO sources (Fig. \ref{fig:1928_etal}), SNR G65.1+0.6, 18 known H\,\textsc{ii} regions, and substantial CO emission in MWISP (Fig. \ref{fig:mwisp_all}). 

\vspace{0.1cm}
\noindent
\paragraph{1LHAASO~J1959+2846u}
contained within the previous source, this one contains a single compact H\,\textsc{ii} region and otherwise no clear associated structure in the radio continuum, although it is coincident with a region of abundant CO emission. 

\vspace{0.1cm}
\noindent
\paragraph{1LHAASO~J2002+3244}
this source is coincident with SNR G69.7+1.0 \citep{reich98}. The LOFAR map shows a bright shell of emission of radius $\sim$$6'$, surrounded by a fainter shell with a distinct rim of radius $\sim$$14'$; only the southern rim of this structure is faintly visible in the 1420~MHz CGPS maps \cite[but not noted in][]{kothes06}. We propose that there might be two SNRs here: SNR G69.7+1.0, with $r\sim6'$, and the SNR candidate formed by the diffuse shell, which we call G69.67+1.01. 

\vspace{0.1cm}
\noindent
\paragraph{1LHAASO~J2005+3050}
this source does not show any clear associated structure in the LOFAR map, although the integrated CO map shows a coincident arc of emission consistent with a wind-blown bubble (see Fig. \ref{fig:1928_etal}, second row, third panel). 

\vspace{0.1cm}
\noindent
\paragraph{1LHAASO~J2005+3415}
this source coincides with the tail of some large-scale, comet-shaped, diffuse emission in the radio continuum map, and substantial molecular emission, as well as 9 H\,\textsc{ii} regions of a much more compact nature than the LOFAR emission. The LHAASO components appear to be nested within a large new SNR candidate, G72.11+1.72 (although this source could well be an uncatalogued thermal shell, since it is also seen clearly in CGPS at 1420~MHz; see Fig. \ref{fig:ancillary_new}, third panel), and the KM2A component is coincident with a radio shell that we propose is SNR candidate G71.18+0.88 (see Fig. \ref{fig:1928_etal}, third row, second and third panels). 

\vspace{0.1cm}
\noindent
\paragraph{1LHAASO~J2018+3643u}
this source is in close proximity to 1LHAASO~J2020+3649u, and both are contained within 1LHAASO~J2020+3638 (see Fig. \ref{fig:1928_etal}, third row, first panel). The bright circular source seen in the LOFAR map in dark blue is one of two H\,\textsc{ii} regions the 1LHAASO source contains. It also overlaps a region of comet-like radio continuum emission whose apex is located in 1LHAASO~J2020+3638. 

\vspace{0.1cm}
\noindent
\paragraph{1LHAASO~J2020+3649u}
the comet-like radio emission described above coincides at its apex with the location of PSR~J2021+3651, the pulsar that powers the Dragonfly PWN, which shows a bow shock in X-rays, and whose association with LHAASO J2018+3651 \citep{cao21} and PeVatron nature has been proposed by \cite{woo23}. LOFAR clearly sees the pulsar as a point source ($4.73\pm1.05$~mJy in the $6''$ map), and the radio pulsar wind nebula \cite[$0.6' \times1.8'$, PA$\approx140^\circ$]{jin23} is seen as an enhancement of the triangular radio emission, although the features described in the \cite{jin23} publication are not resolved. In the LOFAR map the most striking feature, however, is the triangular band of emission that extends to $\sim$$50'$, and that is coincident with a bow shock consistent with the direction of motion of the pulsar. The comet-tail is also seen in the CGPS map at 1420~MHz (Fig. \ref{fig:ancillary_new}, fourth panel), although there it is more confused with the background emission than in the LOFAR map. 

\vspace{0.1cm}
\noindent
\paragraph{1LHAASO~J2020+3638}
in addition to containing the two previous 1LHAASO sources, this source overlaps 28 H\,\textsc{ii} regions, and SNRs CTB~87 and G73.9+0.9. Moreover, there is significant large-scale extended emission encompassing the entirety of the source, with substructures including partial shells and bubbles. 

\vspace{0.1cm}
\noindent
\paragraph{1LHAASO~J2027+3657}
this source also partially overlaps 1LHAASO~J2020+3638; other than a cluster of H\,\textsc{ii} regions at its boundary in the direction of increasing Galactic latitude and longitude (which also shows CO emission in the MWISP map, Fig. \ref{fig:mwisp_all}), there is no clear radio continuum structure at the location of the gamma-ray emission.

\vspace{0.1cm}
\noindent
\paragraph{1LHAASO~J2028+3352}
despite its large angular size ($r_{39}=1.70\pm0.23^\circ$), this source has a relatively high Galactic latitude ($\sim$$3^\circ$) and hence only encompasses a single H\,\textsc{ii} region. It has no clear associated radio continuum structure. We propose a new SNR candidate, G74.49$-$4.12, at the edge of the 1LHAASO source. 

\subsection{1LHAASO sources in the second Galactic quadrant (II)}

The LoTSS Galactic fields are rather sparse in the region around Cygnus A, and so the coverage of 1LHAASO sources is patchy. Maps for the following sources can be found in Fig. \ref{fig:Q1_5}.

\vspace{0.1cm}
\noindent
\paragraph{1LHAASO~J2200+5643u}
this source is not completely covered by the LoTSS maps. It sits in a sparse region with no H\,\textsc{ii} regions. C24 list no association at all. Here we find a radio shell that we identify as new SNR candidate G100.97+1.57, which is also very faintly visible in the CGPS 1420~MHz image (Fig. \ref{fig:ancillary_new}, fifth panel).

\vspace{0.1cm}
\noindent
\paragraph{1LHAASO~J2238+5900}
this source is completely encompassed by 1LHAASO~J2229+5927u. There is no clear extended radio emission in the continuum LOFAR and CGPS maps. 

\vspace{0.1cm}
\noindent
\paragraph{1LHAASO~J2229+5927u}
this is a large ($r_{39} \sim 2^\circ$) source that not only contains the 1LHAASO source above, but also 1LHAASO~J2228+6100u, 29 H\,\textsc{ii} regions, and SNR G106.3+2.7. The SNR further contains a PWN, the Boomerang, which is coincident with 1LHAASO~J2228+5900. The Boomerang sits at the edge of the LoTSS pointing and is truncated in our map. 

\section{Discussion}
\label{sec:discussion}

We list 24 new LOFAR-identified SNR candidates in Table \ref{tab:snr_candidates}. This work did not undertake an exhaustive SNR search; these were discovered simply by examining the footprint of the 1LHAASO sources listed above, and so we claim no completeness limits on the search for new SNR candidates in the LOFAR Galactic fields. We also list 8 shells with a flat spectrum emission and no IR counterpart in Table \ref{tab:irdark_thermal}. Again, this work does not constitute an exhaustive search for these objects.

\subsection{The flat spectrum shells showing faint or no IR emission}
\label{sec:flatshells}

These sources were found from a seemingly flat LOFAR-MeerKAT spectrum, and their non-classification as Galactic H\,\textsc{ii} regions. Radio and mid-infrared maps of each of them are shown in Figs. \ref{fig:shell_cutouts_a} and \ref{fig:shell_cutouts_b}. A possible interpretation is that they are aged H\,\textsc{ii} regions where the photodissociation region has dispersed, or the ionising star has cooled and its UV flux is low. They are also compatible with evolved planetary nebulae. However, we note that the LOFAR-MeerKAT spectral index is only seemingly (and possibly not genuinely) flat, as both the presence of free-free absorption and the $u-v$ mismatch in the images can distort the observed spectral index value. Moreover, if these sources are PWN, they might have genuinely flat spectra and a non-thermal origin \citep{dubner15}. These shells could be synchrotron sources, either aged SNR or PWN, although given the possibility of them belonging to a wider class of sources, we list them separately (Table \ref{tab:irdark_thermal}) from the SNR candidates in Table \ref{tab:snr_candidates}.

\subsection{Notable individual results}
\label{sec:highlights}

\vspace{0.1cm}
\noindent
\paragraph{A first counterpart proposal for two dark sources}
1LHAASO~J2200+5643u has no association of any kind in C24; the radio shell we identify as SNR candidate G100.97+1.57 is, to our knowledge, the first counterpart proposed for this UHE source. Similarly, for 1LHAASO~J0056+6346u we identify faint, diffuse radio emission that leads us to propose the new SNR/PWN candidate G123.4+1.0. Its location at the centre of the $\sim$$3^\circ$ molecular cavity noted by \cite{chen23}, and the absence of any infrared counterpart, are consistent with an evolved remnant or with a PWN powered by an as-yet-undetected pulsar. The latter scenario is favoured by the dedicated LHAASO study of this source \citep{lhaaso25b}, which considers it more likely than gas illuminated by cosmic rays from the nearby SNR candidate G124.0+1.4 \citep{chen23}.

\vspace{0.1cm}
\noindent
\paragraph{A radio candidate inside the TeV shell HESS~J1912+101}
1LHAASO~J1912+1014u corresponds to HESS~J1912+101, one of the very few TeV shells with no confirmed counterpart at any other wavelength \citep{hess18,su17,duvidovich23}. The new SNR candidate G45.26+0.17 lies within the LHAASO emission, and, together with the edgeless synchrotron source we detect near PSR~J1915+1045 provides a possible counterpart to the emission. 
We note that G45.26+0.17 ($16.3'$) is substantially smaller than the TeV shell, so if the association holds it would represent a counterpart to only part of the gamma-ray structure.

\vspace{0.1cm}
\noindent
\paragraph{The SNR hypothesised for 1LHAASO~J1945+2424}
\cite{araya24} discovered an extended GeV counterpart to this source with a hard spectrum, and noted that no radio SNR was known within the gamma-ray footprint. The new SNR candidate G60.2+0.67 ($44.6'$) could provide evidence for their proposed scenario. If confirmed, 1LHAASO~J1945+2424 would go from an unidentified source to a GeV--TeV SNR association.

\vspace{0.1cm}
\noindent
\paragraph{The $50'$ tail of the Dragonfly PWN}
In the field of 1LHAASO~J2020+3649u, the LOFAR map shows the cometary structure trailing PSR~J2021+3651 extending to $\sim$$50'$, much larger than the arcminute-scale PWN measured at higher frequencies \citep{jin23}, and aligned with the direction of motion implied by the X-ray bow shock. 

\vspace{0.1cm}
\noindent
\paragraph{The intricate structure of 1LHAASO~J0428+5531}
The LOFAR data link the previously known SNR candidate G150.3+04.5 and SNR G150.8+03.8 as part of a cohesive, large-scale source with the form of a cluster of soap bubbles, that has excellent positional coincidence with the 1LHAASO source. In fact, \cite{devin20} found a hard GeV spectrum suggestive of a dynamically young SNR, and \cite{li24} present evidence for hybrid emission: hadronic in the south, where the gamma-rays correlate with molecular material at $\sim$$740$~pc \citep{feng24}, and leptonic in the north. The blowout appearance of the WCDA component and the $\sim$$1^\circ$ molecular shell around the KM2A component are consistent with this picture of a remnant shaped by an uneven molecular environment.

\vspace{0.1cm}
\noindent
\paragraph{Two superposed shells at 1LHAASO~J2002+3244}
The LOFAR data resolve what is catalogued as the single SNR G69.7+1.0 into a bright inner shell and a fainter, larger shell (our candidate G69.67+1.01). If the outer shell is confirmed as an independent remnant, the association of the gamma-ray emission with any SNR at this position becomes ambiguous, and any modelling of this source will need to consider two objects with possibly different ages and distances.

\vspace{0.1cm}
\noindent
\paragraph{Pulsar non-detections}
Several of the pulsars that C24 propose as counterparts --- PSR~J1907+0602, PSR~J1849$-$0001, and PSR~J0633+0632 among them --- were discovered in gamma-ray blind searches or in the X-rays, and are radio-quiet or radio-faint \citep{abdo10,gotthelf10,danielenko15}. Our non-detections of these pulsars as point sources are therefore expected and carry no information about the source associations; the meaningful LOFAR result in these fields is instead the surface-brightness limit on any extended PWN emission. In some cases, however, a radio nebula was expected or claimed; relevant results are the compactness of the bright pulsar PSR~B0355+54 down to $0.3''$, and the absence at 144~MHz of the candidate PWN reported at 6~GHz around PSR~J1913+1011 \citep{duvidovich23}, both constraining any nebula to be small or spectrally flat.

\subsection{Statistical assessment of chance coincidence}
\label{sec:bootstrap}

We performed a bootstrap test aimed at determining whether the spatial coincidences between 1LHAASO sources and the radio SNRs are statistically significant. We generated $10^4$ realisations of 50 randomly placed fake sources
within the survey footprint (the statistical sample comprises the 50 covered sources that lie within the Galactic-plane strips; the three high-latitude sources, 1LHAASO~J0703+1405, and Mrk~501 are excluded), resampling $r_{39}$ values with replacement from the real
distribution to preserve the observed size distribution.

To assess the effect of SNR catalogue incompleteness we modelled an underlying
SNR population $f$ times larger than currently catalogued, for $f \in [1, 10]$.
For the \cite{green25} SNRs, 23 of 50 1LHAASO sources (46\%) overlap with a catalogued source. Even for a SNR completeness factor at $f = 10$ (an underlying Galactic SNR population ten times larger
than currently catalogued) the null distribution yields only $10.0 \pm 2.8$
matches. Including the \cite{anderson25} SMGPS candidates alongside Green raises the
$f = 1$ baseline to 676 objects but still reaches only $11.9 \pm 3.0$ at
$f = 10$. In both cases the null curve never reaches the observed count of 23,
indicating that the 1LHAASO--SNR alignment is inconsistent with chance coincidence
under any realistic incompleteness assumption.

The most compelling evidence for a physical association emerges when the sample
is divided by Galactic longitude into an inner-Galaxy subsample
($30^\circ \leq l \leq 84^\circ$; $n = 38$) and an outer-Galaxy subsample
($97^\circ \leq l \leq 209^\circ$; $n = 12$).
In the inner Galaxy, 19 of 38 sources (50\%) have at least one Green SNR
match, whereas the null distribution reaches $5.9 \pm 2.2$ at $f = 1$ and $10.0 \pm 2.7$ at $f = 10$. In the outer Galaxy, 4 of 12 sources (33\%) have at least one Green SNR match, whereas the outer Galaxy has a sparse
SNR background, and the null distribution reaches only $1.4 \pm 1.1$ at $f = 1$
and $2.1 \pm 1.3$ at $f = 10$.

When dividing the sample by the presence of
ultra-high-energy emission, we find that the PeVatron
candidates are neither more nor less likely than the rest of the sample to coincide with a catalogued SNR (10 of 20 versus 13 of 30).

The bootstrap analysis deliberately excludes the new SNR candidates identified in this work (Table~\ref{tab:snr_candidates}) from the background density estimate in order to avoid circularity. 
A statistically
significant excess of known SNRs at 1LHAASO positions, and the independent discovery
of new SNR candidates in the same fields strongly supports a physical
association between a large fraction of the 1LHAASO source population and the
Galactic SNR environment.
The most natural interpretation is that 1LHAASO source population
in the Galactic plane comprises PWN, SNR-PWN composite
systems, sources illuminated by cosmic rays escaping from nearby SNRs, or pulsar TeV halos. All of
these require recent core-collapse supernovae as progenitors and are therefore
expected to be spatially correlated with the SNR population.

Our bootstrap analysis allows a statistical estimate of the fraction of 1LHAASO
Galactic-plane sources genuinely residing in SNR environments by subtracting the
null expectation from the observed count. This gives an excess of $\sim$17 sources at
$f=1$ and $\sim$13 at $f=10$, corresponding to roughly $26$--$33\%$ of the 50-source
sample. When the new SNR candidates from this work are included in the observed
count ($N_\mathrm{obs}=37$, null $\lesssim 12$ at $f=10$), the estimated genuine
fraction rises to $\sim50\%$.

\section{Summary and conclusions}
\label{sec:conclusions}

We have presented the Galactic fields of the LoTSS, and used them, together with ancillary radio, infrared, and molecular-line data, to search for low-frequency counterparts to the 55 sources of the First LHAASO Catalogue that fall within the survey footprint. Our main results are as follows.

\begin{enumerate}
    \item We provide an overview of the 144~MHz emission towards all 55 1LHAASO sources, complemented, where available, by MeerKAT 1.3~GHz maps, which we used to make $144~\mathrm{MHz}-1.3$~GHz spectral index maps. We identify, for each source, the radio structures potentially related to the gamma-ray emission.
    \item We identify 24 new SNR candidates (Table~\ref{tab:snr_candidates}), 18 of which lie within the $r_{39}$ region of a 1LHAASO source, found by inspecting the 1LHAASO fields, as well as eight flat-spectrum shells with faint or no infrared counterpart (Table~\ref{tab:irdark_thermal}), whose nature --- evolved H\,\textsc{ii} regions, planetary nebulae, absorbed synchrotron shells, or PWNe --- cannot be established with the current data.
    \item A bootstrap analysis shows that the coincidence between 1LHAASO sources and catalogued SNRs cannot be attributed to chance: 23 of 50 sources (46\%) overlap a \cite{green25} SNR, against a null expectation of $10.0\pm2.8$ even for an underlying SNR population ten times larger than catalogued. In the outer Galaxy this excess is particularly noteworthy.
    \item Subtracting the null expectation from the observed counts, we estimate that at least $\sim$$26-33\%$ of the 1LHAASO Galactic-plane sources genuinely reside in SNR environments, a fraction that rises to $\sim$$50\%$ when the new SNR candidates of this work are included. This is consistent with a population consisting of PWNe, SNR--PWN composites, pulsar TeV halos, and clouds illuminated by cosmic rays escaping nearby remnants: all end products of recent core-collapse supernovae.
\end{enumerate}

This work demonstrates that the low-frequency radio band has a central role to play in identifying the counterparts of the gamma-ray source population. The start of operations of the Cherenkov Telescope Array \cite[CTA,][]{hofmann23} will open the era of high-resolution gamma-ray astronomy, and will provide clarity on the candidate associations proposed here.

\section*{Data availability}
The LoTSS DR-3 mosaics are publicly available at \url{https://lofar-surveys.org/}. The full atlas of per-source LOFAR maps and the ancillary archival maps referenced in the text are reproduced in Appendix~\ref{app:atlas}; they are also available as online supplementary material on Zenodo at \url{https://doi.org/10.5281/zenodo.22536839}.

\begin{acknowledgements}
MA acknowledges financial support from grants CEX2021-001131-S, PID2021-123930OB-C21 and PID2024-155817OB-I00, funded by MICIU/AEI/ 10.13039/501100011033 and by ERDF/EU, and from the coordination of the participation in SKA-SPAIN, funded by the Ministry of Science, Innovation and Universities (MICIU). RT is grateful for support from the UKRI Future Leaders Fellowship (grant MR/Y020405/1). This work was supported by the STFC [grants ST/T000244/1, ST/V002406/1, ST/Y001249/1].

This paper uses data obtained with the LOFAR telescope (LOFAR-ERIC, projects LT14\_004 and LT16\_004). LOFAR \citep{vanhaarlem13} is the Low Frequency Array designed and constructed by ASTRON, with facilities in several countries owned by various parties (each with their own funding sources), collectively operated by the LOFAR European Research Infrastructure Consortium (LOFAR-ERIC) under a joint scientific policy. The LOFAR-ERIC resources have benefited from the following recent major funding sources: CNRS-INSU, Observatoire de Paris and Universit\'e d'Orl\'eans, France; Istituto Nazionale di Astrofisica (INAF), Italy; BMBF, MIWF-NRW, MPG, Germany; Science Foundation Ireland (SFI), Department of Business, Enterprise and Innovation (DBEI), Ireland; NWO, The Netherlands; The Science and Technology Facilities Council, UK; Ministry of Science and Higher Education, Poland.

Computing resources were provided by the Dutch national e-infrastructure with support of the SURF Cooperative (e-infra 180169) and the LOFAR e-infra group; by the J\"ulich Supercomputing Centre (JSC), which coordinates and operates the J\"ulich LOFAR Long Term Archive and the German LOFAR network, with time on JUWELS granted by the Gauss Centre for Supercomputing e.V. (grant CHTB00) through the John von Neumann Institute for Computing (NIC); by the University of Hertfordshire high-performance computing facility and the LOFAR-UK computing facility, supported by STFC [ST/P000096/1]; and by the Italian LOFAR-IT infrastructure supported and operated by INAF, including the PLEIADI special ``LOFAR'' project by USC-C of INAF and the C3S Supercomputing Centre of Turin university (under an agreement with Consorzio Interuniversitario per la Fisica Spaziale), Italy.

The MeerKAT telescope is operated by the South African Radio Astronomy Observatory, which is a facility of the National Research Foundation, an agency of the Department of Science and Innovation. This research made use of data from the Milky Way Imaging Scroll Painting (MWISP) project, a multi-line survey in $^{12}$CO/$^{13}$CO/C$^{18}$O along the northern Galactic plane with the PMO-13.7m telescope, and from the Canadian Galactic Plane Survey, a Canadian project with international partners, supported by the Natural Sciences and Engineering Research Council; the Dominion Radio Astrophysical Observatory is a National Facility operated by the National Research Council Canada. The CGPS mosaics were obtained through the Canadian Astronomy Data Centre, operated by the National Research Council of Canada with the support of the Canadian Space Agency.
\end{acknowledgements}

\begin{appendix}

\section{New SNR candidates and flat spectrum shells showing faint or no IR emission}

\begin{table}[!ht]
\centering
\caption{New SNR candidates identified in this work.}
\label{tab:snr_candidates}
\begin{tabular}{lrrr}
\toprule
Name & Diameter ($'$) & $l$ (deg) & $b$ (deg) \\
\midrule
G32.64+0.55   &  12.5 &  32.641 &  0.550 \\
G35.38$-$0.27 &  12.2 &  35.386 & $-$0.274 \\
G35.42$-$0.08 &  13.8 &  35.428 & $-$0.085 \\
G36.02$-$0.03 &  38.6 &  36.017 & $-$0.027 \\
G37.05$-$0.49 &  27.4 &  37.046 & $-$0.494 \\
G40.20+0.77   &  10.5 &  40.197 &  0.768 \\
G41.02+0.06   &  43.0 &  41.017 &  0.059 \\
G42.09$-$0.21 &   8.6 &  42.095 & $-$0.208 \\
G45.26+0.17   &  16.3 &  45.264 &  0.174 \\
G51.21$-$0.18 &  14.8 &  51.214 & $-$0.184 \\
G53.23+0.60   &  33.2 &  53.233 &  0.604 \\
G55.74+0.22   &  24.8 &  55.744 &  0.220 \\
G56.49$-$0.78 &  14.4 &  56.487 & $-$0.778 \\
G56.71+0.35   &  10.2 &  56.710 &  0.352 \\
G57.11+0.38   &  22.2 &  57.112 &  0.384 \\
G60.2+0.67    &  44.6 &  60.223 &  0.667 \\
G69.67+1.01   &  28.0 &  69.670 &  1.006 \\
G71.18+0.88   &  41.2 &  71.179 &  0.883 \\
G72.01$-$0.31 &  30.4 &  72.013 & $-$0.308 \\
G72.11+1.72   & 142.8 &  72.117 &  1.719 \\
G72.74+0.21   &  45.0 &  72.739 &  0.210 \\
G74.49$-$4.12 &  40.2 &  74.488 & $-$4.124 \\
G100.97+1.57  &  27.0 & 100.966 &  1.567 \\
G123.4+1.0    &  26.0 & 123.400 &  1.000 \\
\bottomrule
\end{tabular}
\end{table}

\begin{table}[!ht]
\centering
\caption{Flat spectrum shells showing faint or no IR emission identified in this work.}
\label{tab:irdark_thermal}
\begin{tabular}{lrrr}
\toprule
Name & Diameter ($'$) & $l$ (deg) & $b$ (deg) \\
\midrule
G32.24$-$0.22\tablefootmark{a} & 7.6 &  32.237 & $-$0.216 \\
G33.85$+$0.06\tablefootmark{a} & 1.2 &  33.848 & 0.062 \\
G36.67$-$0.14 & 17.6 &  36.669 & $-$0.138 \\
G42.62+0.16\tablefootmark{a}   &  7.6 &  42.623 &  0.162 \\
G42.25+0.19   &  6.2 &  42.255 &  0.191 \\
G46.53$-$0.00 & 14.0 &  46.535 & $-$0.000 \\
G53.79$+$0.88 & 22.1 &  53.794 & 0.880 \\
G53.57$+$0.75 & 13.6 &  53.566 & 0.750 \\
\bottomrule
\end{tabular}
\tablefoot{
\tablefoottext{a}{Listed as SNR candidates: G032.229$-$00.210, G033.847+00.062, G042.623+00.174 by \cite{anderson25}.}
}
\end{table}

\clearpage

\section{Atlas of per-source LOFAR maps}
\label{app:atlas}

This appendix collects the per-source maps that accompany the article. In the
journal version they are online-only supplementary material, deposited on
Zenodo at \url{https://doi.org/10.5281/zenodo.22536839}; they are reproduced
here, with the same numbering, so that the references in the text resolve
within this document.

\begin{figure*}[p]
\centering
\includegraphics[width=\textwidth]{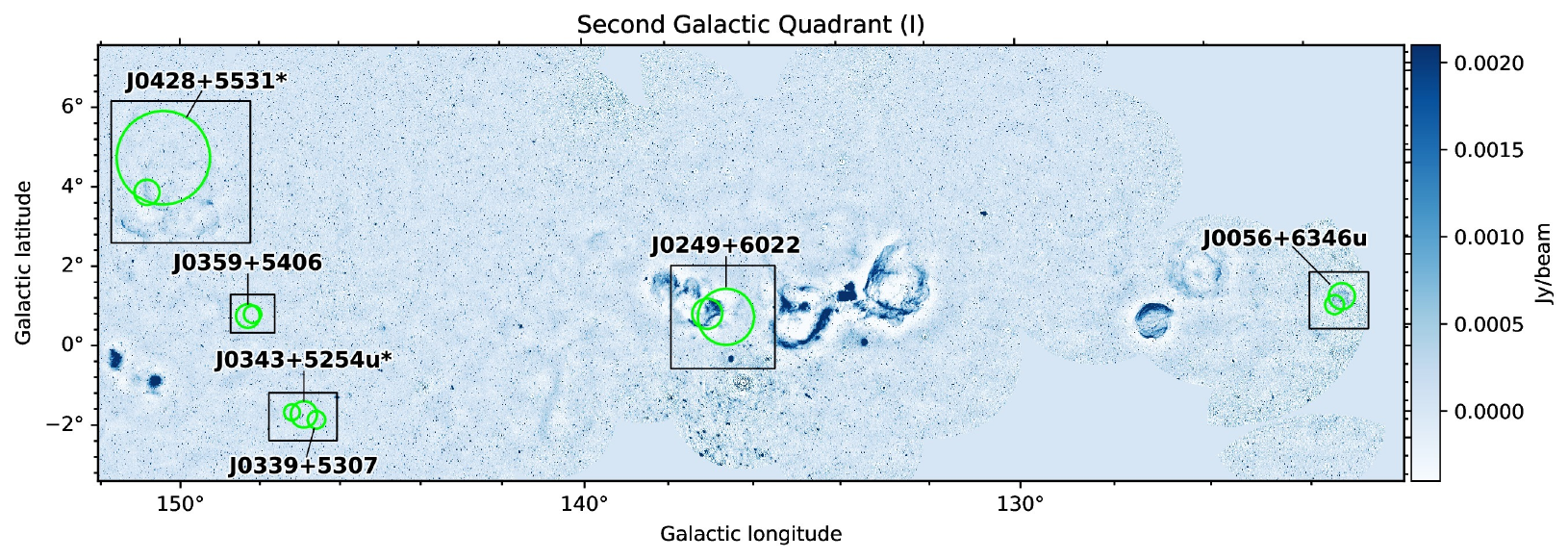}\\[4pt]
\includegraphics[width=\textwidth]{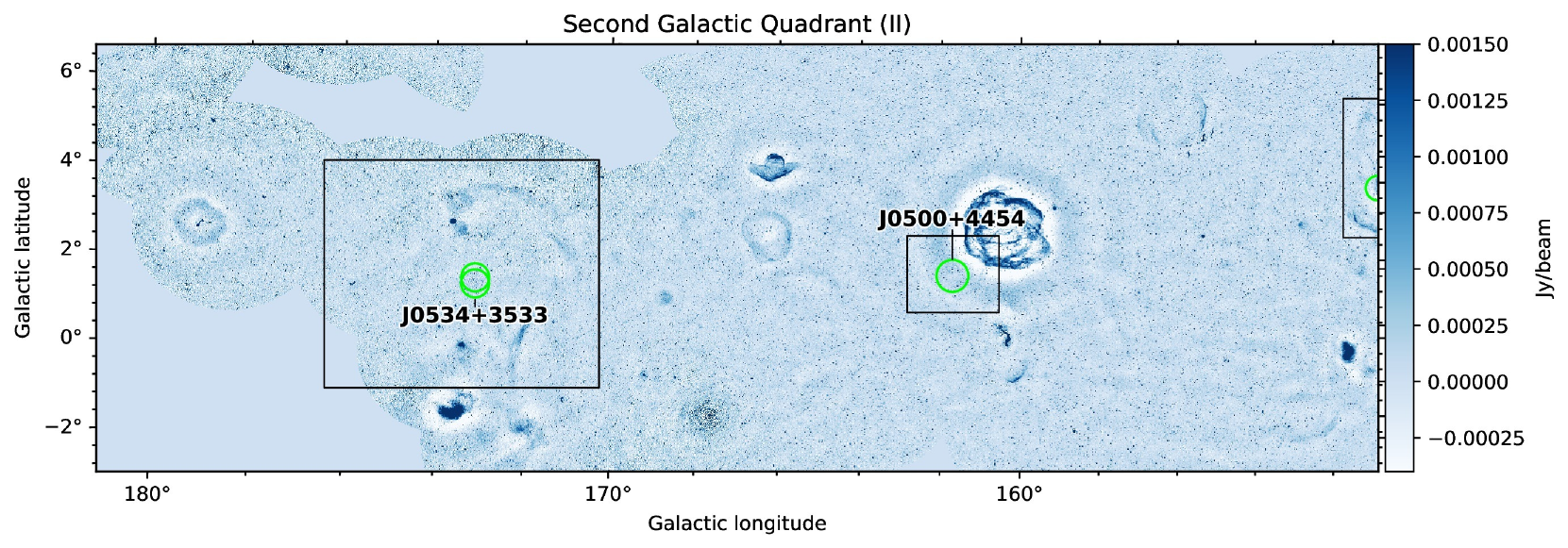}\\[4pt]
\includegraphics[width=0.75\textwidth]{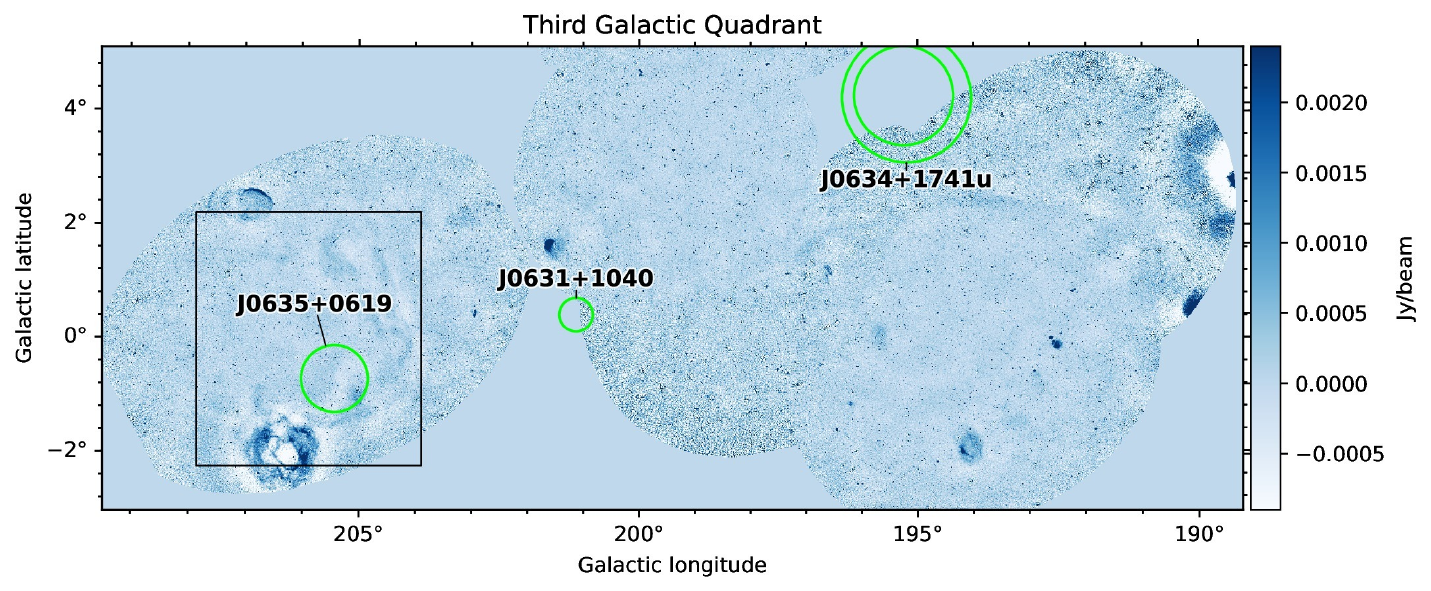}
\caption{LOFAR maps of the second (top two panels) and third (bottom panel) Galactic quadrants. The black boxes correspond to the cutouts shown in Fig. \ref{fig:secondQ_1}. The radius of the 1LHAASO source regions is $r_{39}$, the 39\% containment radius of the two-dimensional Gaussian model given in Cao et al. (2024).}
\label{fig:finder_outer}
\end{figure*}

\begin{figure*}[p]
\centering
\includegraphics[width=0.32\textwidth]{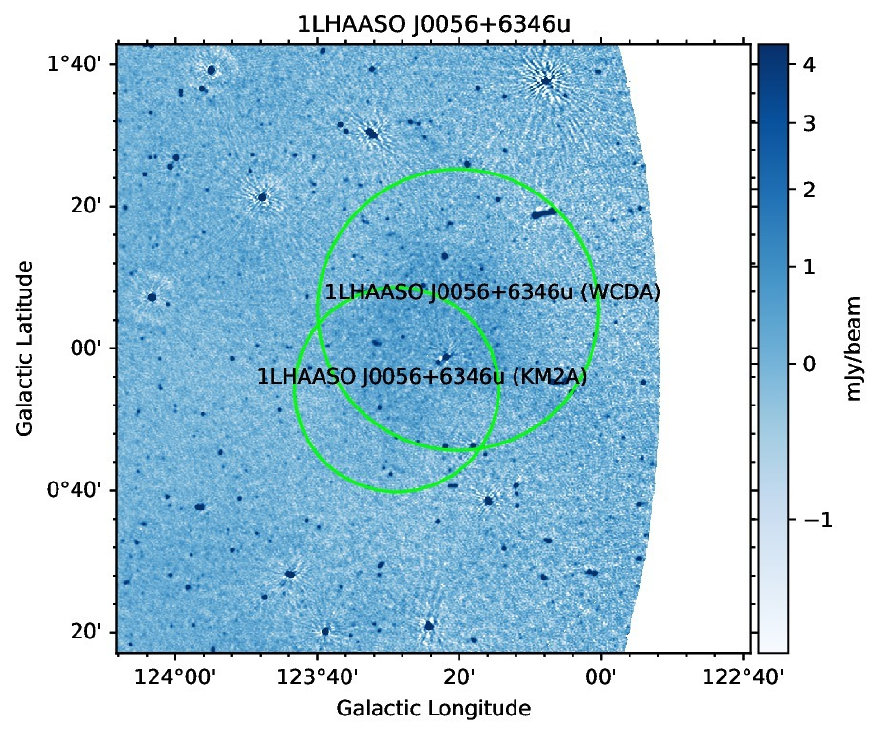}\hfill
\includegraphics[width=0.32\textwidth]{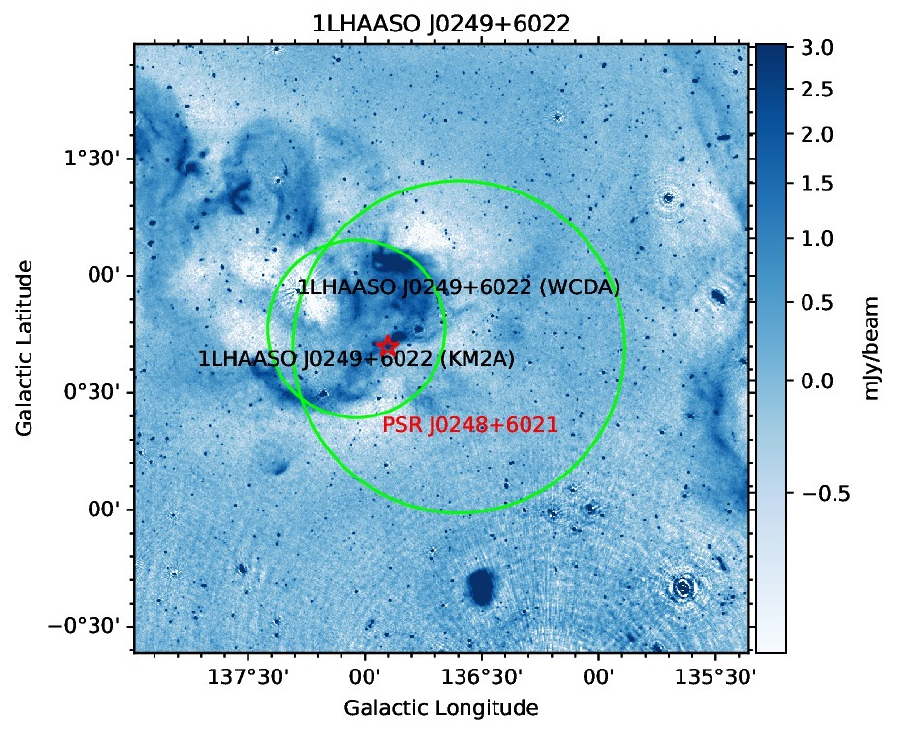}\hfill
\includegraphics[width=0.32\textwidth]{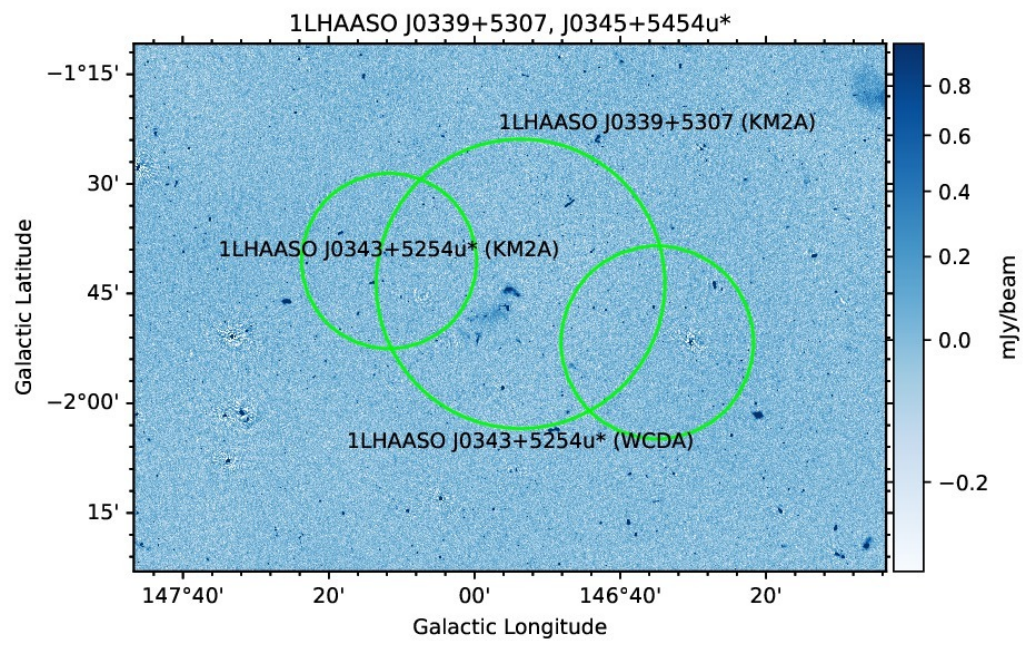}\\[3pt]
\includegraphics[width=0.32\textwidth]{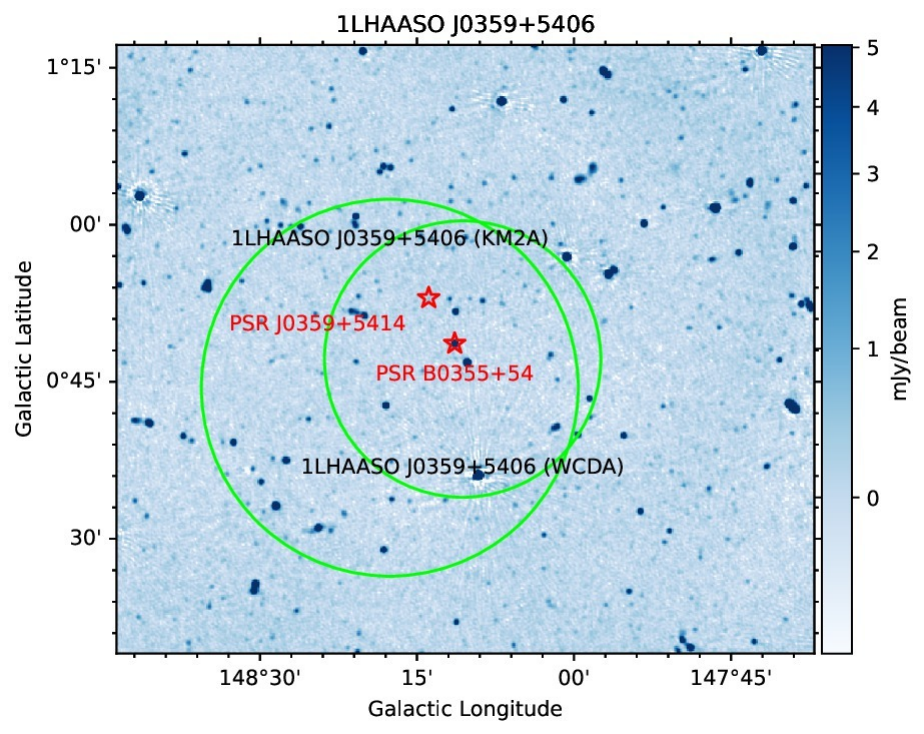}\hfill
\includegraphics[width=0.32\textwidth]{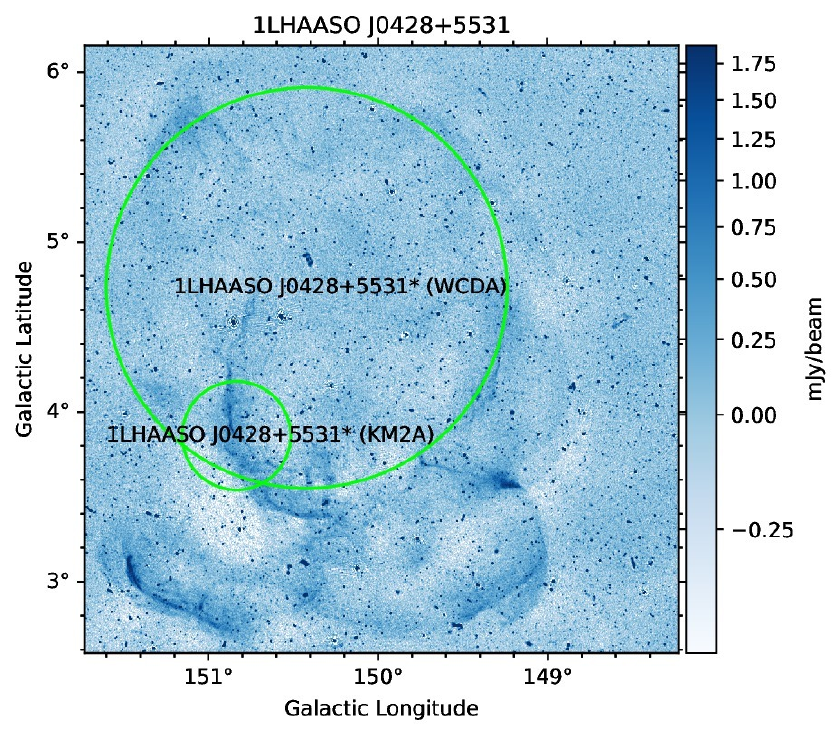}\hfill
\includegraphics[width=0.32\textwidth]{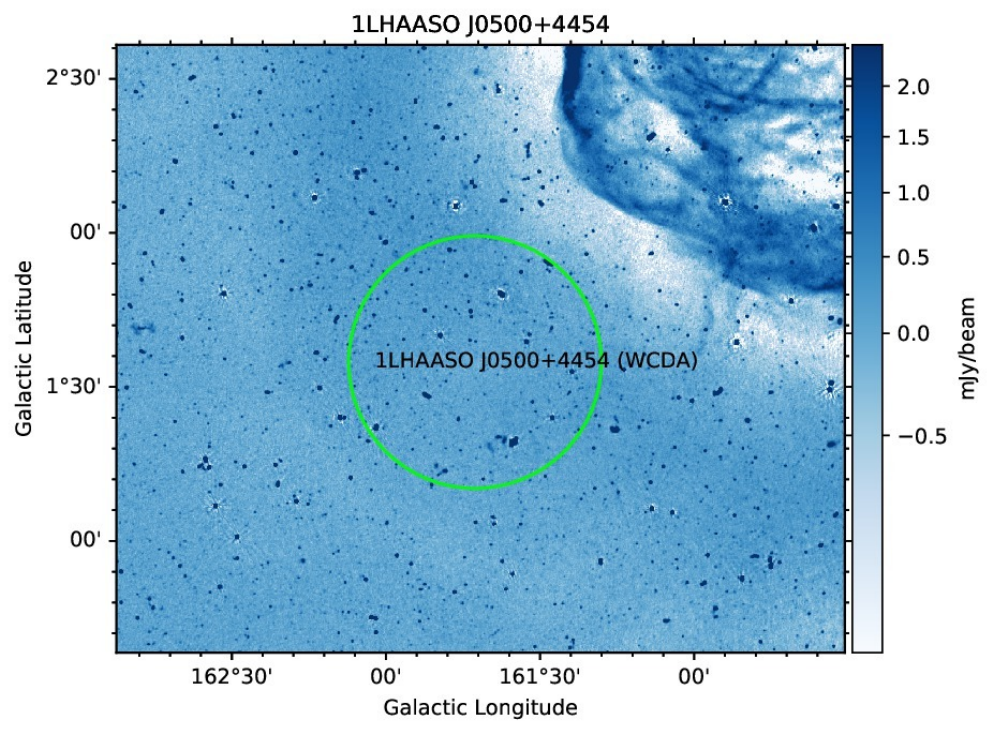}\\[3pt]
\includegraphics[width=0.32\textwidth]{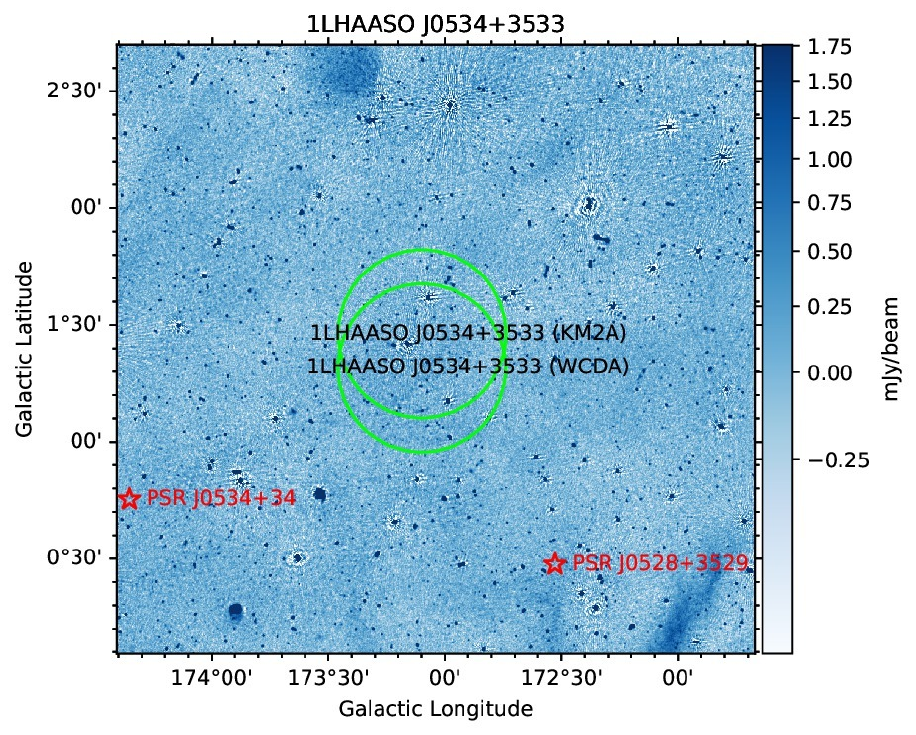}\hfill
\includegraphics[width=0.32\textwidth]{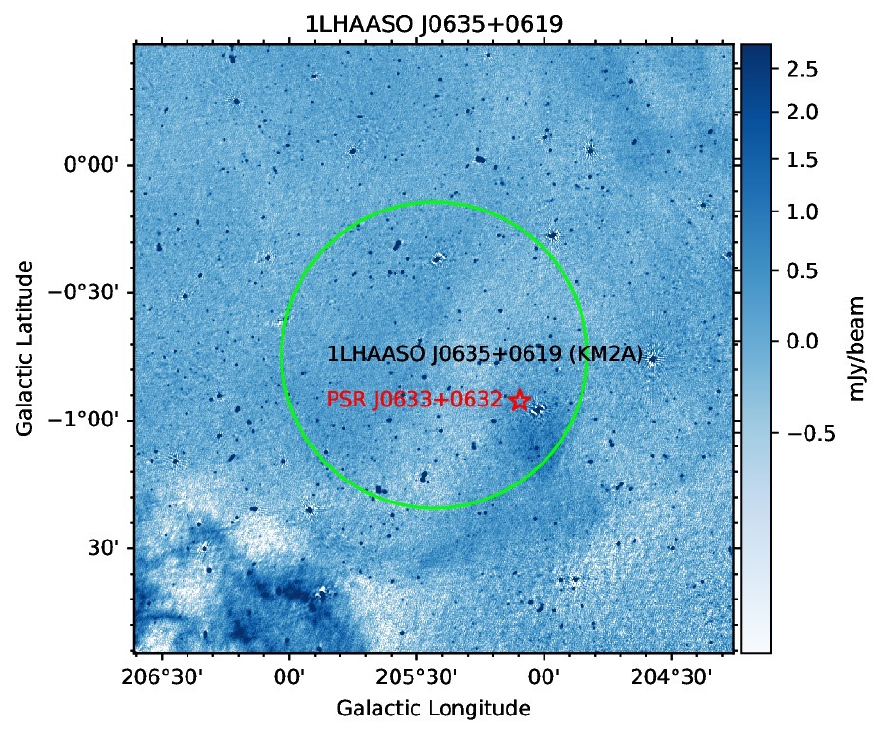}\hfill
\includegraphics[width=0.32\textwidth]{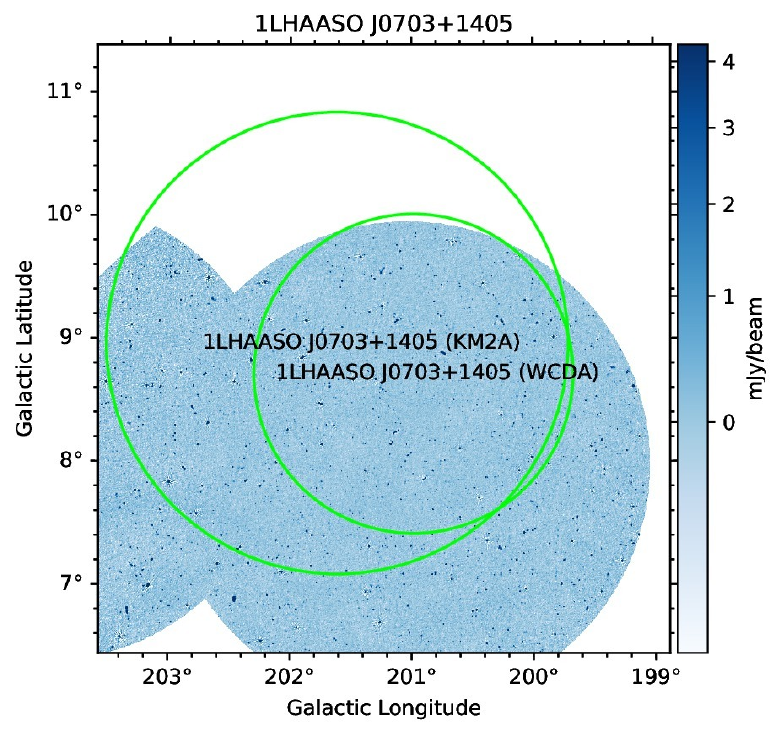}\\[3pt]
\includegraphics[width=0.32\textwidth,valign=t]{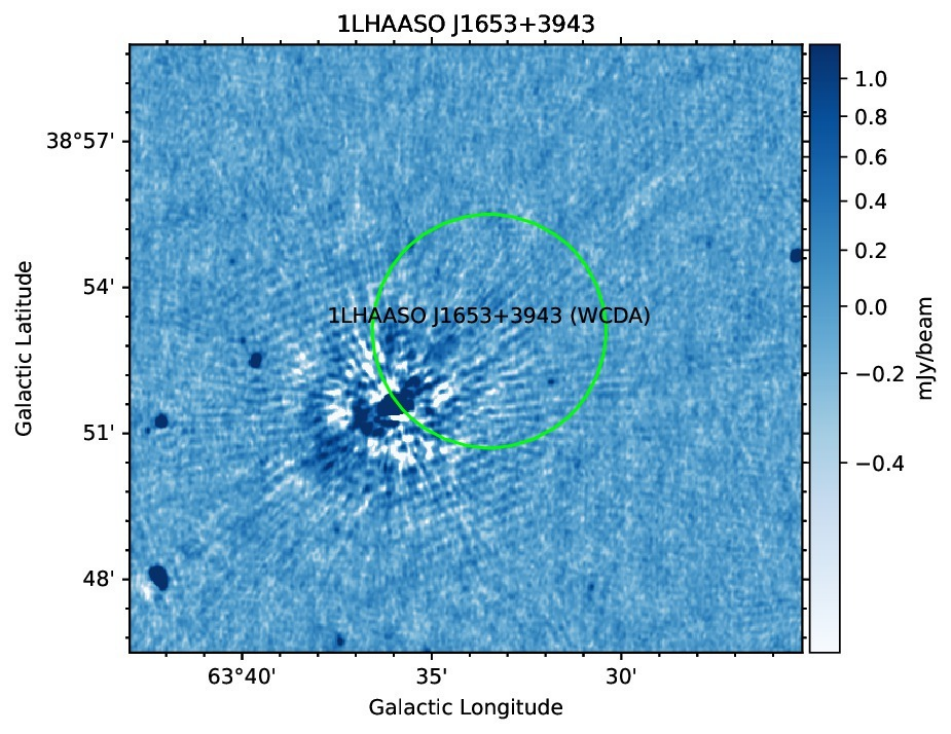}\hfill
\includegraphics[width=0.55\textwidth,valign=t]{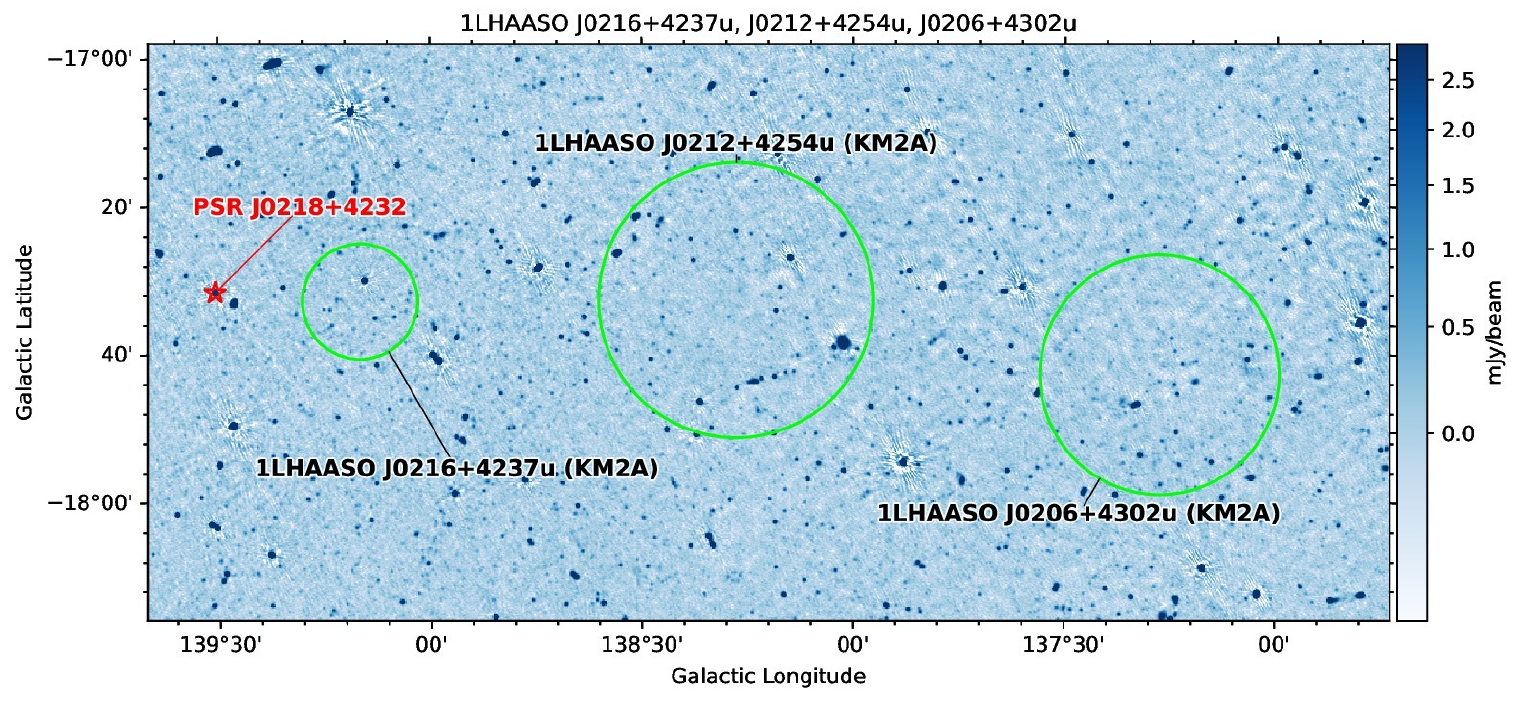}
\caption{LOFAR $20''$ (144~MHz) cutouts of the 1LHAASO sources in the second and third Galactic quadrants, in reading order:
1LHAASO~J0056+6346u;
1LHAASO~J0249+6022;
1LHAASO~J0339+5307 and 1LHAASO~J0343+5254u;
1LHAASO~J0359+5406;
1LHAASO~J0428+5531;
1LHAASO~J0500+4454;
1LHAASO~J0534+3533;
1LHAASO~J0635+0619;
1LHAASO~J0703+1405;
1LHAASO~J1653+3943; and, in the wide bottom-right panel, the three high
Galactic latitude ($b\approx-17.6^\circ$) sources 1LHAASO~J0216+4237u,
1LHAASO~J0212+4254u and 1LHAASO~J0206+4302u. The radius of the 1LHAASO source regions is $r_{39}$, the 39\% containment radius of the two-dimensional Gaussian model given in Cao et al. (2024). Red stars mark pulsars discussed in the main text (positions from ATNF, Manchester et al. 2005).}
\label{fig:secondQ_1}
\label{fig:thirdQ}
\end{figure*}

\begin{figure*}[p]
\centering
\includegraphics[width=0.8\textwidth]{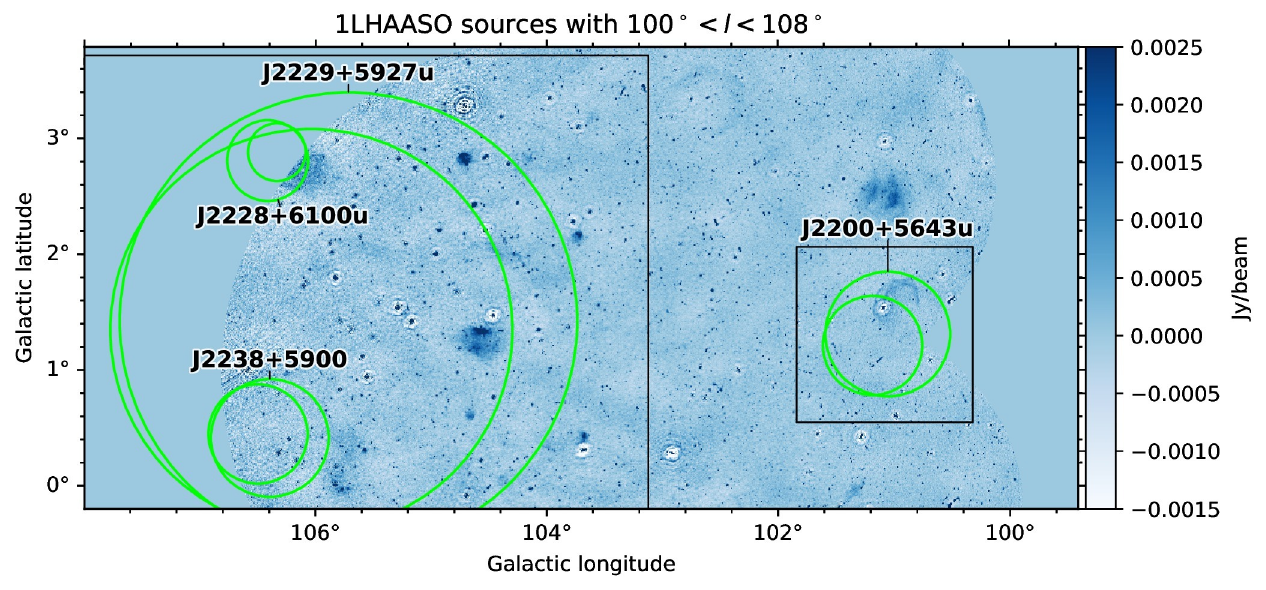}\\[4pt]
\includegraphics[width=0.32\textwidth]{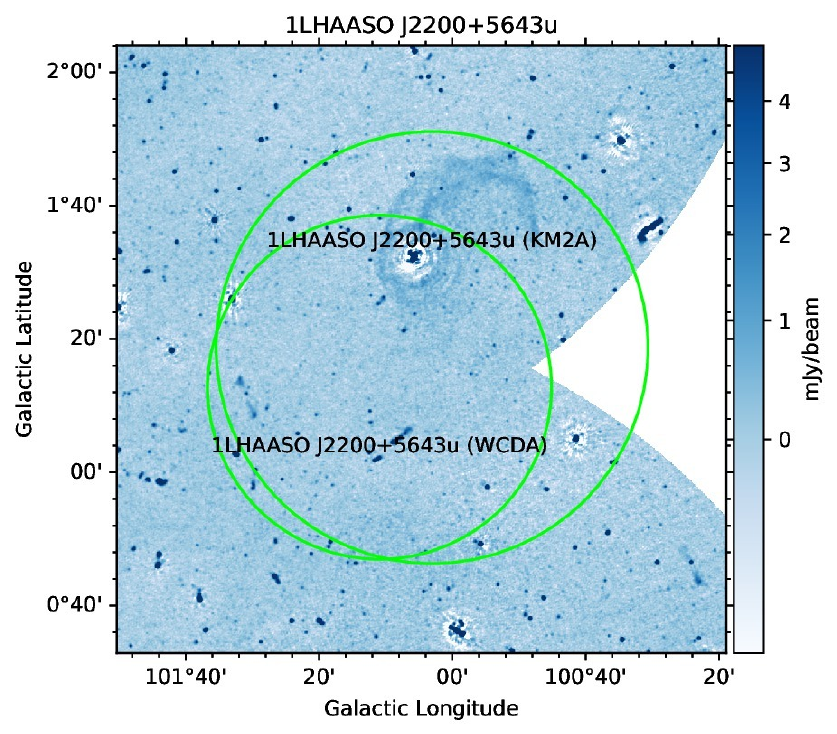}\hfill
\includegraphics[width=0.32\textwidth]{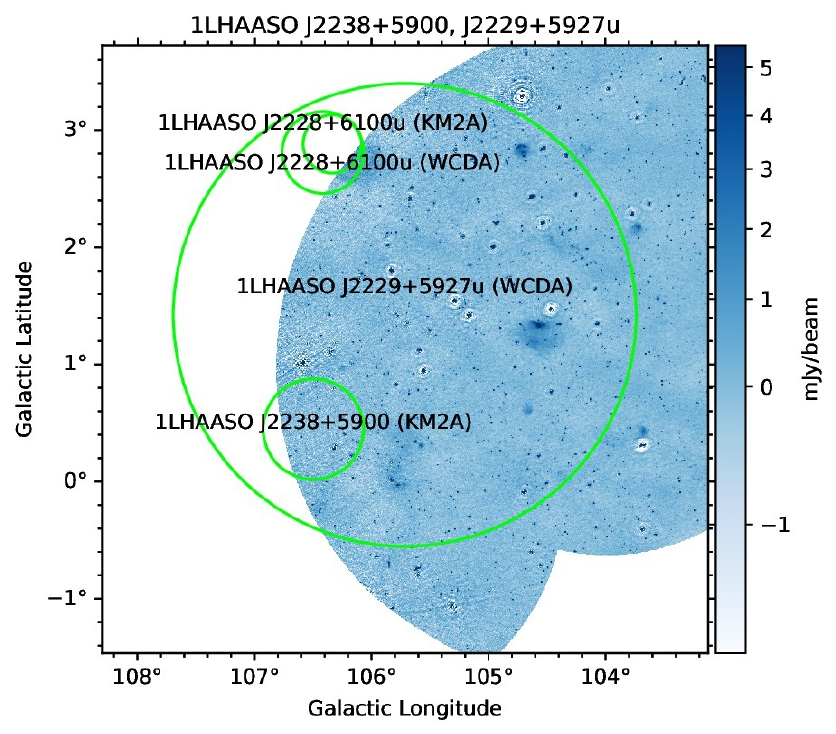}\hfill
\includegraphics[width=0.32\textwidth]{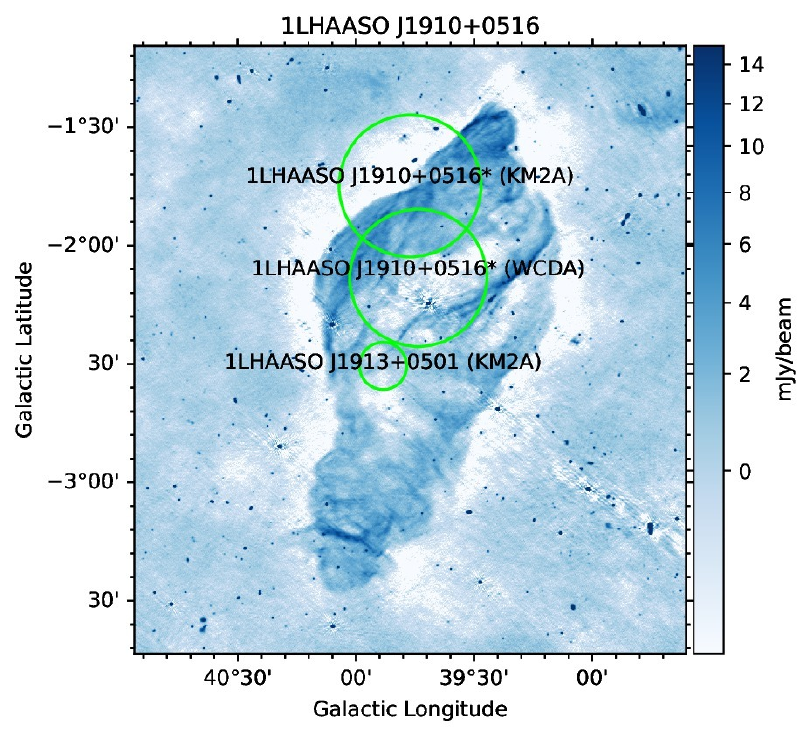}
\caption{\textit{Top:} LOFAR map of the region around Cygnus~A; the black boxes correspond to the cutouts below. \textit{Bottom}, LOFAR $20''$ (144~MHz) cutouts of:
1LHAASO~J2200+5643u;
1LHAASO~J2238+5900 and 1LHAASO~J2229+5927u, whose $r_{39}$ also contains 1LHAASO~J2228+6100u; and
1LHAASO~J1910+0516 and 1LHAASO~J1913+0501 (the SS433/W50 system, which lies outside the regions covered by the finder charts above). The radius of the 1LHAASO source regions is $r_{39}$, the 39\% containment radius of the two-dimensional Gaussian model given in Cao et al. (2024).}
\label{fig:Q1_5}
\end{figure*}

\begin{figure*}[p]
\centering
\includegraphics[width=0.32\textwidth,height=0.19\textheight,keepaspectratio]{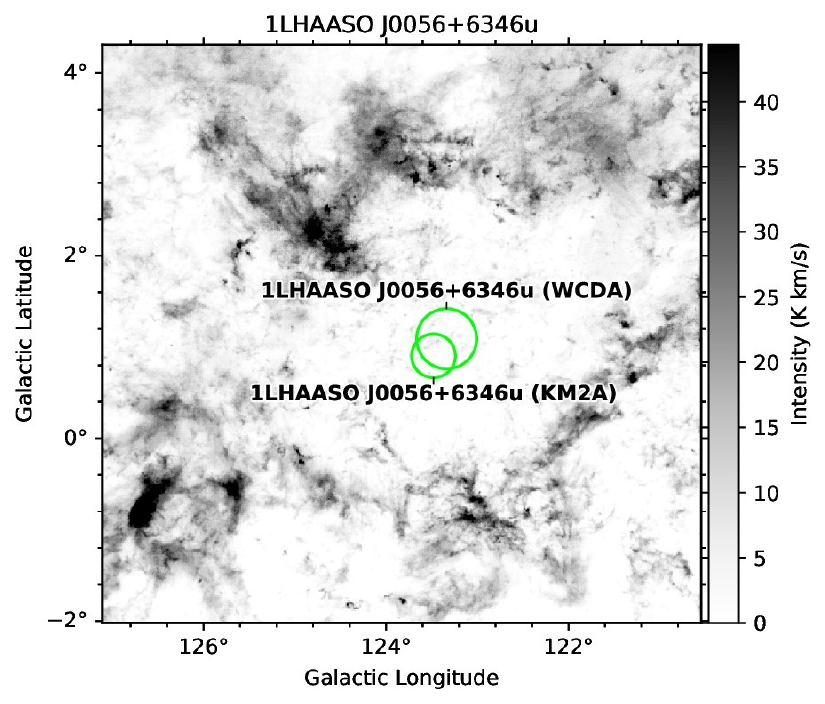}\hfill
\includegraphics[width=0.32\textwidth,height=0.19\textheight,keepaspectratio]{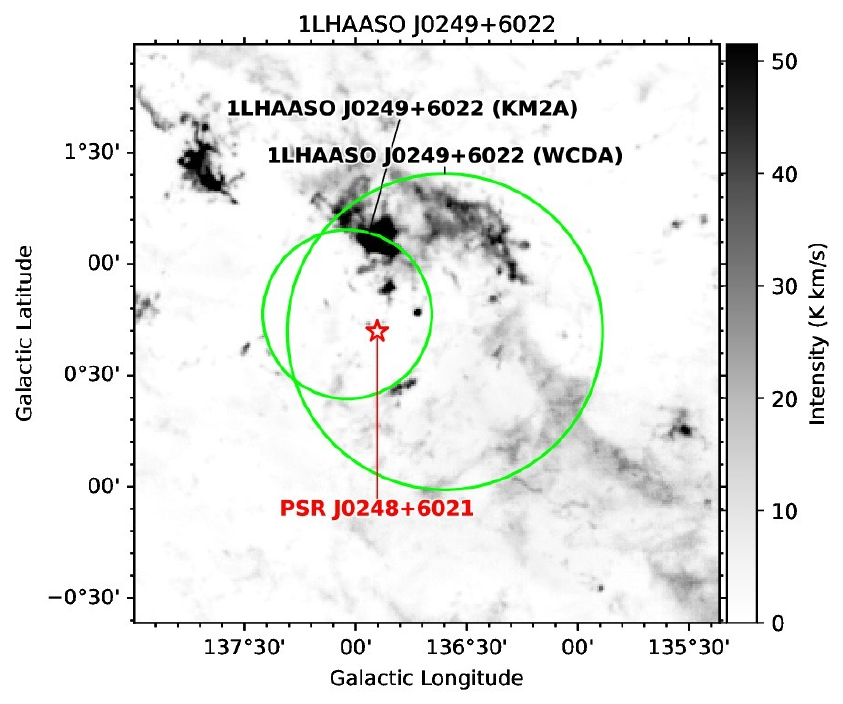}\hfill
\includegraphics[width=0.32\textwidth,height=0.19\textheight,keepaspectratio]{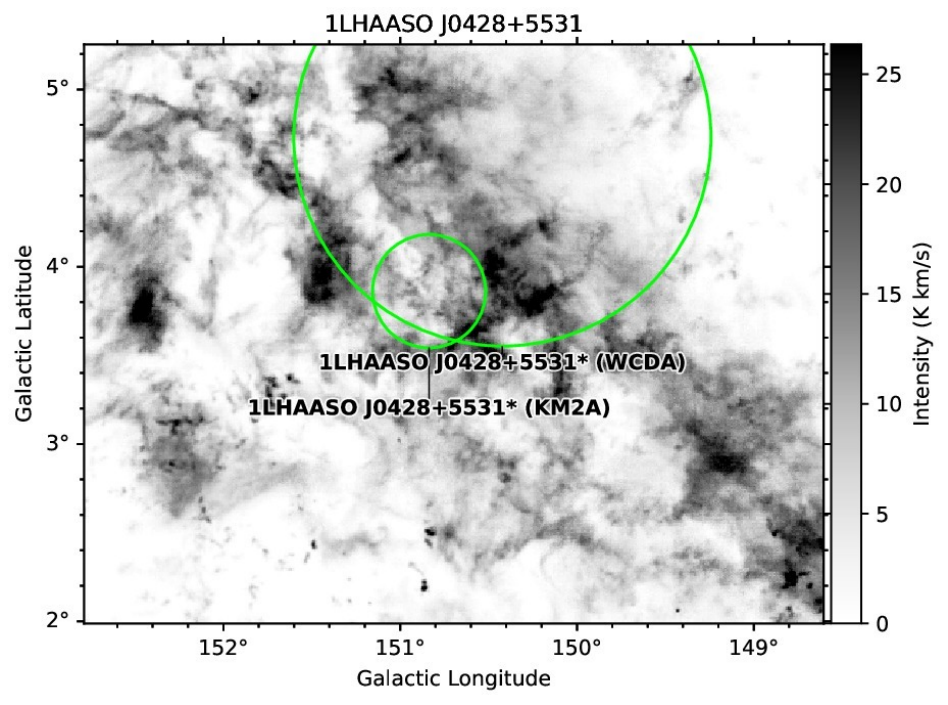}\\[3pt]
\includegraphics[width=0.32\textwidth,height=0.19\textheight,keepaspectratio]{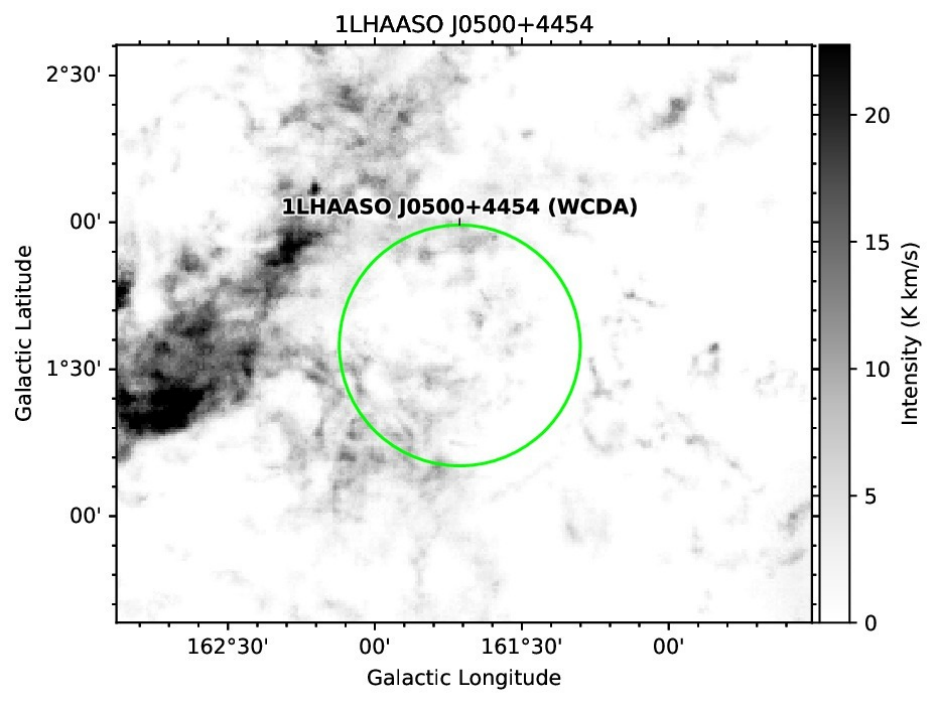}\hfill
\includegraphics[width=0.32\textwidth,height=0.19\textheight,keepaspectratio]{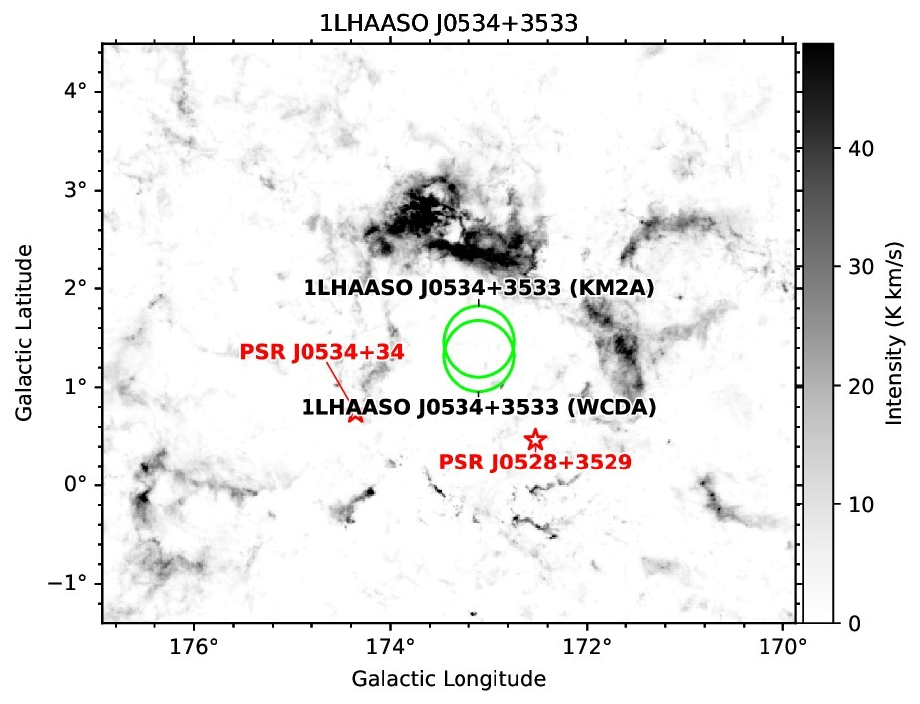}\hfill
\includegraphics[width=0.32\textwidth,height=0.19\textheight,keepaspectratio]{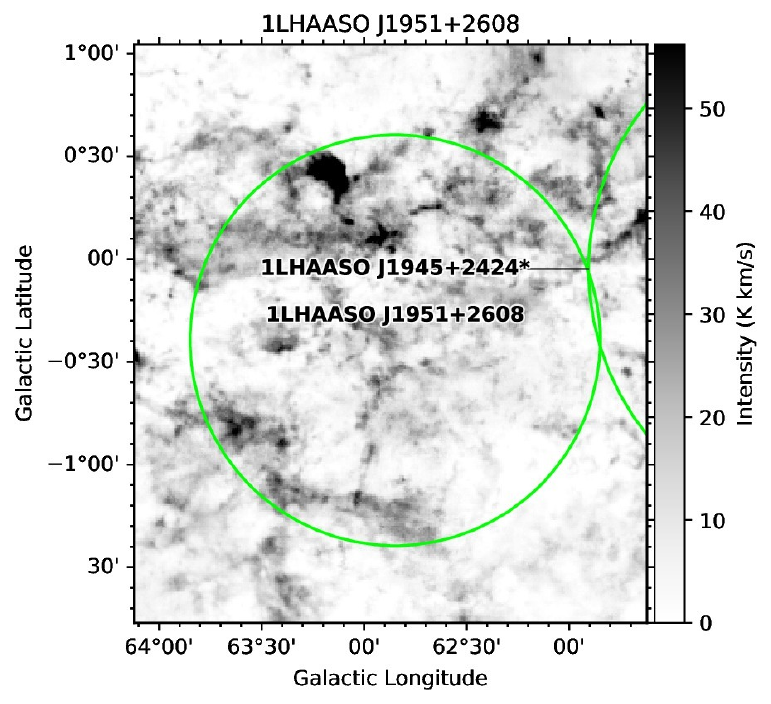}\\[3pt]
\includegraphics[width=0.32\textwidth,height=0.19\textheight,keepaspectratio]{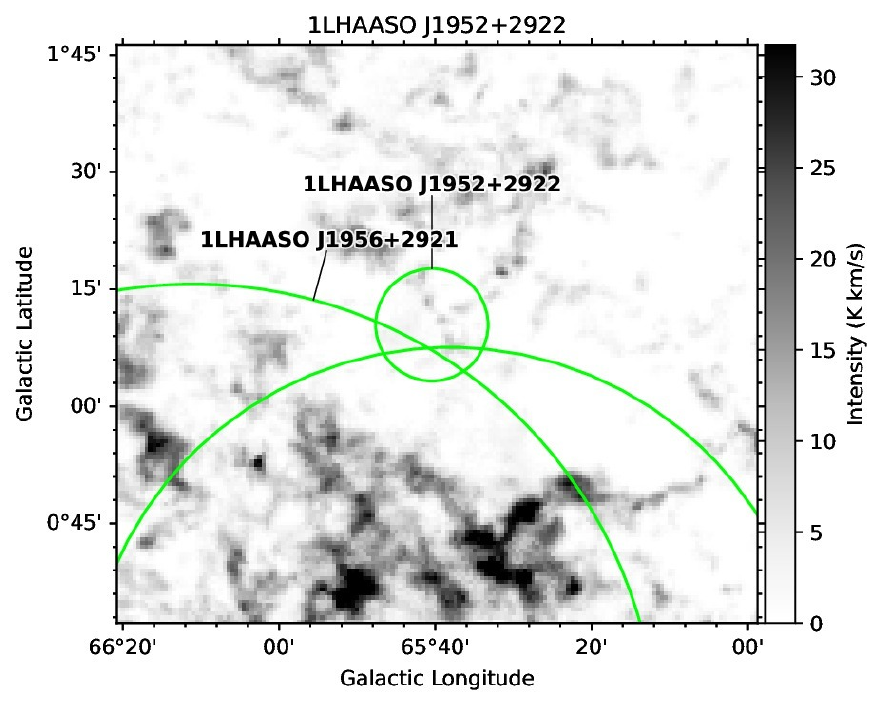}\hfill
\includegraphics[width=0.32\textwidth,height=0.19\textheight,keepaspectratio]{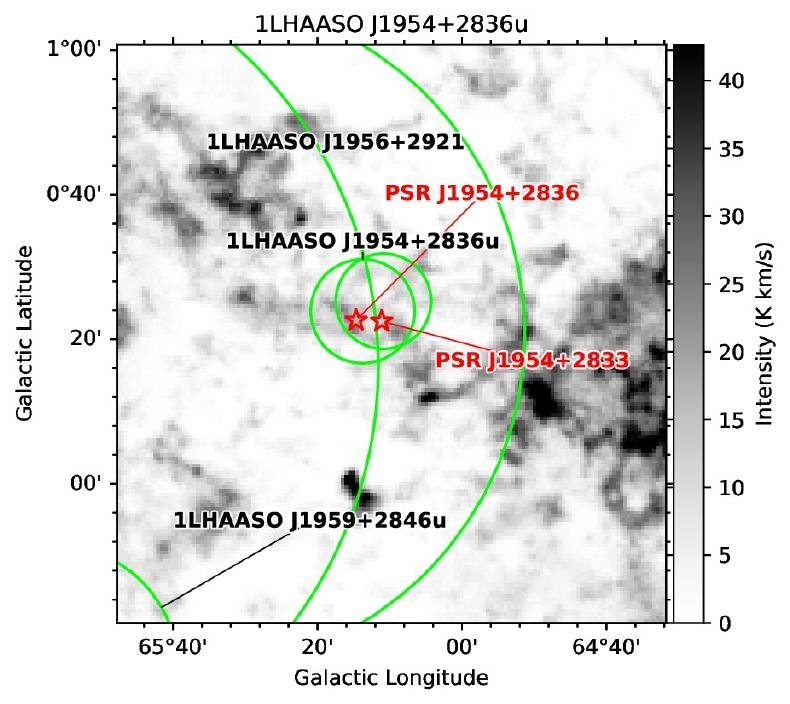}\hfill
\includegraphics[width=0.32\textwidth,height=0.19\textheight,keepaspectratio]{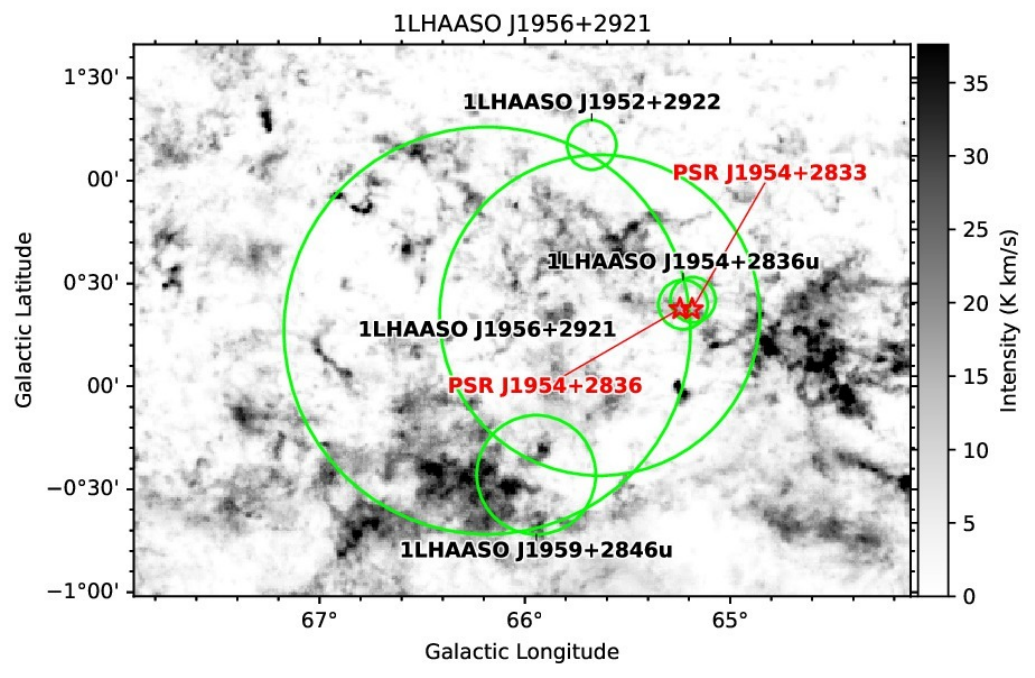}\\[3pt]
\includegraphics[width=0.32\textwidth,height=0.19\textheight,keepaspectratio]{mwisp/J2005+3050_mwisp.pdf}\hfill
\includegraphics[width=0.32\textwidth,height=0.19\textheight,keepaspectratio]{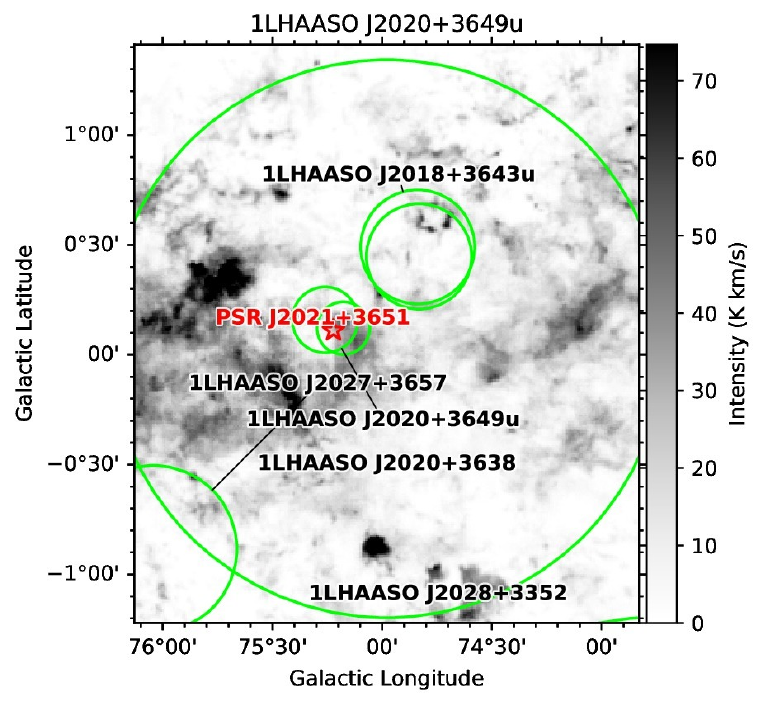}\hfill
\includegraphics[width=0.32\textwidth,height=0.19\textheight,keepaspectratio]{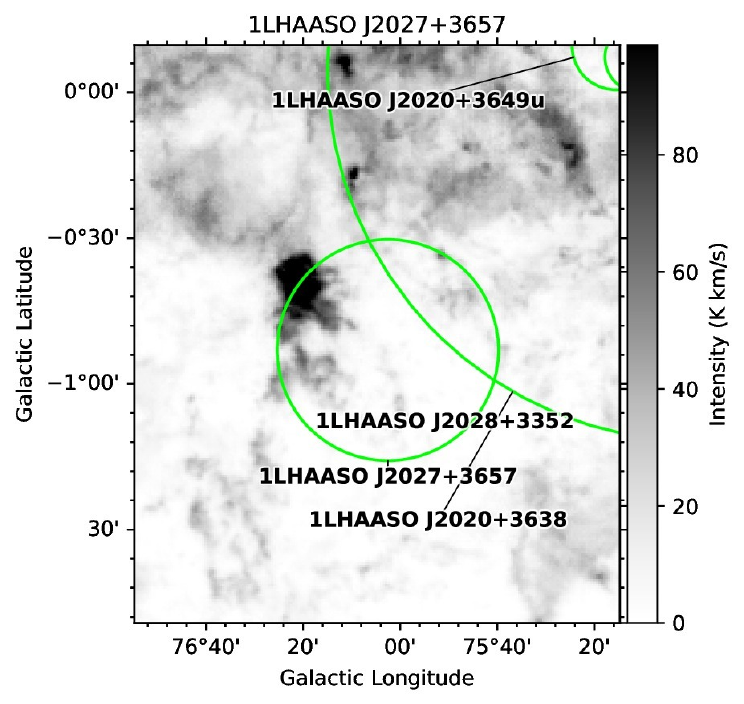}
\caption{MWISP $^{12}$CO $J=1-0$ maps, integrated over $-94$ to
$+24$~km~s$^{-1}$, of the 1LHAASO sources with CO coverage. Green circles are
the 1LHAASO components; red stars mark pulsars discussed in the main text
(positions from ATNF, Manchester et al. 2005). The radius of the 1LHAASO
regions is the 39\% containment radius of the two-dimensional Gaussian model
given in Cao et al. (2024). In reading order: 1LHAASO~J0056+6346u;
1LHAASO~J0249+6022; 1LHAASO~J0428+5531; 1LHAASO~J0500+4454;
1LHAASO~J0534+3533; 1LHAASO~J1951+2608; 1LHAASO~J1952+2922;
1LHAASO~J1954+2836u; 1LHAASO~J1956+2921; 1LHAASO~J2005+3050;
1LHAASO~J2020+3649u; and 1LHAASO~J2027+3657.}
\label{fig:mwisp_all}
\end{figure*}

\begin{figure*}[p]
\centering
\includegraphics[width=0.32\textwidth,height=0.19\textheight,keepaspectratio]{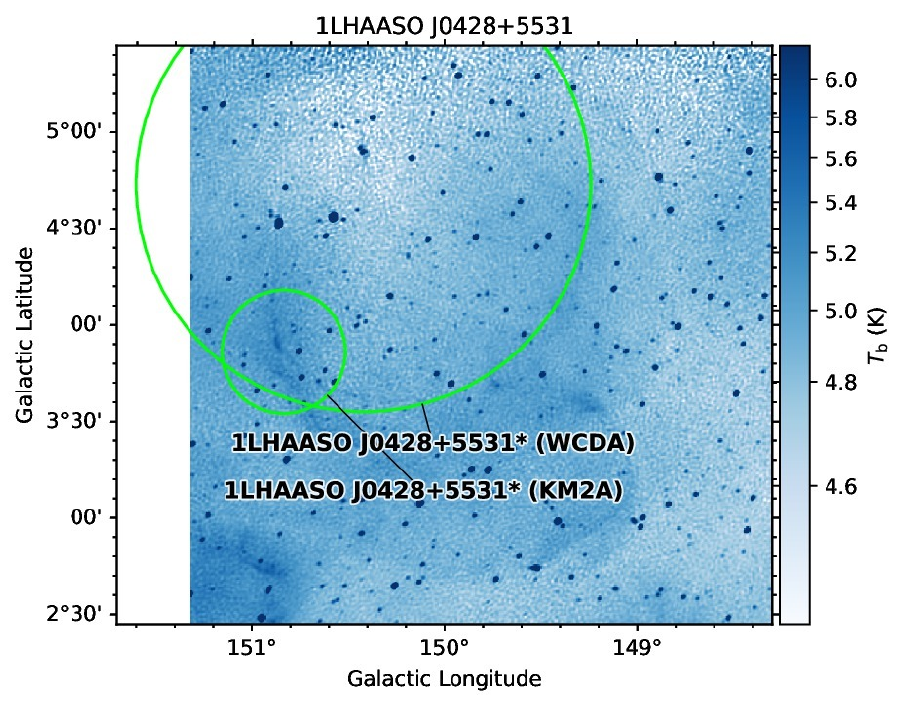}\hfill
\includegraphics[width=0.32\textwidth,height=0.19\textheight,keepaspectratio]{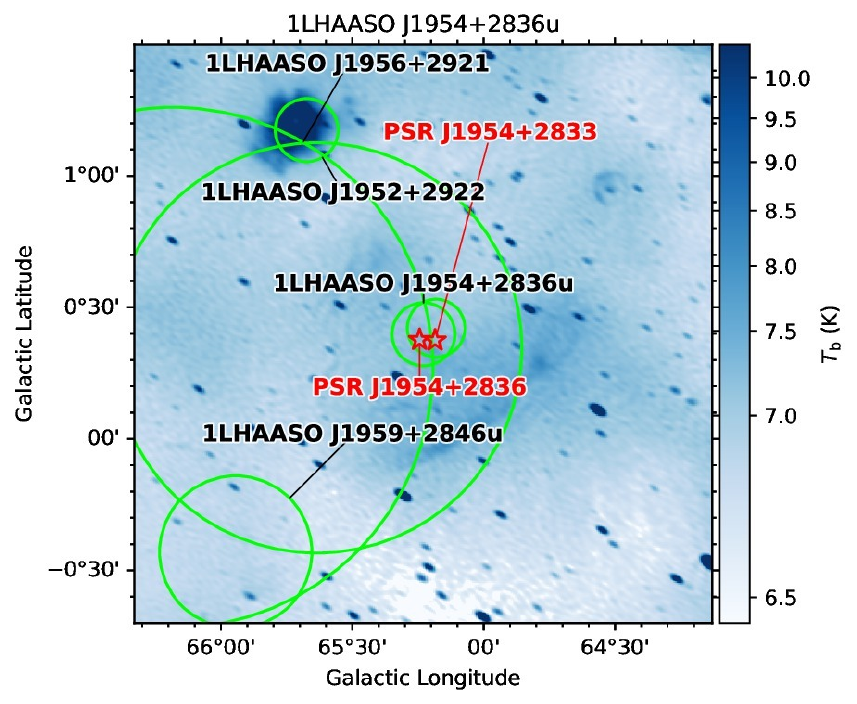}\hfill
\includegraphics[width=0.32\textwidth,height=0.19\textheight,keepaspectratio]{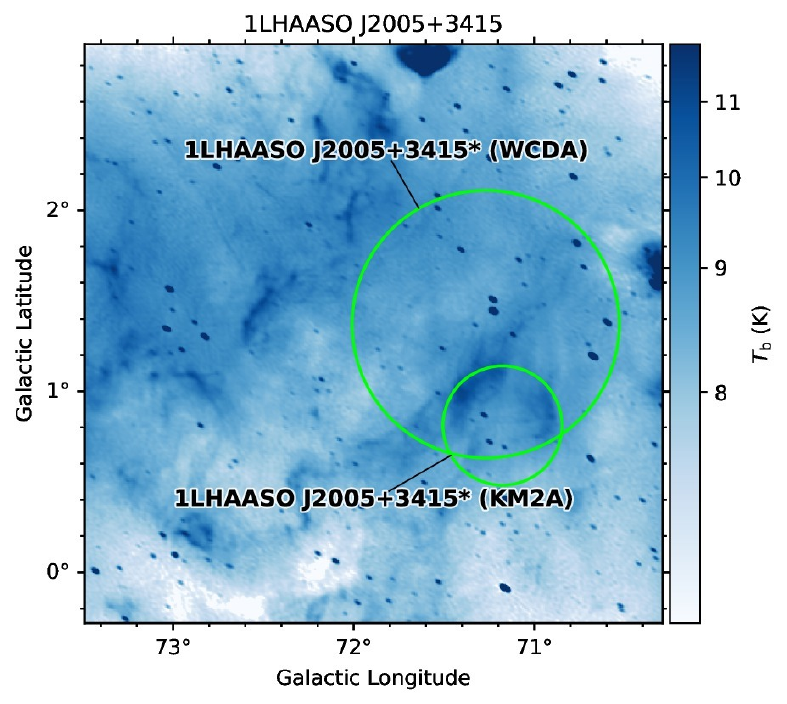}\\[3pt]
\includegraphics[width=0.32\textwidth,height=0.19\textheight,keepaspectratio]{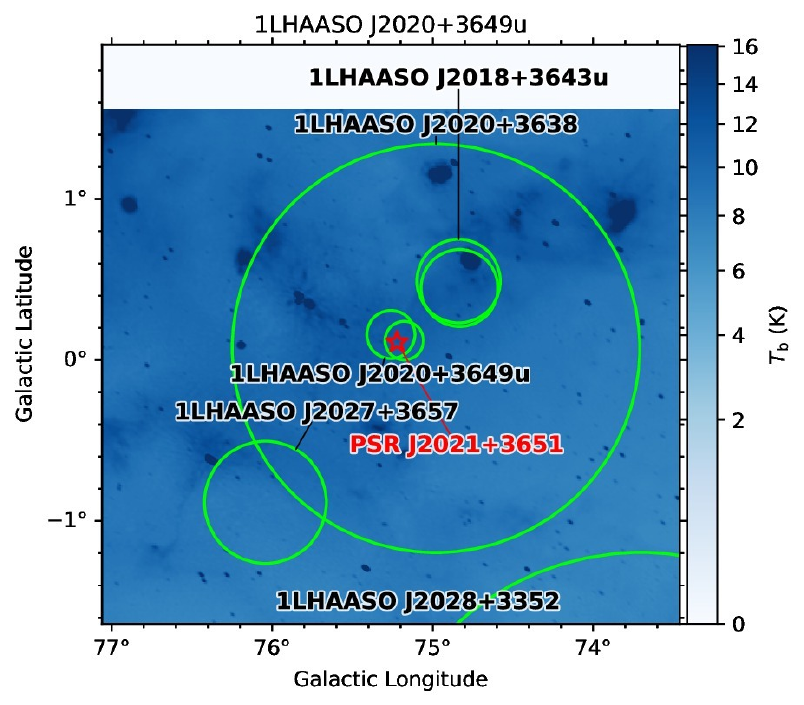}\hfill
\includegraphics[width=0.32\textwidth,height=0.19\textheight,keepaspectratio]{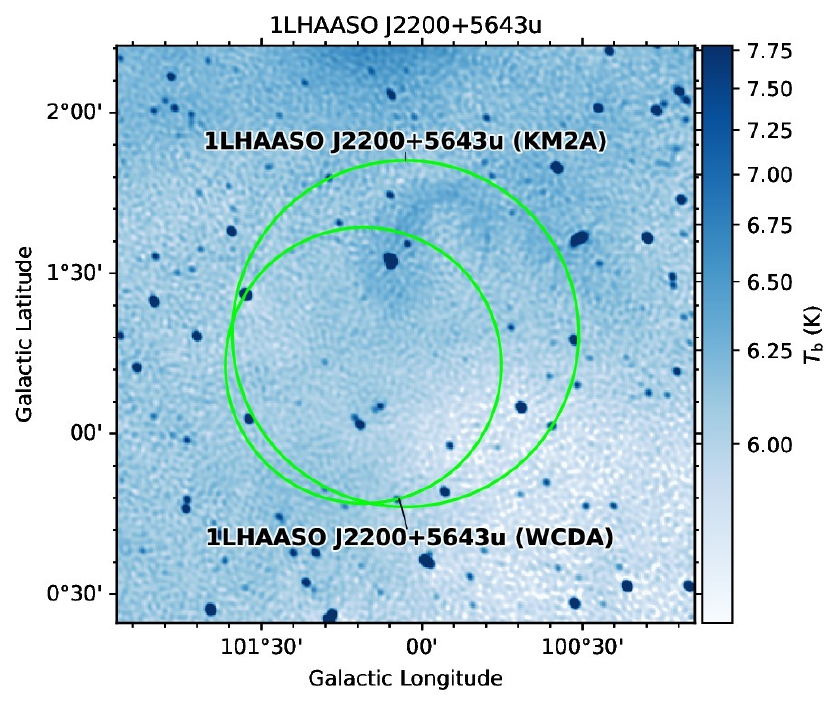}\hspace{0.32\textwidth}
\caption{CGPS 1420~MHz continuum maps of the fields referenced in the main text. In reading order:
1LHAASO~J0428+5531;
1LHAASO~J1954+2836u;
1LHAASO~J2005+3415;
1LHAASO~J2020+3649u; and 1LHAASO~J2200+5643u. Green circles are the $r_{39}$ regions of the 1LHAASO components,
labelled by source name; red stars mark pulsars discussed in the text. The
radius of the 1LHAASO regions is the 39\% containment radius of the
two-dimensional Gaussian model given in Cao et al. (2024). The data are the
CGPS mosaics from the Canadian Astronomy Data Centre; the blank corner of the J0428+5531 panel lies beyond the CGPS
limit of $b=+5.5^\circ$.}
\label{fig:ancillary_new}
\end{figure*}

\begin{figure*}[p]
\centering
\includegraphics[width=\textwidth]{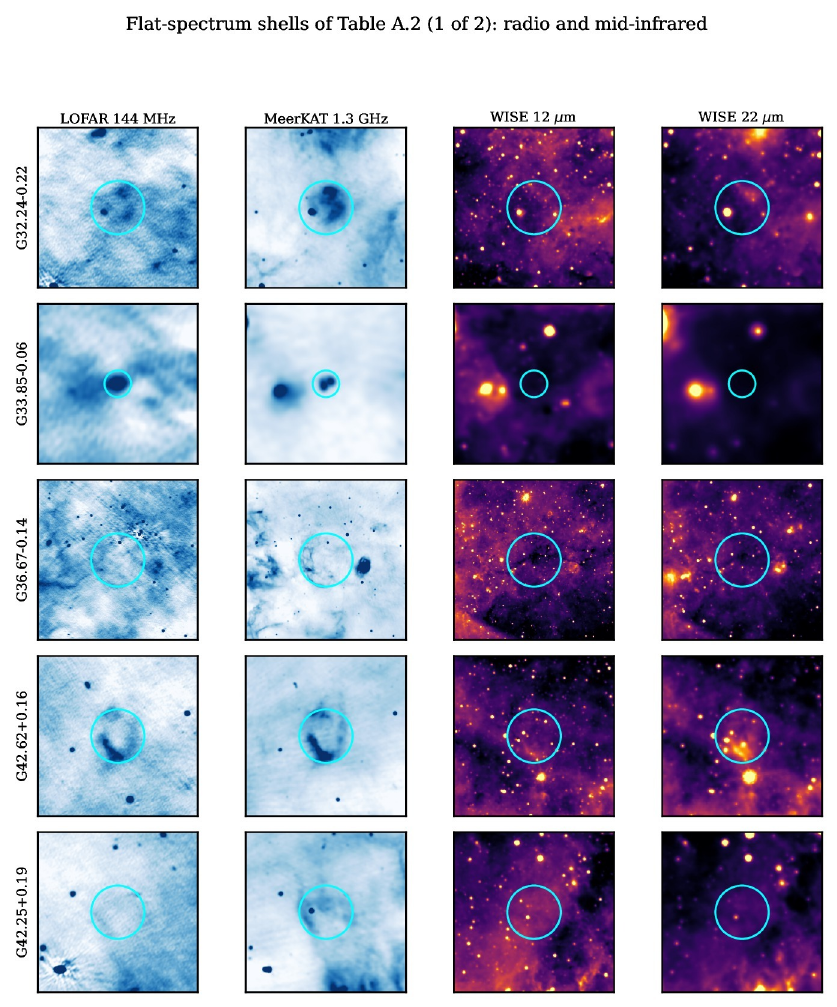}
\caption{The flat spectrum shells of Table \ref{tab:irdark_thermal} in the radio and in the mid-infrared, one row per source, all panels of a given row
on a common sky grid with the tabulated aperture drawn. From left to right:
LOFAR 144~MHz, MeerKAT 1.3~GHz, AllWISE $12~\mu$m and AllWISE $22~\mu$m. The
WISE panels are built from resampled HiPS tiles and are shown for morphology
only. This part: the first five shells of the table.}
\label{fig:shell_cutouts_a}
\end{figure*}

\begin{figure*}[p]
\centering
\includegraphics[width=\textwidth]{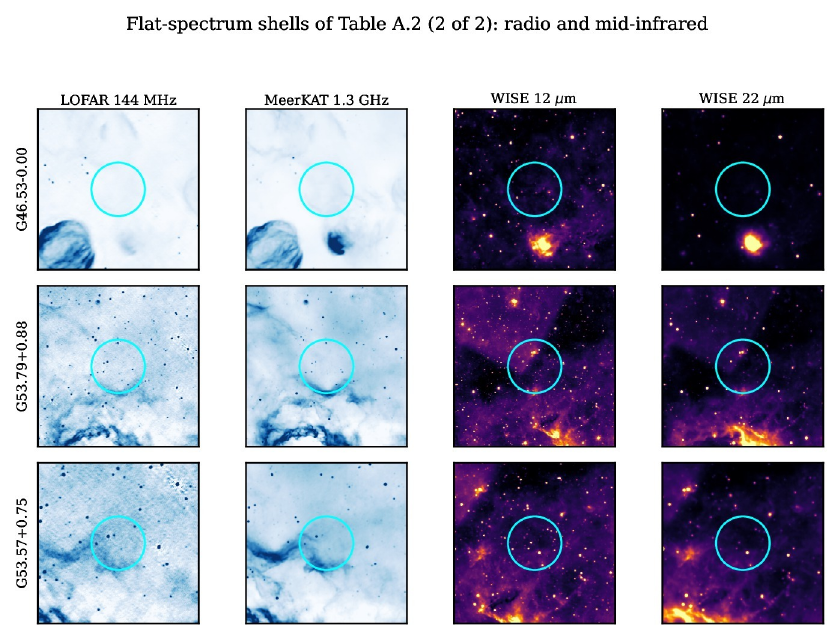}
\caption{Same as Fig. \ref{fig:shell_cutouts_a}, for the remaining three
shells.}
\label{fig:shell_cutouts_b}
\end{figure*}

Figure \ref{fig:finder_outer} shows the LoTSS coverage of the second and third Galactic quadrants, with the footprints of the per-source cutouts indicated; the cutouts themselves are shown in Fig. \ref{fig:secondQ_1}. Figures \ref{fig:Q1_1}--\ref{fig:Q1_5} show the corresponding finder charts and per-source maps for the first Galactic quadrant. The three high Galactic latitude sources, which fall outside the Galactic fields, are shown in the wide bottom-right panel of Fig. \ref{fig:secondQ_1}. Figure \ref{fig:mwisp_all} shows the MWISP $^{12}$CO maps of every source with CO coverage; Fig. \ref{fig:ancillary_new} shows the ancillary CGPS data referred to in the main text; and Figs. \ref{fig:shell_cutouts_a} and \ref{fig:shell_cutouts_b} show the flat spectrum shells of Table \ref{tab:irdark_thermal} in the radio and the mid-infrared. 


\begin{figure*}[p]
\centering
\includegraphics[width=0.8\textwidth]{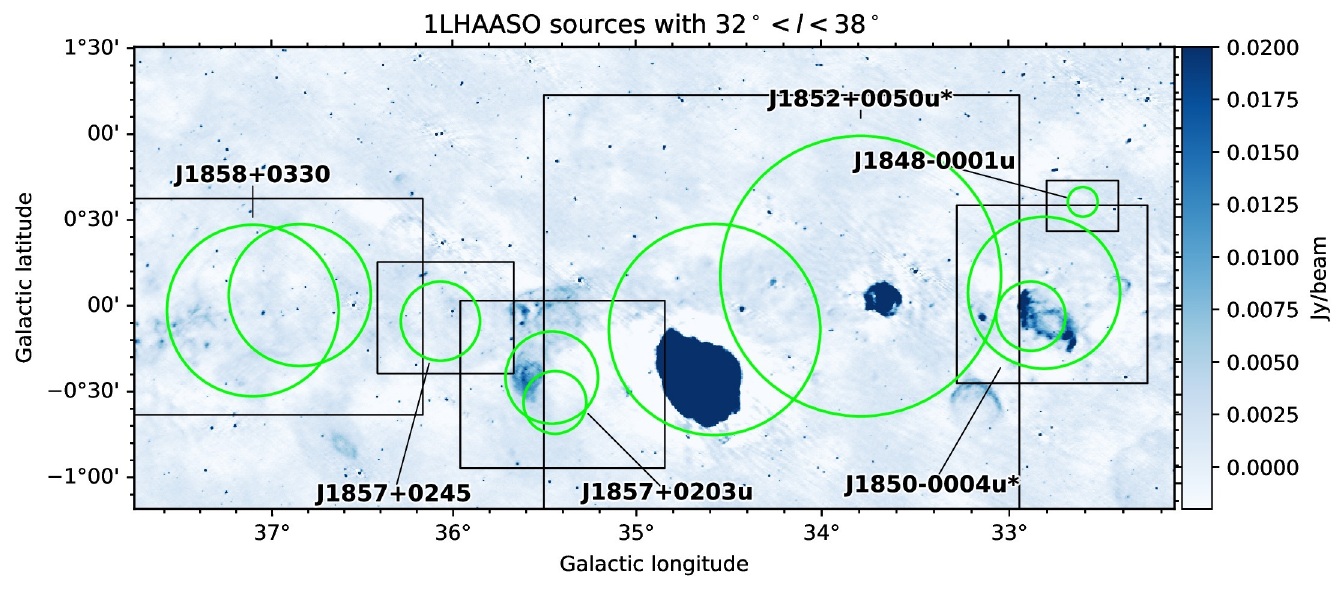}\\[4pt]
\includegraphics[width=\textwidth]{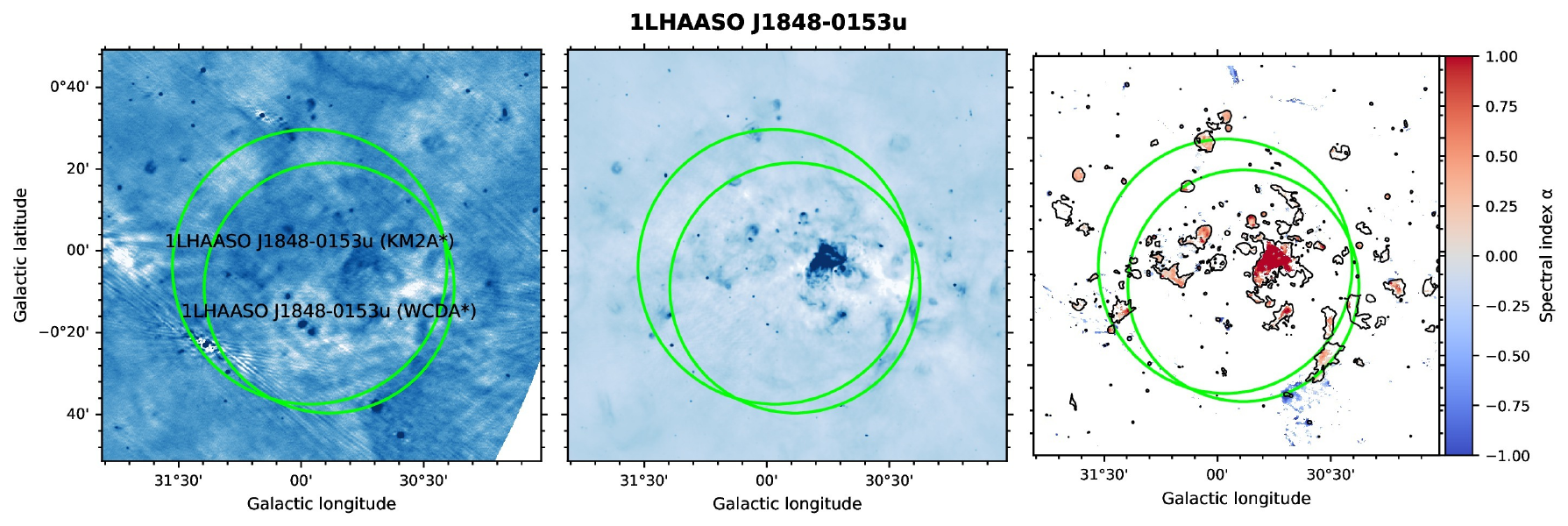}\\[3pt]
\includegraphics[width=\textwidth]{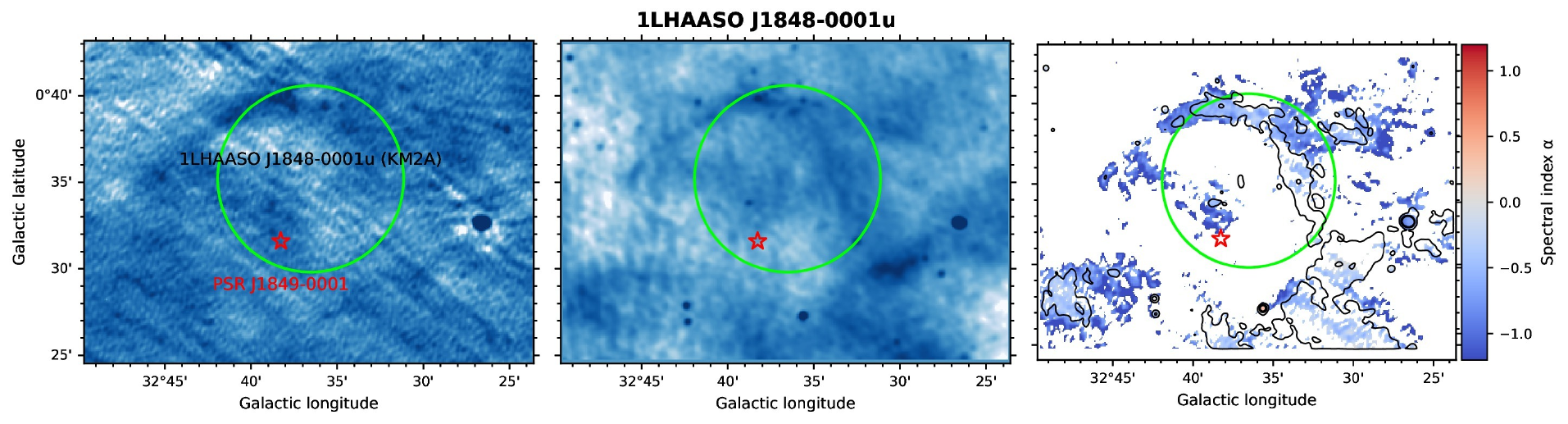}
\caption{\textit{Top:} LOFAR map of the region $32^\circ<l<38^\circ$; the black boxes correspond to the cutouts shown below and in the continued parts of this figure. \textit{Below:} tripanel LOFAR $20''$ (144~MHz), MeerKAT (1.3~GHz), and $144~\mathrm{MHz}-1.3$~GHz spectral index maps; each tripanel is labelled with the corresponding 1LHAASO source(s). This part: 1LHAASO~J1848$-$0153u and 1LHAASO~J1848$-$0001u. The radius of the 1LHAASO source regions is $r_{39}$, the 39\% containment radius of the two-dimensional Gaussian model given in Cao et al. (2024). Red stars mark pulsars discussed in the main text (positions from ATNF, Manchester et al. 2005).}
\label{fig:Q1_1}
\end{figure*}

\begin{figure*}[p]\ContinuedFloat
\centering
\includegraphics[width=\textwidth]{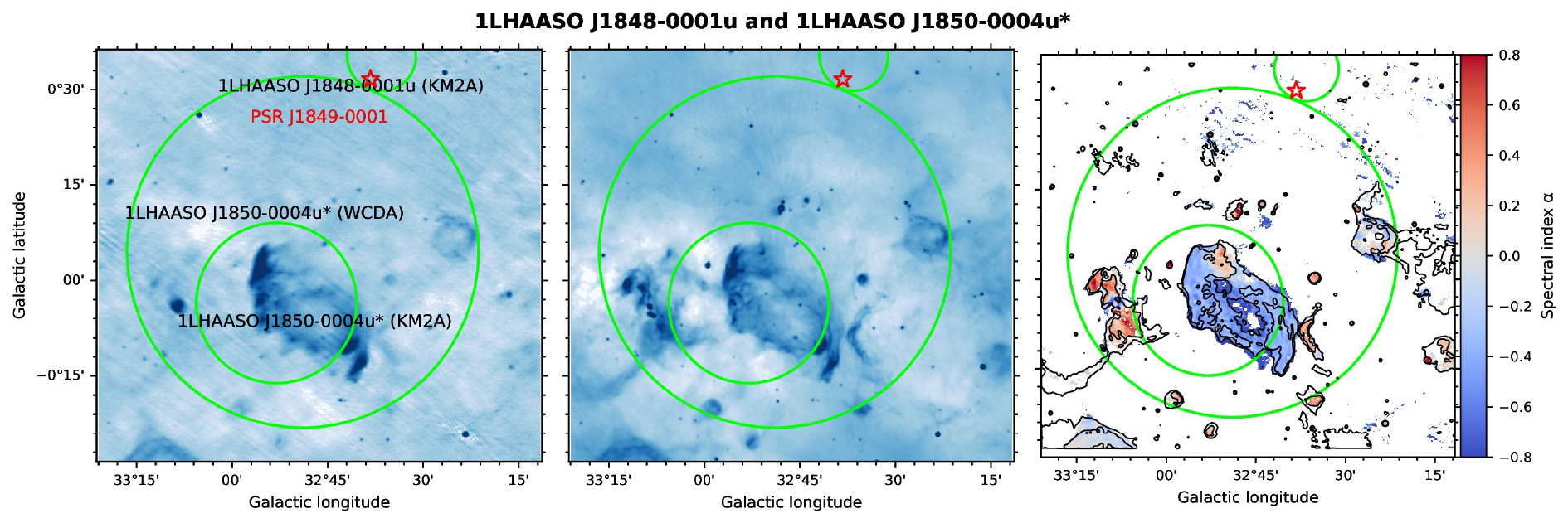}\\[3pt]
\includegraphics[width=\textwidth]{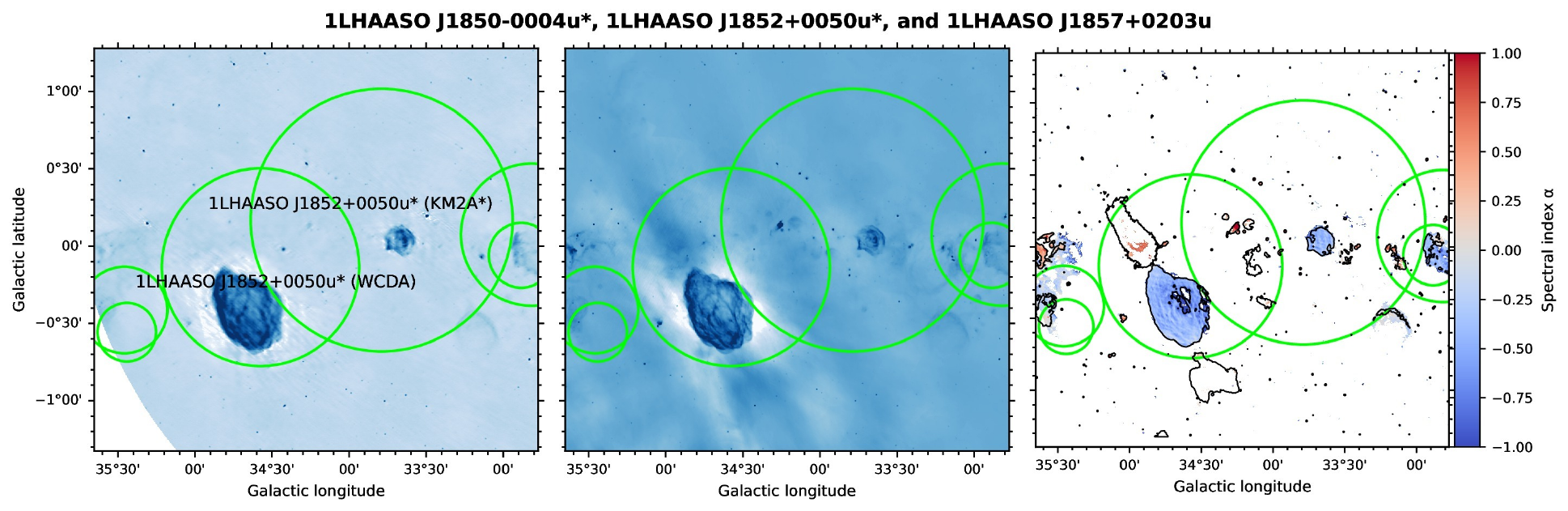}\\[3pt]
\includegraphics[width=\textwidth]{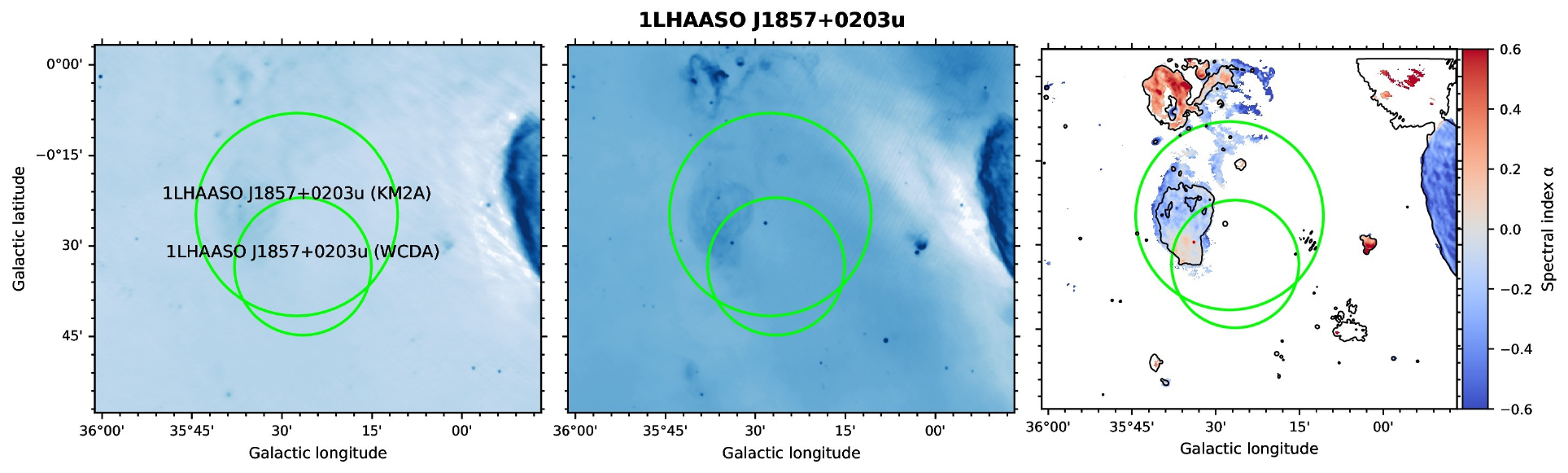}
\caption{Continued. 1LHAASO~J1850$-$0004u, 1LHAASO~J1852+0050u, and 1LHAASO~J1857+0203u.}
\end{figure*}

\begin{figure*}[p]\ContinuedFloat
\centering
\includegraphics[width=\textwidth]{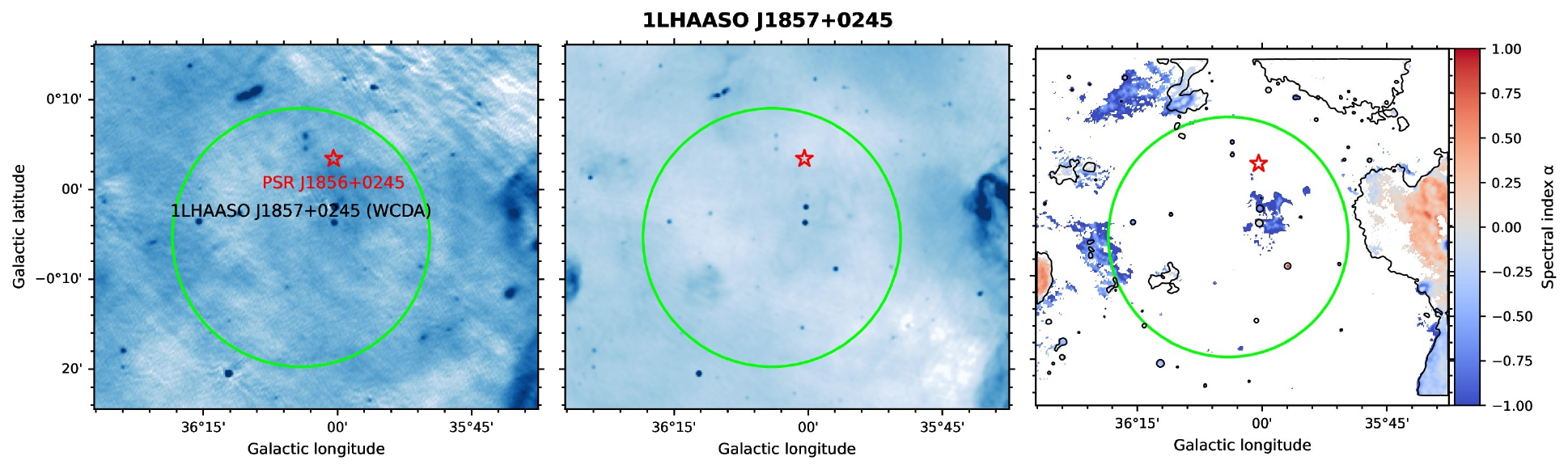}\\[3pt]
\includegraphics[width=\textwidth]{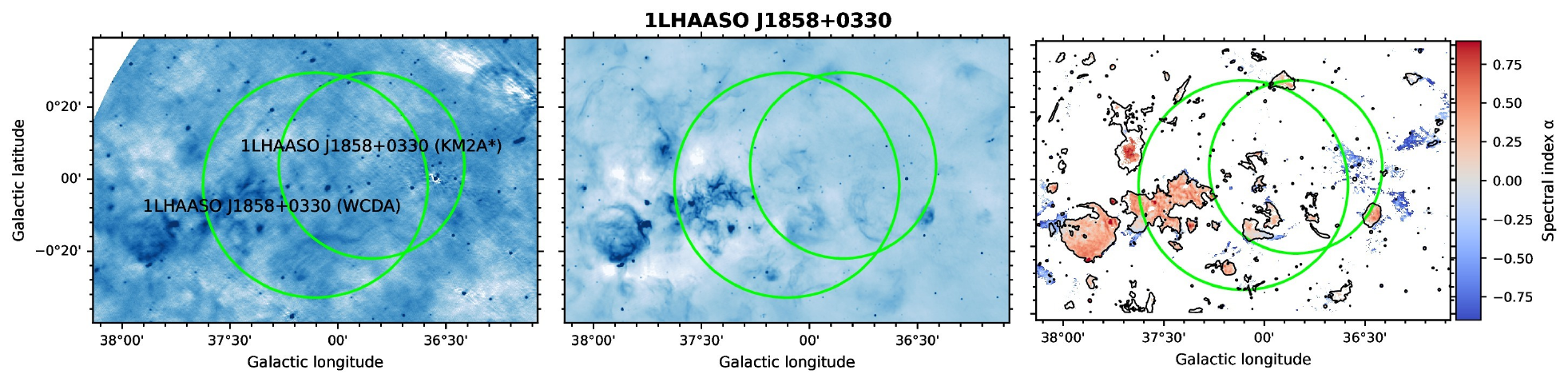}
\caption{Continued. 1LHAASO~J1857+0245 and 1LHAASO~J1858+0330.}
\end{figure*}

\begin{figure*}[p]
\centering
\includegraphics[width=0.8\textwidth]{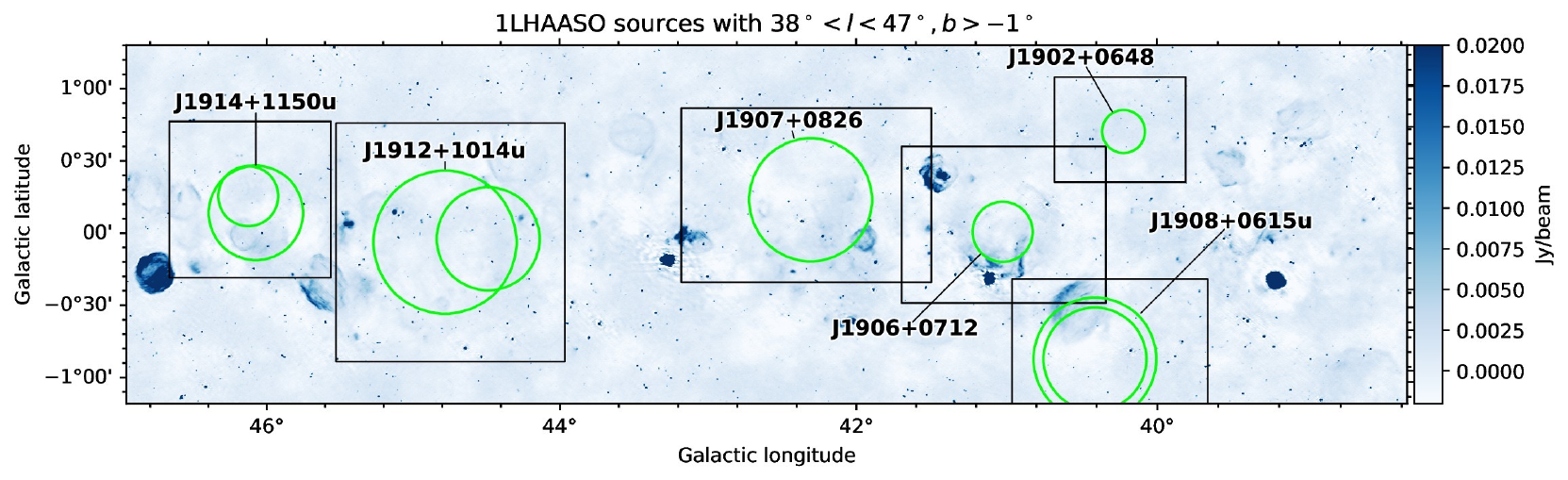}\\[4pt]
\includegraphics[width=\textwidth]{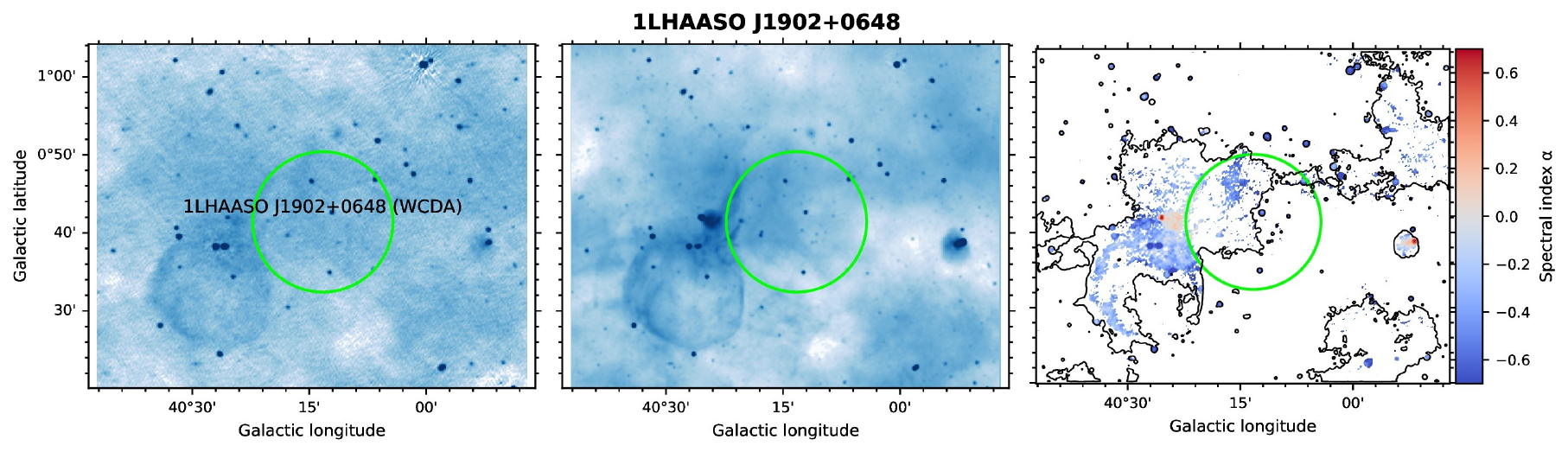}\\[3pt]
\includegraphics[width=\textwidth]{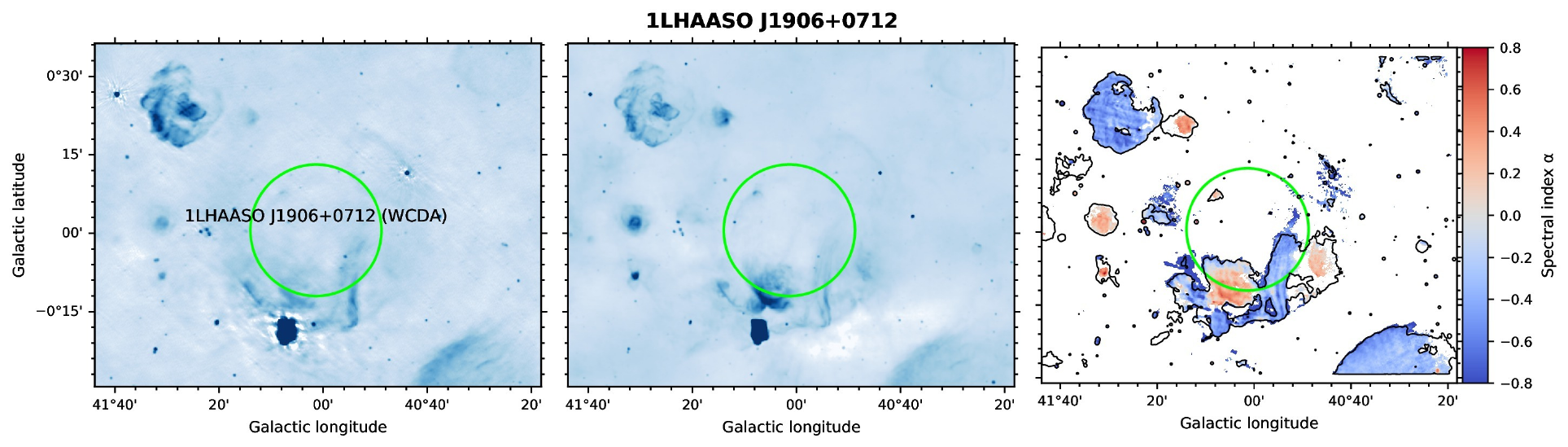}
\caption{Same as Fig. \ref{fig:Q1_1}, for the next region of the first Galactic quadrant. This part: 1LHAASO~J1902+0648 and 1LHAASO~J1906+0712. The radius of the 1LHAASO source regions is $r_{39}$, the 39\% containment radius of the two-dimensional Gaussian model given in Cao et al. (2024).}
\label{fig:Q1_2}
\end{figure*}

\begin{figure*}[p]\ContinuedFloat
\centering
\includegraphics[width=\textwidth]{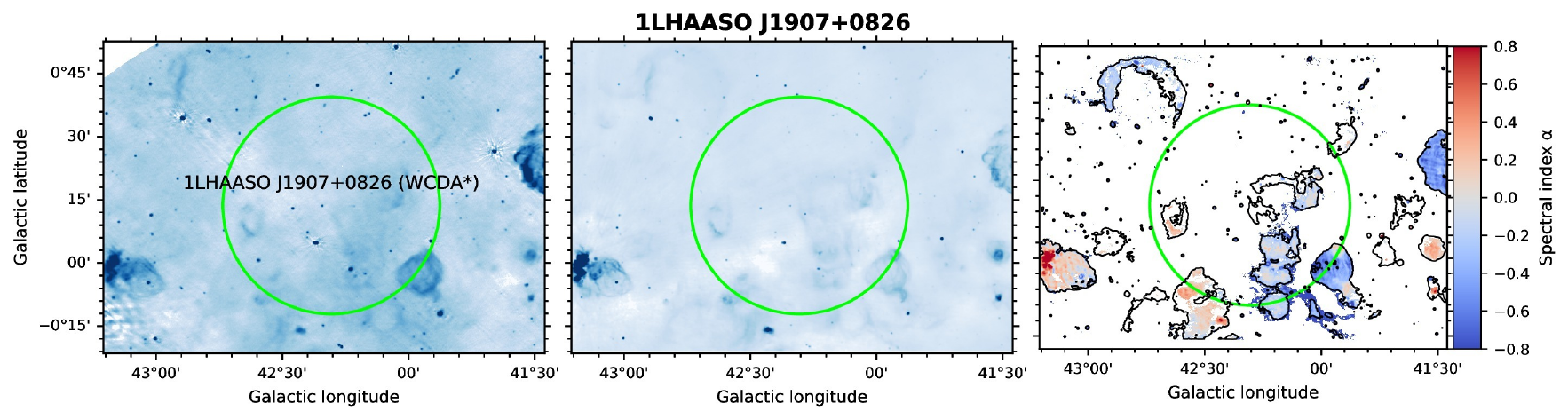}\\[3pt]
\includegraphics[width=\textwidth]{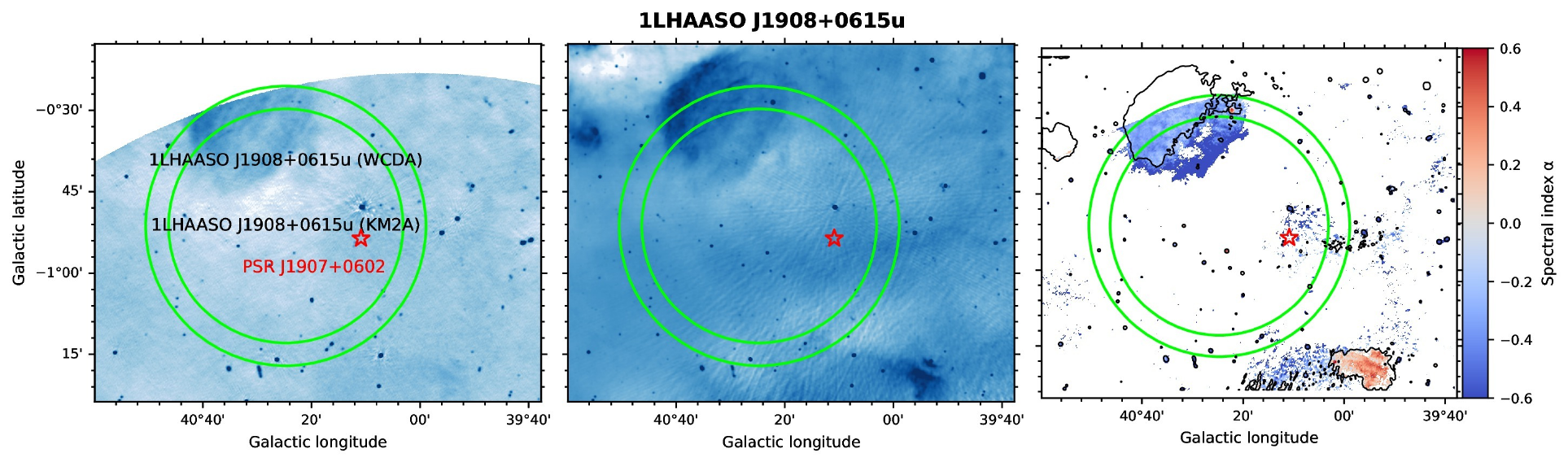}
\caption{Continued. 1LHAASO~J1907+0826 and 1LHAASO~J1908+0615u.}
\end{figure*}

\begin{figure*}[p]\ContinuedFloat
\centering
\includegraphics[width=\textwidth]{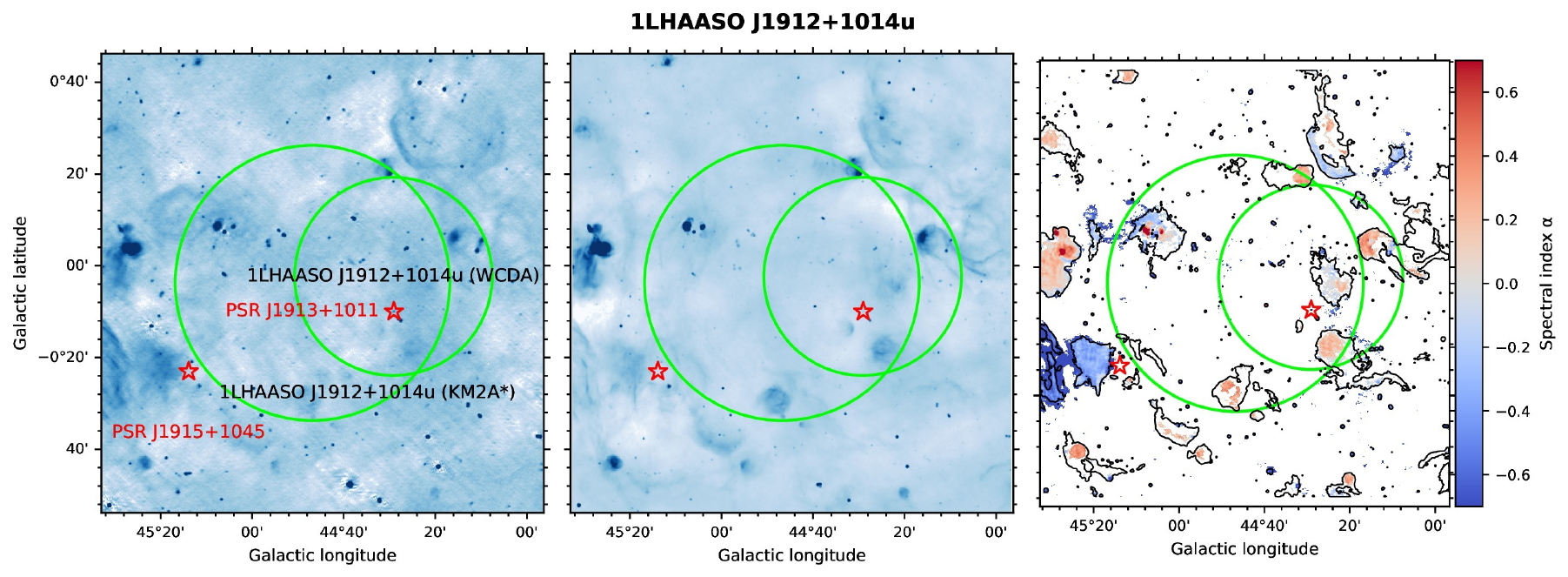}\\[3pt]
\includegraphics[width=\textwidth]{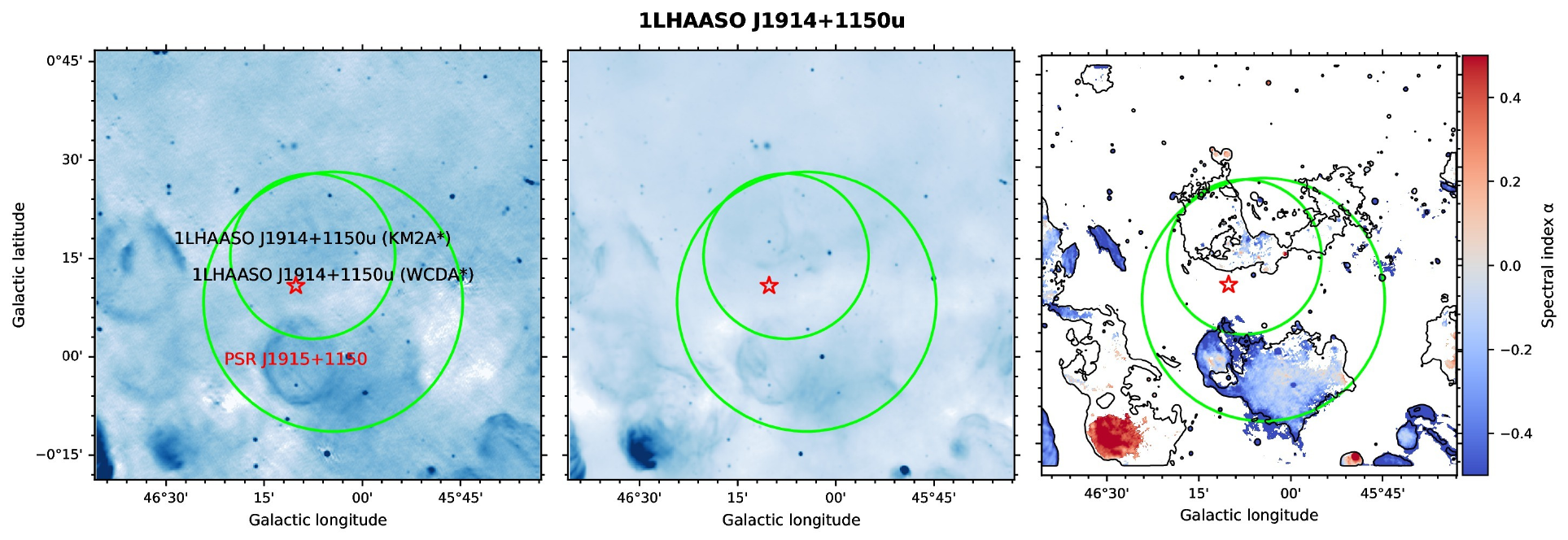}
\caption{Continued. 1LHAASO~J1912+1014u and 1LHAASO~J1914+1150u.}
\end{figure*}

\begin{figure*}[p]
\centering
\includegraphics[width=0.8\textwidth]{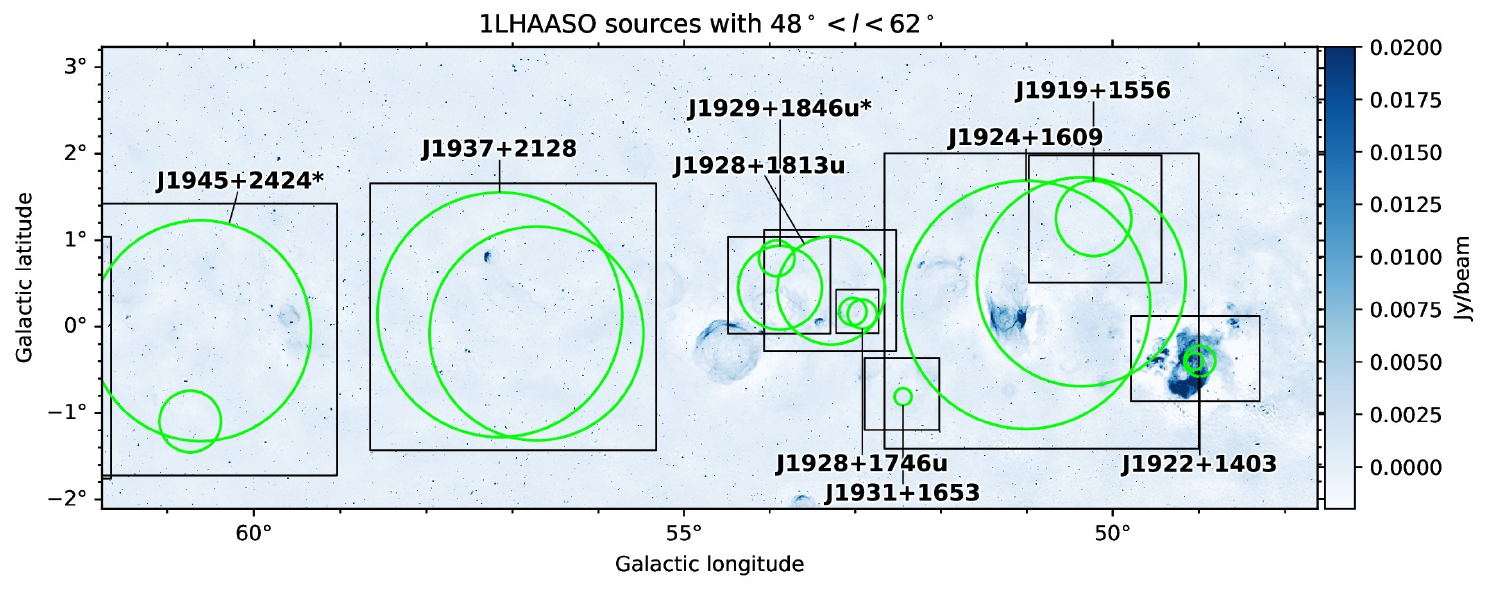}\\[4pt]
\includegraphics[width=\textwidth]{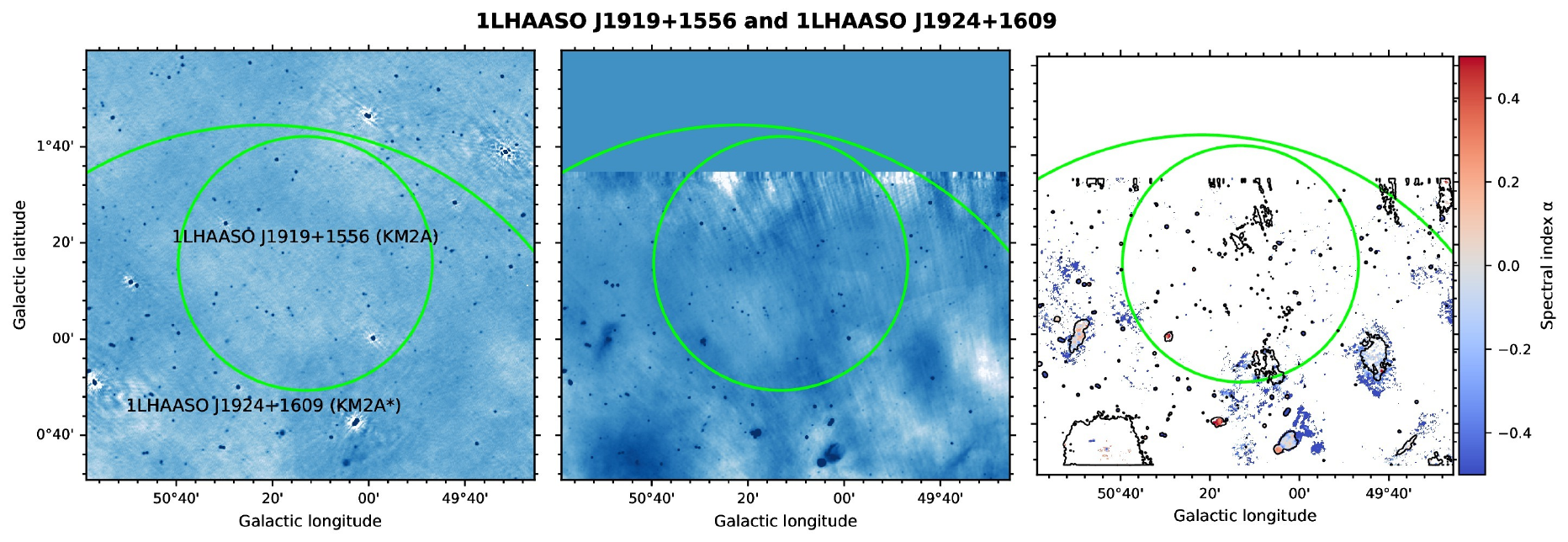}\\[3pt]
\includegraphics[width=\textwidth]{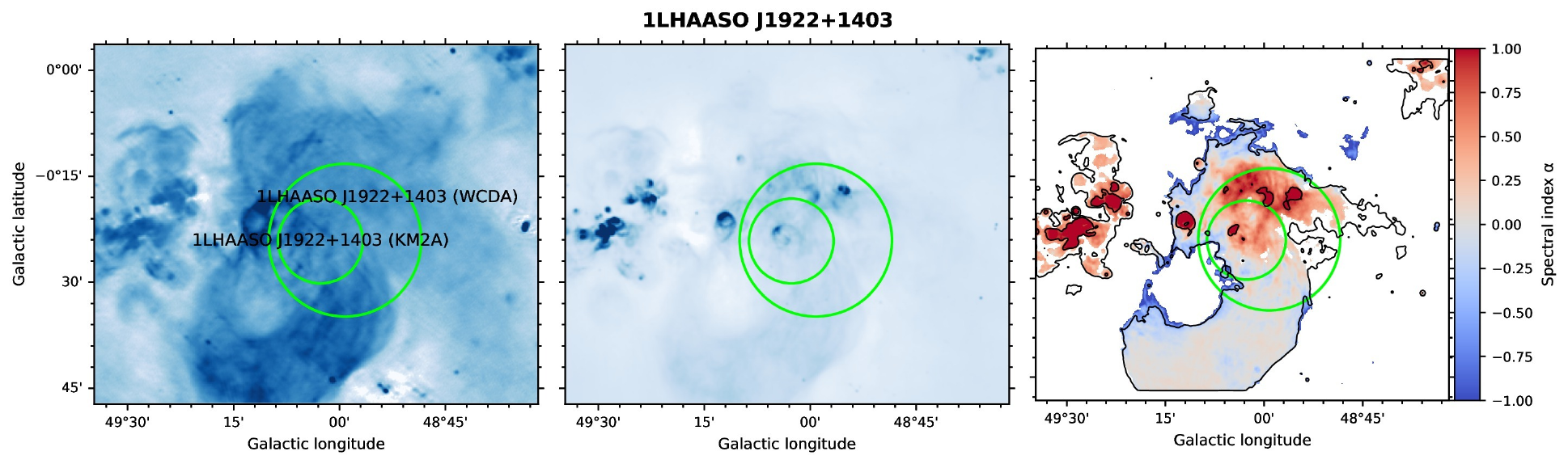}
\caption{Same as Fig. \ref{fig:Q1_1}, for the next region of the first Galactic quadrant. This part: 1LHAASO~J1919+1556 and 1LHAASO~J1922+1403. The radius of the 1LHAASO source regions is $r_{39}$, the 39\% containment radius of the two-dimensional Gaussian model given in Cao et al. (2024).}
\label{fig:Q1_3}
\end{figure*}

\begin{figure*}[p]\ContinuedFloat
\centering
\includegraphics[width=\textwidth]{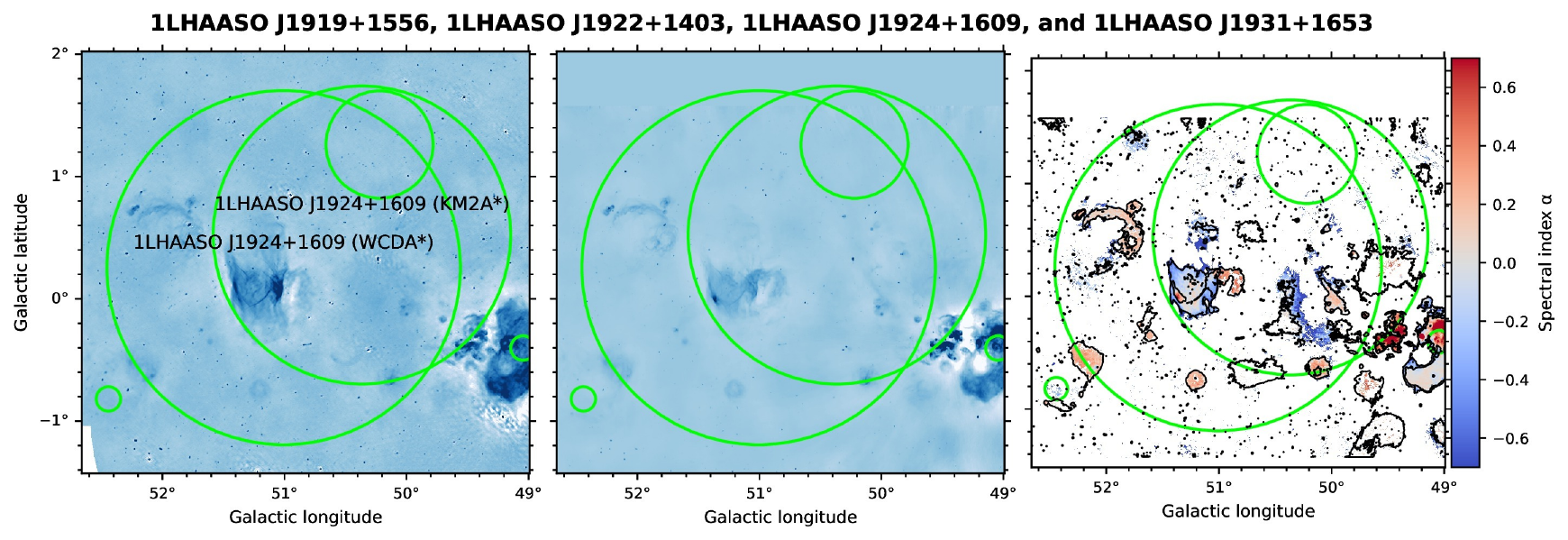}\\[3pt]
\includegraphics[width=\textwidth]{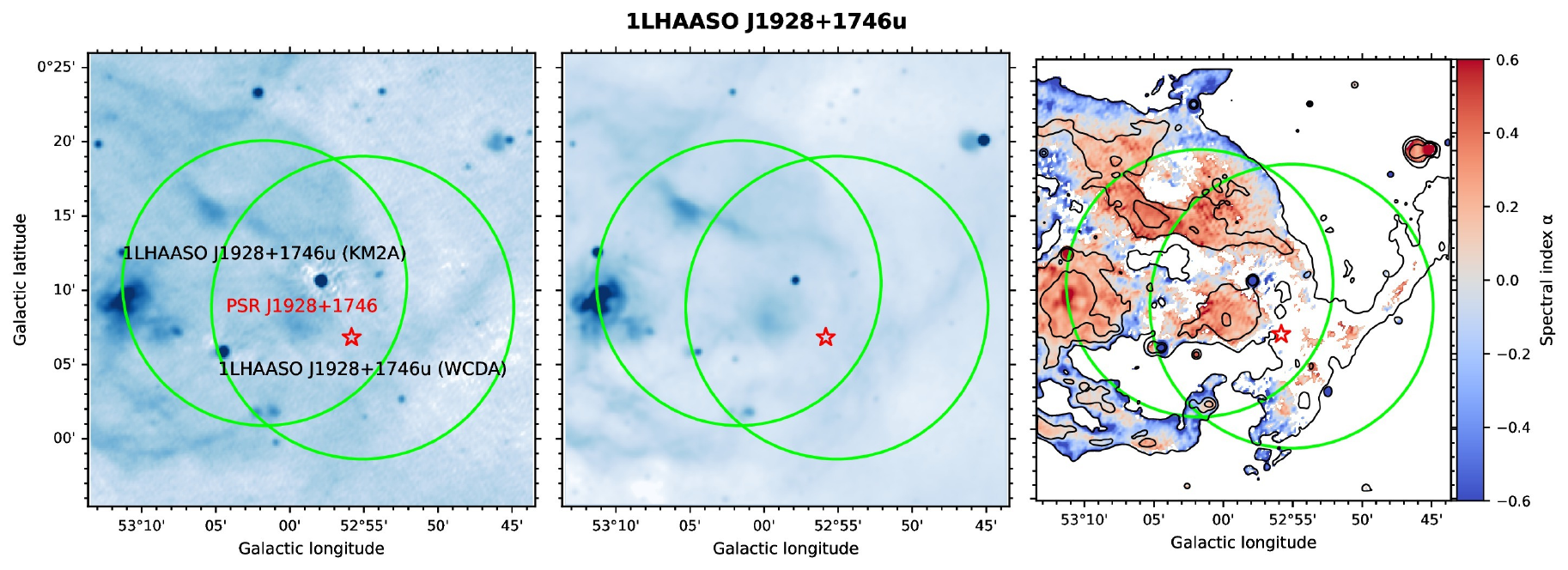}\\[3pt]
\includegraphics[width=\textwidth]{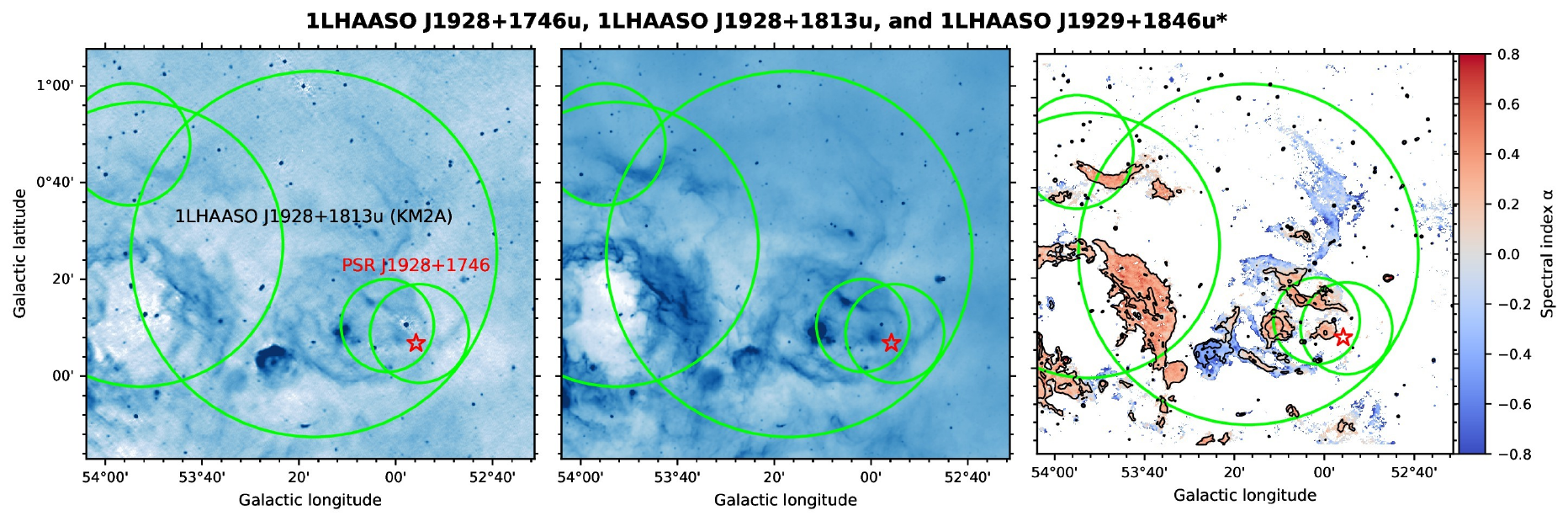}
\caption{Continued. 1LHAASO~J1924+1609, 1LHAASO~J1928+1746u, and 1LHAASO~J1928+1813u.}
\end{figure*}

\begin{figure*}[p]\ContinuedFloat
\centering
\includegraphics[width=\textwidth]{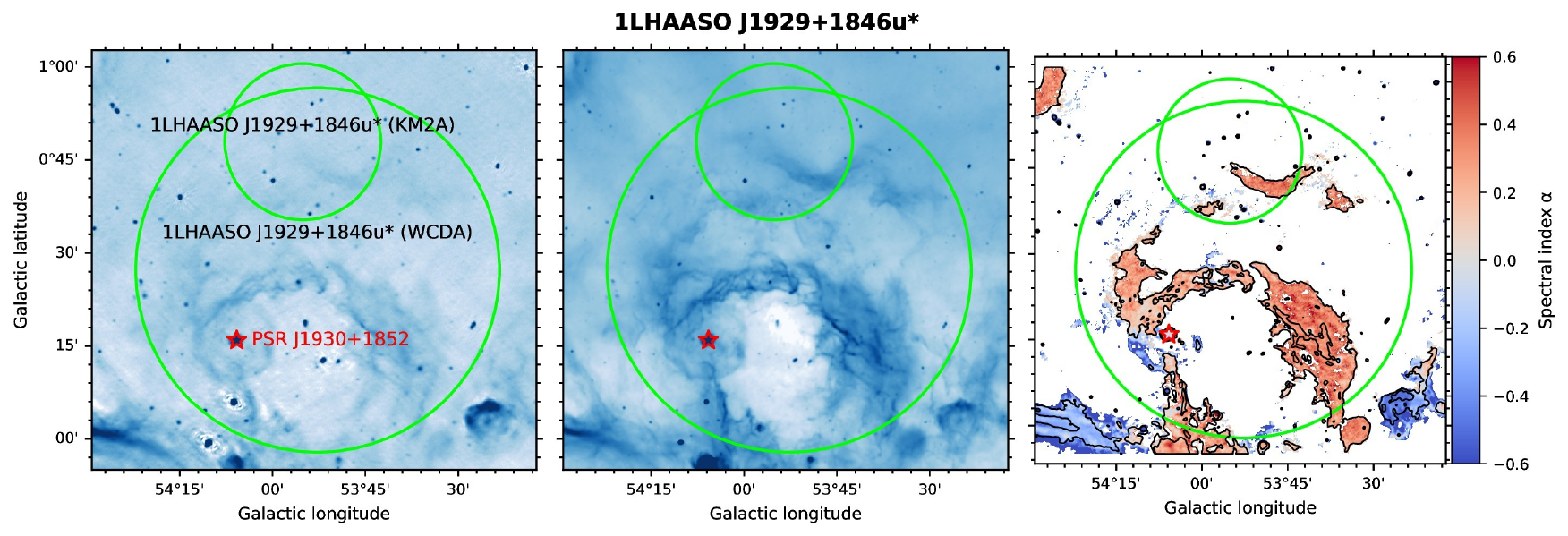}\\[3pt]
\includegraphics[width=\textwidth]{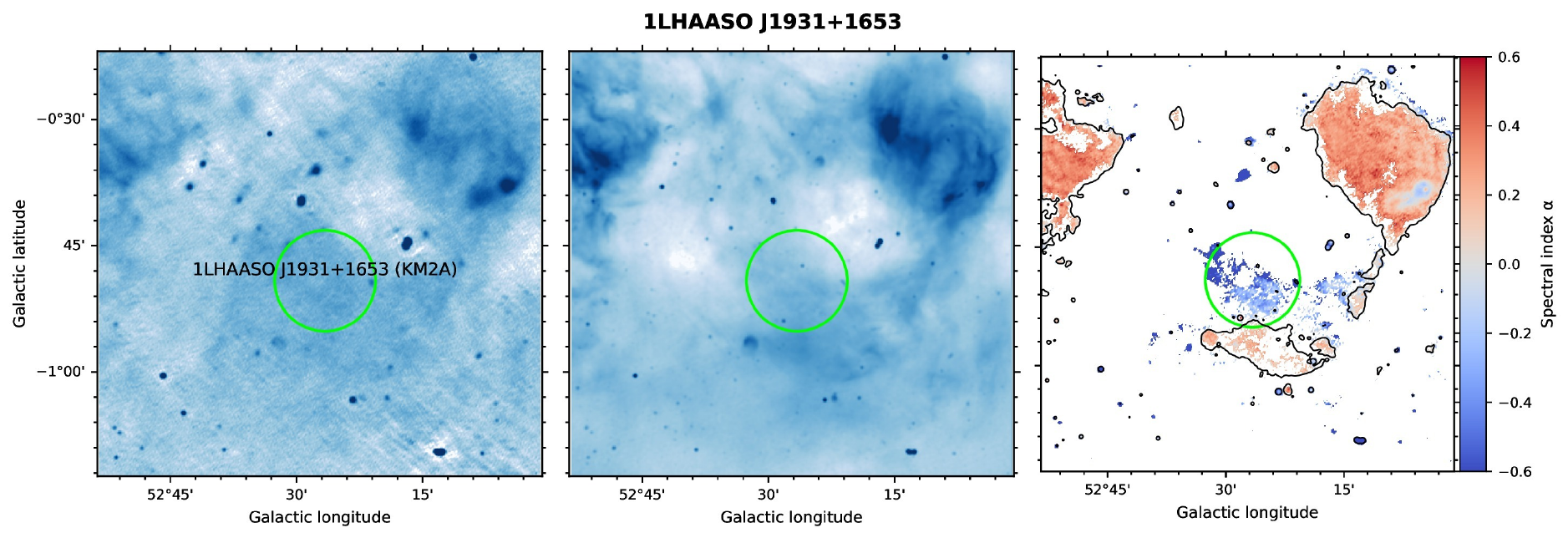}
\caption{Continued. 1LHAASO~J1929+1846u and 1LHAASO~J1931+1653.}
\end{figure*}

\begin{figure*}[p]\ContinuedFloat
\centering
\includegraphics[width=\textwidth]{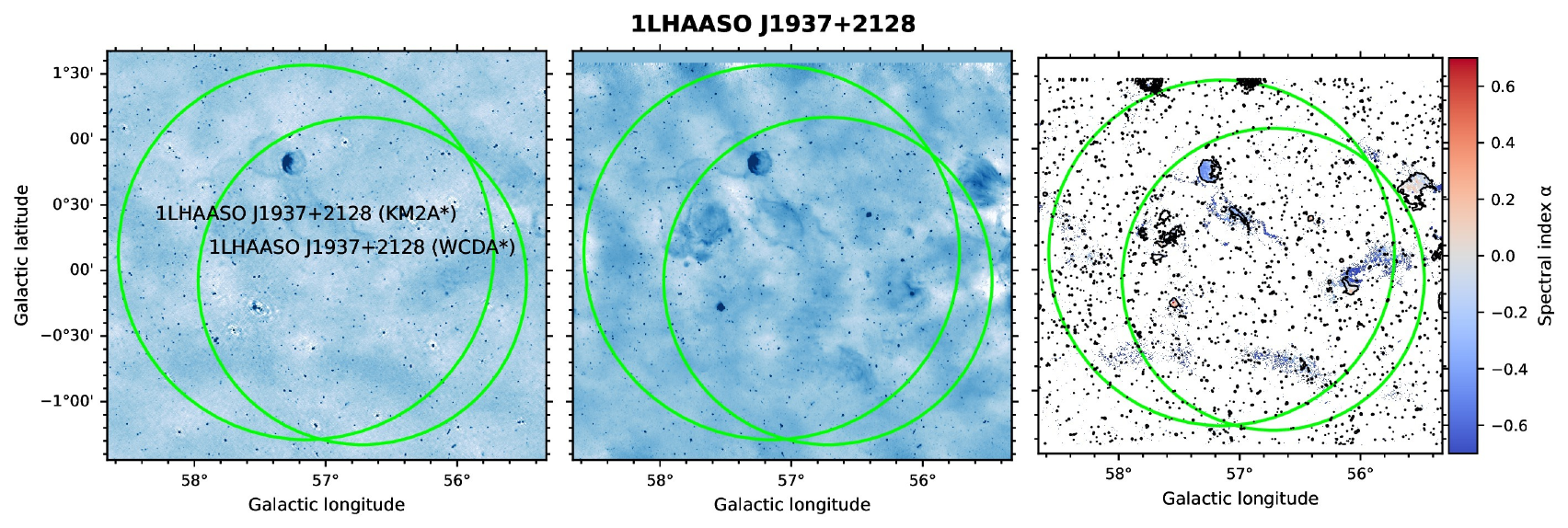}\\[3pt]
\includegraphics[width=\textwidth]{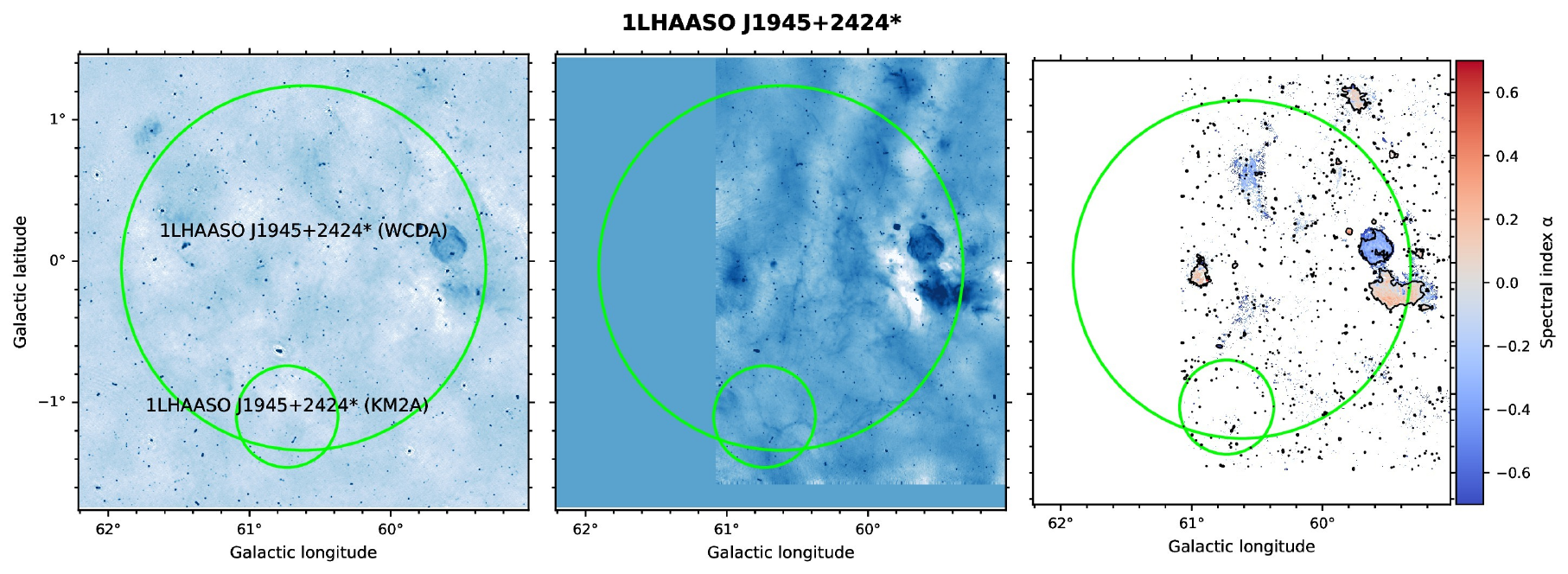}
\caption{Continued. 1LHAASO~J1937+2128 and 1LHAASO~J1945+2424.}
\end{figure*}

\begin{figure*}[p]
\centering
\includegraphics[width=0.8\textwidth]{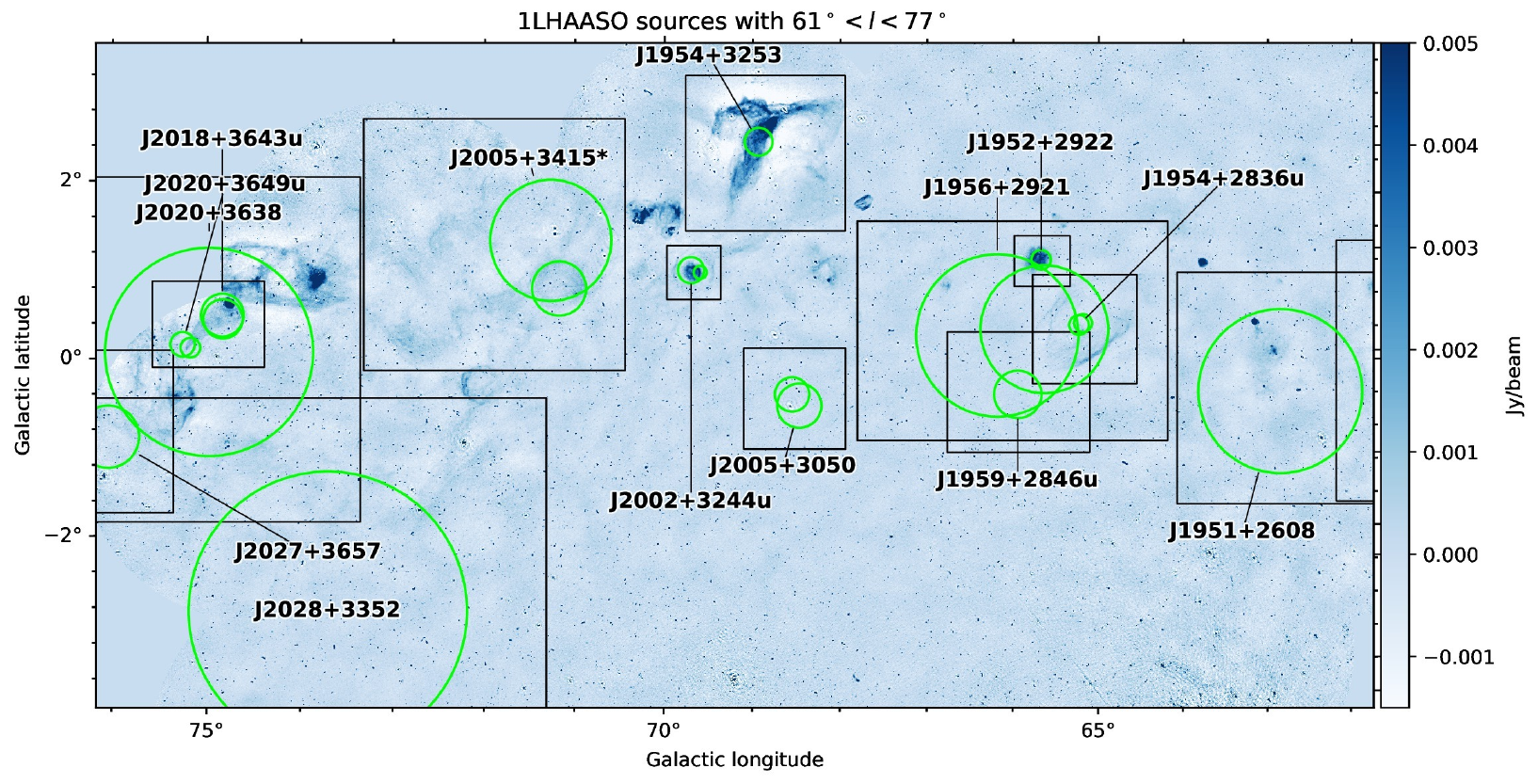}\\[4pt]
\includegraphics[width=0.32\textwidth]{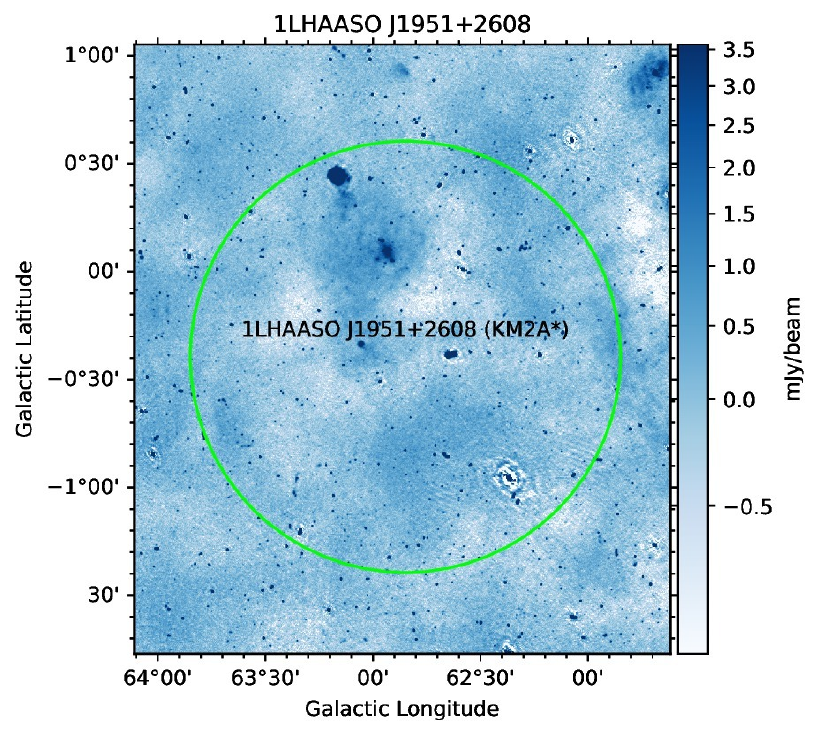}\hfill
\includegraphics[width=0.32\textwidth]{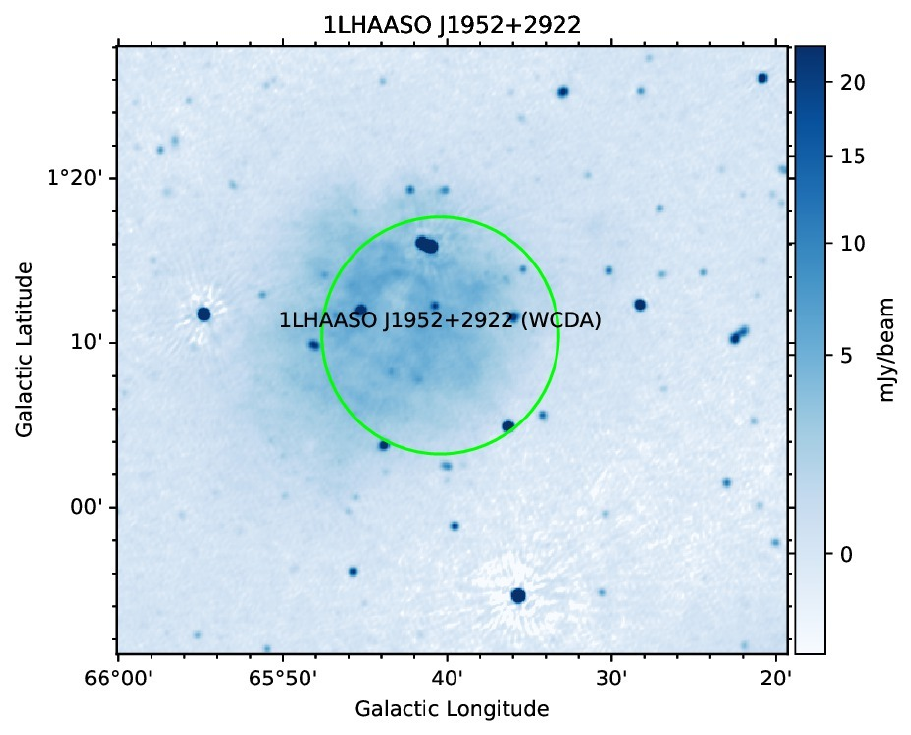}\hfill
\includegraphics[width=0.32\textwidth]{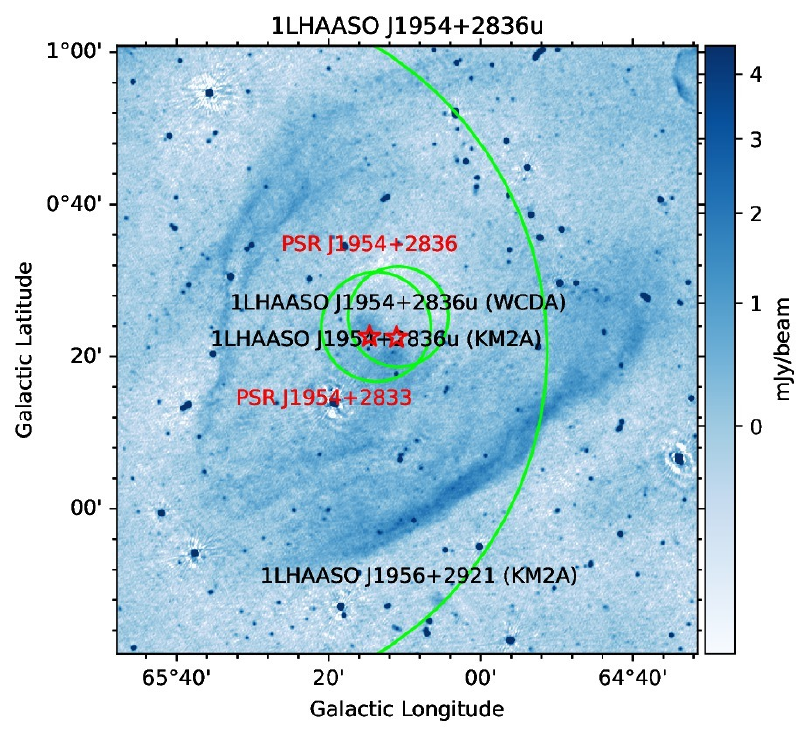}\\[3pt]
\includegraphics[width=0.32\textwidth]{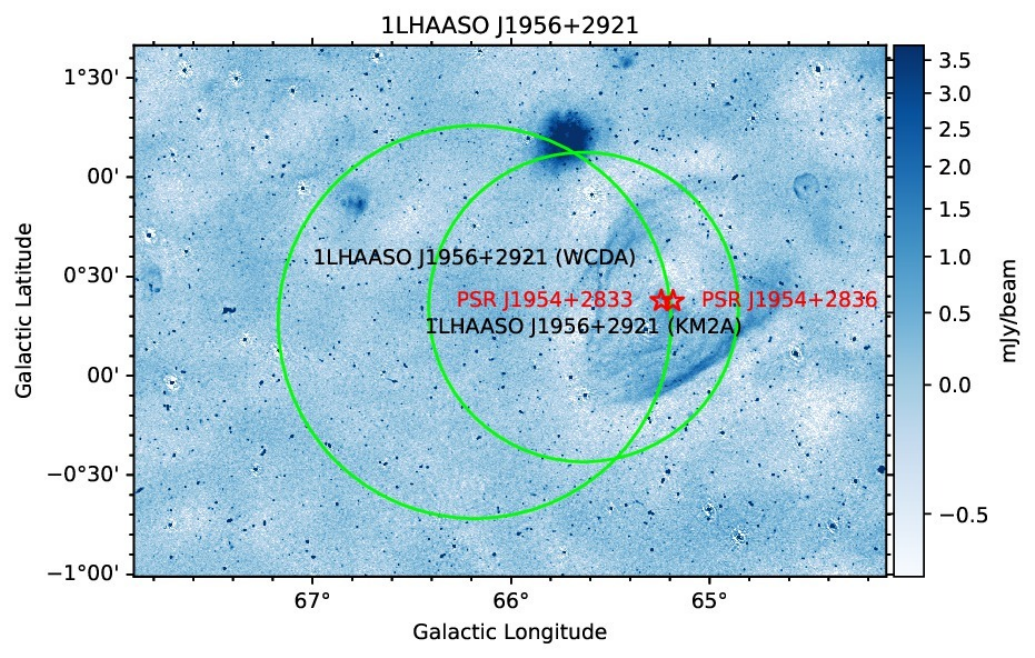}\hfill
\includegraphics[width=0.32\textwidth]{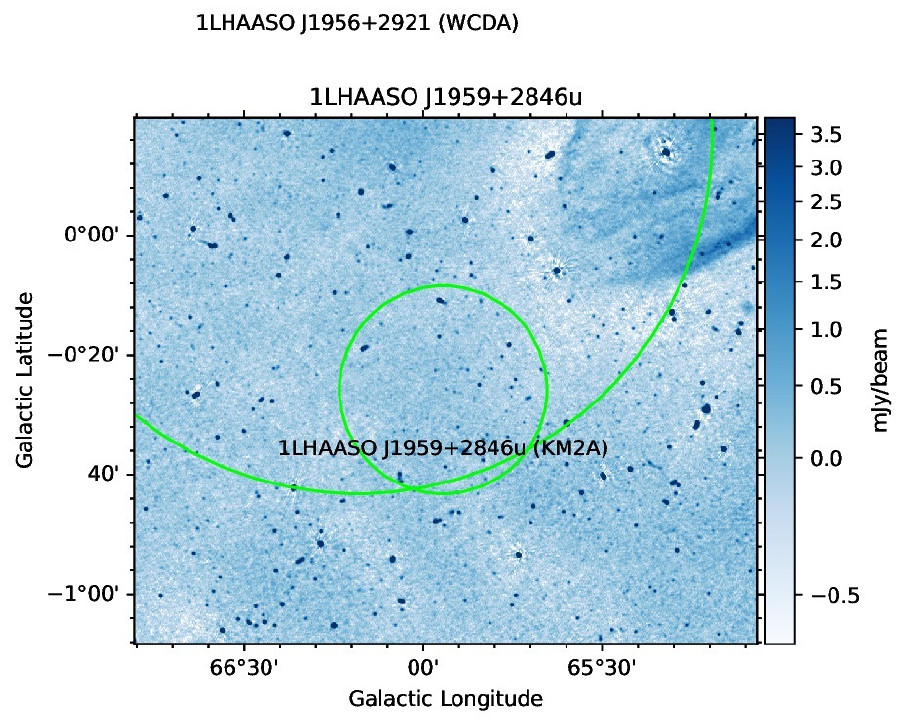}\hfill
\includegraphics[width=0.32\textwidth]{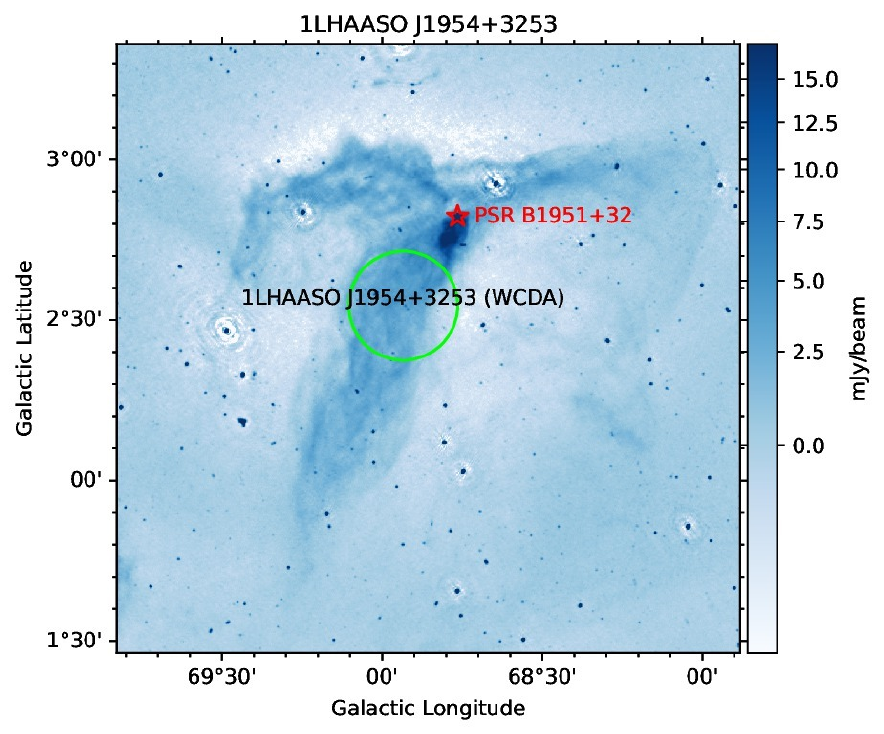}
\caption{\textit{Top:} LOFAR map of the region around the 1LHAASO sources with no SMGPS coverage; the black boxes correspond to the cutouts shown below and in the continued part of this figure. \textit{Below}, LOFAR $20''$ (144~MHz) cutouts of, in reading order:
1LHAASO~J1951+2608;
1LHAASO~J1952+2922;
1LHAASO~J1954+2836u;
1LHAASO~J1956+2921;
1LHAASO~J1959+2846u; and
1LHAASO~J1954+3253. The radius of the 1LHAASO source regions is $r_{39}$, the 39\% containment radius of the two-dimensional Gaussian model given in Cao et al. (2024). Red stars mark pulsars discussed in the main text (positions from ATNF, Manchester et al. 2005).}
\label{fig:Q1_4}
\end{figure*}

\begin{figure*}[p]\ContinuedFloat
\centering
\includegraphics[width=0.32\textwidth]{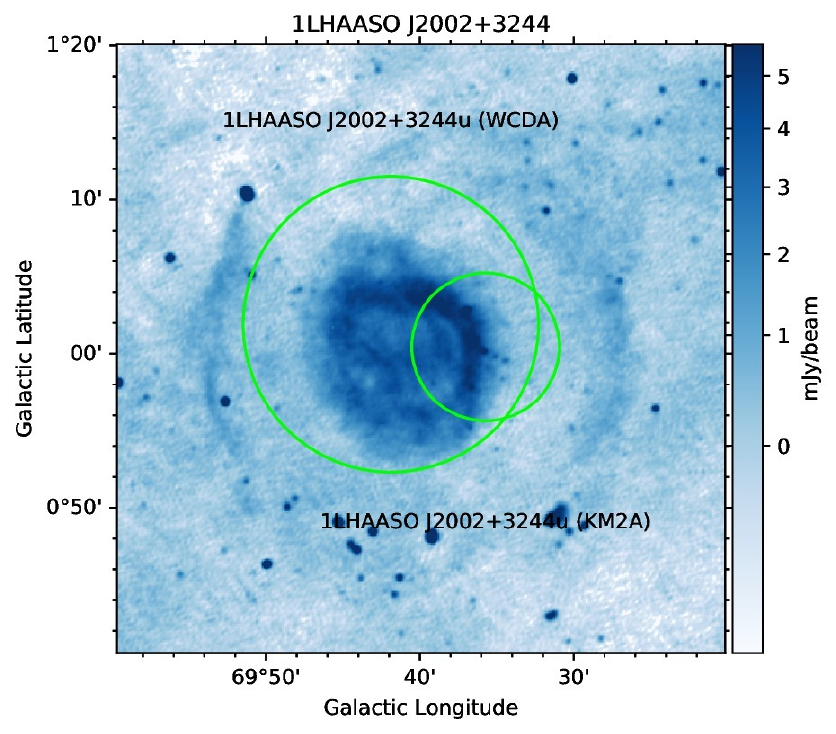}\hfill
\includegraphics[width=0.32\textwidth]{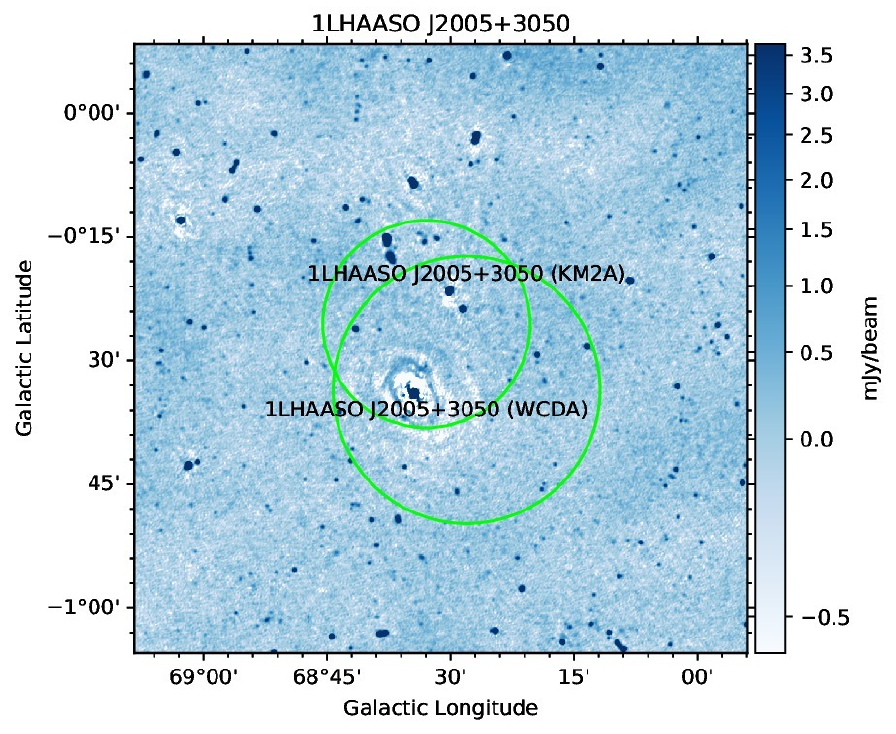}\hfill
\includegraphics[width=0.32\textwidth]{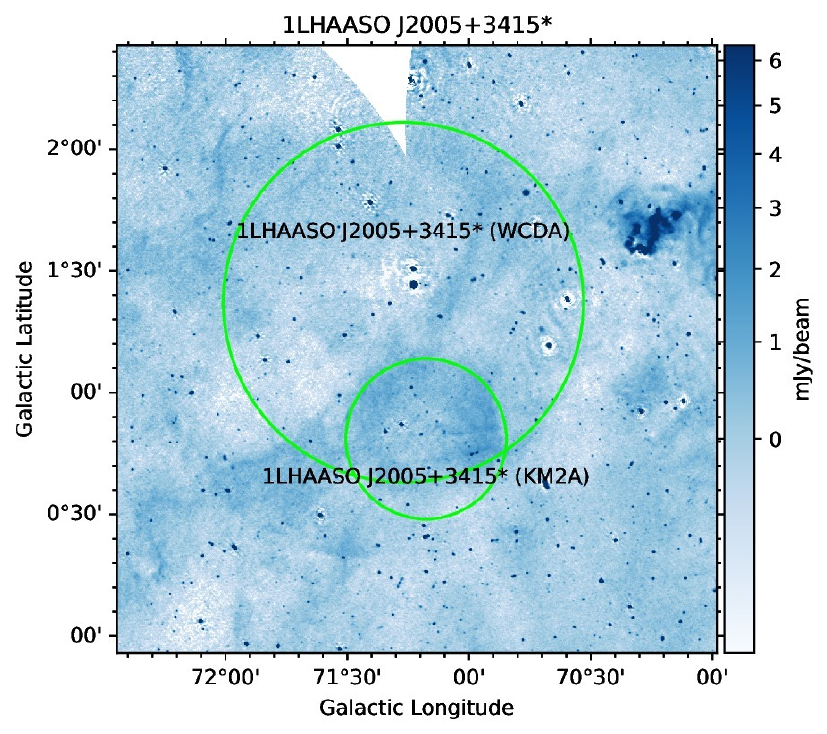}\\[3pt]
\includegraphics[width=0.32\textwidth]{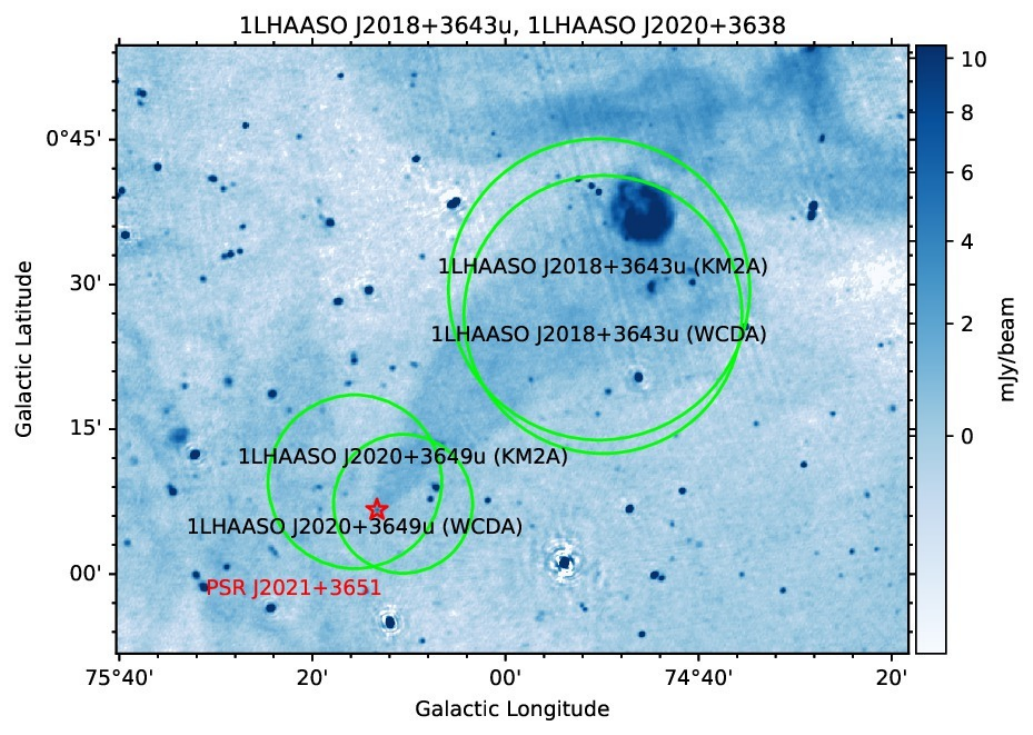}\hfill
\includegraphics[width=0.32\textwidth]{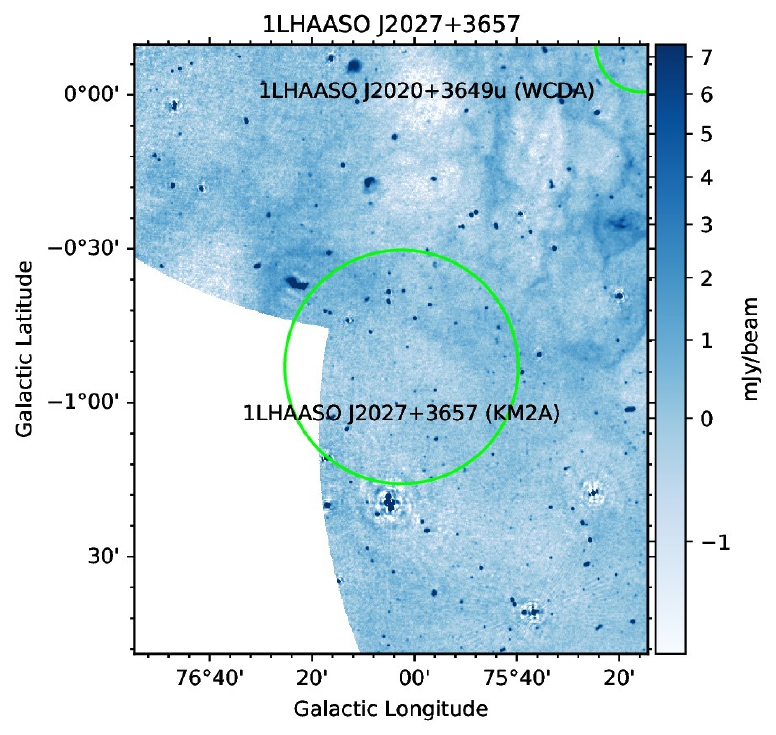}\hfill
\includegraphics[width=0.32\textwidth]{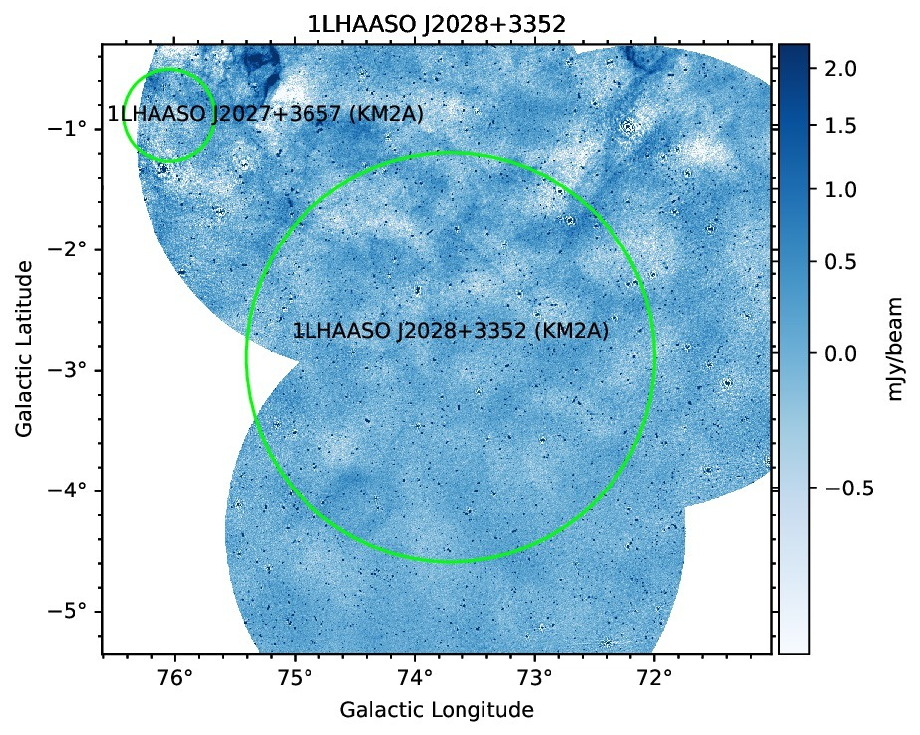}
\caption{Continued.
1LHAASO~J2002+3244;
1LHAASO~J2005+3050;
1LHAASO~J2005+3415;
1LHAASO~J2018+3643u, shown together with 1LHAASO~J2020+3649u and 1LHAASO~J2020+3638, which fall in the same field and are labelled on the panel;
1LHAASO~J2027+3657; and
1LHAASO~J2028+3352.}
\end{figure*}


\end{appendix}

\end{document}